# How the Voynich Manuscript was created

Torsten Timm
torsten.timm@kereti.de

**Abstract:** The Voynich manuscript is a medieval book written in an unknown script. This paper studies the relation between similarly spelled words in the Voynich manuscript. By means of a detailed analysis of similar spelled words it was possible to reveal the text generation method used for the Voynich manuscript.

## 1 Introduction

The Voynich manuscript, rediscovered by Wilfrid Voynich in 1912, contains a text in an unknown script (see figure 1) as well as exotic illustrations of unidentifiable plants, cosmological charts, astronomical symbols and bathing women. The author, the purpose and the origin of the manuscript are still unknown. The manuscript consists of 240 parchment pages. The parchment was carbon-dated to the early 15th century [see Stolte]. The style of the illustrations is that of medieval Europe. The script uses 20-30 different glyphs. The exact number is uncertain since it is unclear whether some of the glyphs are distinct characters or a ligature of two other characters. The text is written from left to right and apparently divided by spaces into word tokens. For a more detailed introduction see the paper "What We Know About The Voynich Manuscript" by Reddy and Knight [see Reddy].

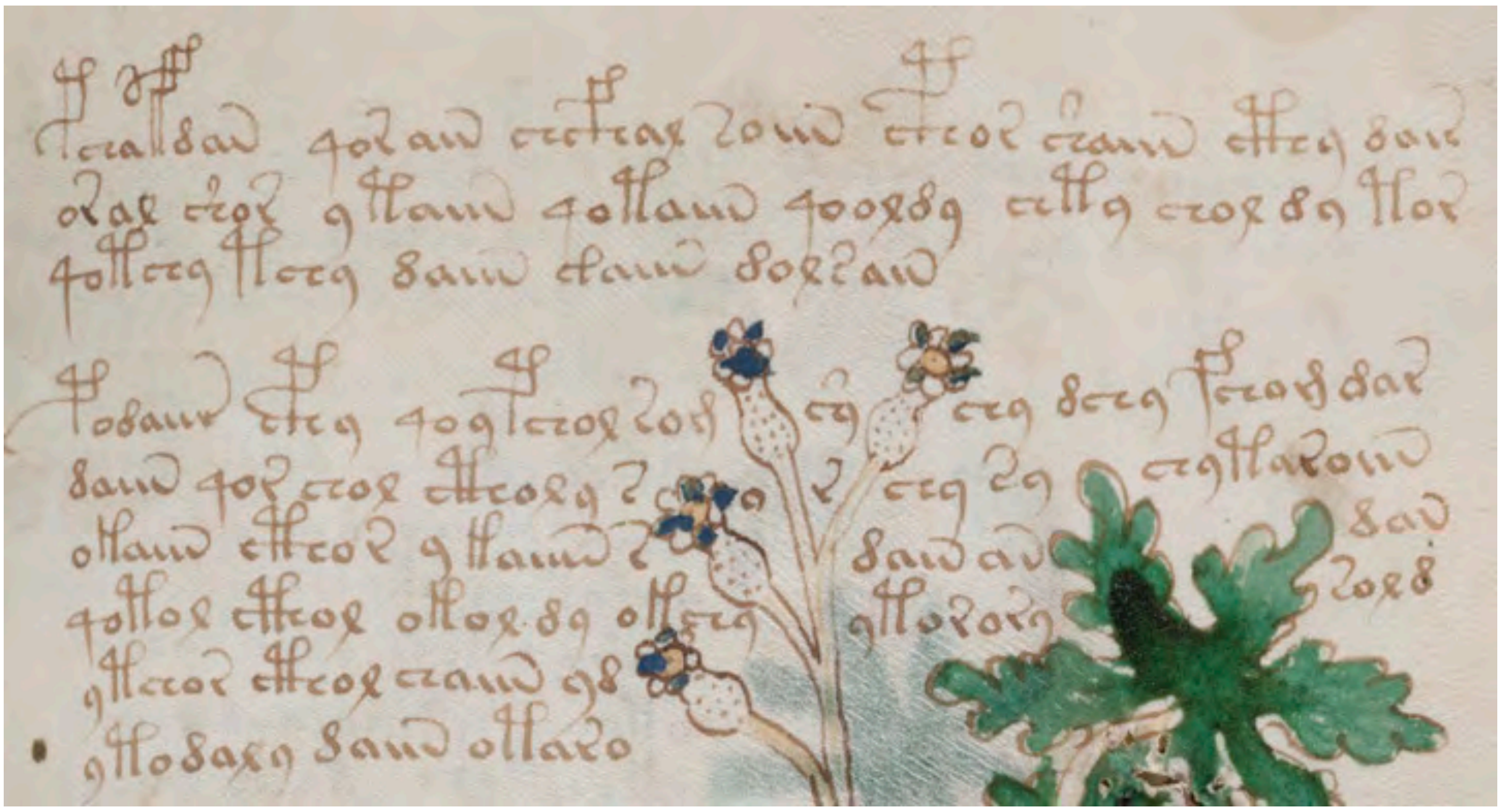

Figure 1: part of page <f36r>[1]

[1] High-resolution scans are available at http://www.jasondavies.com/voynich/

The Voynich manuscript, also referred to as the VMS, raises several unanswered questions. The main question is whether the text within the manuscript contains a message in an unknown or constructed language, an encoded message using an unknown cipher system or whether it is a pseudo text containing no message at all. Some statistical features of the text correspond to a natural language whereas others do not. For instance, the number of different glyphs and their frequency distribution indicate an alphabetical script. The word frequency distribution also behaves as expected according to Zipf's law. This law was proposed by Georg Zipf in 1935 and makes predictions about the word frequencies for a text using natural language. For instance, it predicts that the second most frequent word will occur approximately half as frequently as the most frequent word. The word length distribution in the manuscript follows a binomial distribution with an underrepresentation of short and long words, an unusual characteristic in a natural language.[2] One possible explanation for this is, indeed, that the manuscript contains text in a constructed language.[3] It could, however, feasibly be argued that the manuscript does contain a natural language but that short words were not always given their own boundaries and long words are rare because vowels were omitted.

Although the text reveals numerous semantic patterns, no one has been able to read it or to decipher the script in the last hundred years. One hypothesis, therefore, is that the VMS is a medieval hoax containing a meaningless text. However, there are some good reasons for refuting such a hypothesis. It seems unlikely that anybody in the 15th century would have had the knowledge to simulate numerous language-like patterns. Furthermore, there would have been no need for a medieval forger to simulate statistical patterns of which nobody at the time was aware.

## 2 Repeated word sequences

An important initial task when deciphering an unknown script is to describe the most characteristic properties and to determine interesting patterns. Gunther Ipsen described this task in 1954 as:

> "Deciphering always begins with a determination of the object. This is initially a simple description of what is.

[2] The length of the words is equally distributed around the arithmetical mean [see Stolfi]. The average word length is 5.5. Such a distribution is unusual because natural languages tend to make frequent use of short words. Typical for a natural language is therefore an asymmetric distribution.

[3] The cryptographer William F. Friedman (1891-1969) suggested: "The Voynich MSS was an early attempt to construct an artificial or universal language of the a priori type."

> If this description is conducted with care and attention and observes not only that which is close and obvious but also the small and delicate features of the object, then this description becomes a determination which recognizes the individual element as something distinctive; it penetrates." [Ipsen: p. 421, own translation].[4]

The text of the VMS seems to be unique because repetitive phrases are missing. On the one hand, it is possible to describe adjacency rules on the letter level. There are, for instance, typical letter combinations but also letter combinations that never occur. On the other hand, there are no similar rules on the word level. Only a few repetitive phrases can be found. There are only 35 word sequences which use at least three words and appear at least three times.[5] Only for five of these sequences is the word order unchanged for the whole manuscript, whereas for 30 out of 35 phrases the word order does change.[6] Does this mean that it is simply coincidence that these five sequences did not appear in a different order? An additional observation is that in 24 out of 35 cases these repeated sequences use at least two words which are either spelled the same or very similarly. Does that mean that they are only repeated because they use similar words? For instance, the words ("chol")[7], ("shol") and ("cthol") occur together three times.[8] Each time, the word order is different:

<f1v.P.6> ("chol cthol shol")

<f4r.P.2> ("chol shol cthol")

<f42r.P2.10> ("cthol chol shol")

In a text using human language grammatical relations should exist between words, and these relations should result in words used together multiple times. Therefore, the lack of repetitive phrases is surprising for a whole book containing more than 37,000 words. Is it possible to find an explanation for the observed weak word order in the VMS?

---

4 Original: "Jede Entzifferung hebt mit der Bestimmung ihres Gegenstandes an. Das ist zunächst schlichte Beschreibung dessen, was ist. Wenn dies Beschreiben sorgfältig und aufmerksam erfolgt, daß es nicht nur das Nächste bemerkt, das offen liegt, sondern auch die feinen und kleinen Züge am Sosein des Gegenstandes, dann geht die Beschreibung in Bestimmung über, die das Einzelne als Besonderes erkennt; sie wird eindringlich." [Ipsen: p. 421].

5 Transcription by Takeshi Takahashi [Takahashi].

6 See addendum: I. Repeated sequences using the same words (p. 44).

7 By using the EVA alphabet it is possible to write the VMS-word as "chol". The EVA alphabet, created by René Zandbergen and Gabriel Landini, can be used to analyse the text and to name the VMS-words [see Zandbergen]. The letters do not give any information about the meaning of the corresponding VMS glyph.

8 Two more similar phrases exist: ("chor shor cphor") in line <f17r.P.5> and ("chol cphol shol") in line <f100r.P2.6>. This observation fits the hypothesis that phrases are repeated because they contain words similar to each other.

## 3 A word grid

A closer look at the VMS gives the impression that similarly spelled words occur frequently above each other.[9] Does this mean that there is a relation between words spelled similarly? The starting point for the following considerations is the examination of the connection between words spelled similarly. The most frequently used words are ("daiin" occurring 863 times), ("ol" occurring 537 times) and ("chedy" occurring 501 times). Hence, it is possible to group the words according to their similarity and frequency:

-series

daiin (863 times) | aiin (469 times) | dain (211 times) ...
qokain (279) | qokaiin (262) | okaiin (212) | otaiin (154) ...

-series

ol (537) | or (363) | dol (117) | dor (73) ...
ar (350) | dar (318) | al (260) | dal (253) ...
chol (396) | chor (219) | cheol (172) | cheor (100) ...

-series

chedy (501) | chey (344) | cheey (174) | chy (155) ...
shedy (426) | shey (283) | sheey (144) | shy (104) ...
qokeedy (305) | qokedy (272) | otedy (155) | okedy (118) ...
qokeey (308) | okeey (177) | qoky (147) | qokey (107) ...

At first glance, it seems possible to describe at least two different groups of glyphs. The first group contains glyphs handled like vowels, while the second group of glyphs is handled like consonants:

I) (o, a, e, i, y)

II) (ch, sh, n, r, s, l, d, m, q, k, t, p, f)

An interesting observation is that glyphs from group II are rarely consecutive. There is at least one exception to this rule. Before and after ("l") glyphs from group II are also allowed.

In most cases, similarly spelled variants of a glyph group exist. Sometimes the only difference between two glyph groups is an additional quill stroke. This is the case, for instance, for ("daiin") and ("dain"). In other cases, similarly shaped glyphs replace each other.[10] One example of such a case is ("okaiin") and ("otaiin"). It is interesting that for ("k") and ("t") two other similarly shaped glyphs occur within the VMS. These glyphs are ("f") and ("p").[11] Surprisingly, words spelled similarly to and and containing and such as ("ofaiin") and ("opaiin") also occur within the VMS. This raises the question as to whether there is a connection between similarly spelled words?

[9] Example: See , and above each other in figure 1 (p. 1) lower left half.

[10] See also the glyph relation chart by Sean B. Palmer [Palmer].

Based on the observation that it is possible to generate other words, which exist in the VMS, by replacing similar shaped glyphs, it is possible to list the following rules:

"in", "iin" and "iiin" can replace each other

"e", "ee" and "eee" can replace each other

"ee" and "ch" can replace each other

"ch" and "sh" can replace each other

"k", "t", "p" and "f" can replace each other

"chk", "ckh" and "eke" can replace each other

"o" and "a" can replace each other

"o" and "y" can replace each other

"n", "r" and "m" can replace each other

"l", "r" and "m" can replace each other

"r" and "s" can replace each other

"s" and "d" can replace each other

These rules for similar glyphs only apply with some restrictions. For instance and can replace each other only as the first or as the last sign. Another example is that ("o") is interchangeable with ("a") before ("l") and ("r") but not after ("q") or before ("k").[12] Additionally, prefixes such as ("qo"), ("l") or ("ch") are common for the VMS-words. After such a prefix, the following glyph ("ch") or ("d") commonly changes into ("k"). Furthermore, it is possible to delete a glyph, if this leads to a valid VMS-word. But although it is possible to describe such rules, in most cases it is also possible to find at least one exception to them. For instance it is possible to find strange words like ("daisn"), ("oqaiin"), ("okeokeokeody") or ("otkchedy").[13] Such words usually occur only once.

With this set of rules, it is possible to build a "grid" for the VMS-words. For a grid containing all words occurring at least four times and covering 80% of the VMS-text see

[11] Regarding and an interesting observation was published by D'Imperio in 1980. A sequence of glyphs is repeated four times on folio <f57r>. "In two instances with only one loop occurs … while in the other two, we see with two clear loops in the corresponding position. Since all the other symbols appear identical, the conclusion seems inescapable that the single- and double-looped forms are functionally the same" [D'Imperio: p. 24f].

[12] The -glyph does occur before (46.5%), (22.6%), (21.6%) and (6.2%).

[13] occurs in line <f102v1.P1.2>, in line <f103v.P.18>, in line <f71r.R1.1> and can be found in line <f104.P.17>.

addendum: V. Grid (p. 66). It is surprising that in most cases all conceivable spelling permutations of a glyph group exist. If there is a gap, such as the missing word ("doir") for the -series, this can be explained by the fact that words spelled similarly to also occur only a limited number of times. Moreover, it is possible to describe relations for words using similar glyphs. For instance, in most cases, words with ("sh") are less frequent than the corresponding variant using ("ch"). Also words using ("p") or ("f") instead of ("k") and ("t") are generally less frequent. Similar relations can be described for words using ("a") instead of ("o"), ("ee") instead of ("e") etc. Furthermore, it seems that if a word is spelled similarly to ("daiin"), ("ol") or ("chedy"), it is more frequent than a word, which is spelled in a less similar way. To quantify this effect, the edit distance (ED), defined as the number of steps required to transform two words into each other, can be used.[14] For instance, it is possible to transform ("daiin") into ("dain") by deleting one -glyph and into ("dair") by deleting an ("i") and by replacing ("n") with ("r"). Therefore, the edit distance between ("daiin") and ("dain") is one, whereas the edit distance between ("daiin") and ("dair") is two.

**- or -series**

| | | | | |
|---|---|---|---|---|
| daiin (863) | aiin (469) | dain (211) | ain (89) | ED=0/1/1/2 |
| daiir ( 23) | aiir ( 23) | dair (106) | air (74) | ED=1/2/2/3 |
| daiim ( 5) | aiim ( 3) | daim ( 11) | aim ( 7) | ED=1/2/2/3 |
| daiis ( 5) | aiis ( 3) | dais ( 4) | ais ( 1) | ED=1/2/2/3 |
| daiil ( 1) | aiil ( 1) | dail ( 2) | ail ( 5) | ED=1/2/2/3 |

Obviously the relationship between edit distance and frequency of a word is not without exceptions. For instance, ("dair" ED=2) is more frequent than ("daiir" ED=1). However, this exception also applies to ("daim" ED=2), since is also more frequent than ("daiim"). A similar behavior can be observed for the -series, where it is also not possible to explain that ("ar") is more frequent than ("al") based on the edit instance alone. And this pattern is also repeated for similarly spelled glyph groups. For instance, ("dar") is also more frequent than ("dal"), and ("sar") is more frequent than ("sal").

These observations make it possible to predict the occurrence and the frequency of similarly spelled words. For instance, if it is known that chedy ("chedy") is frequent, it is possible to predict that shedy ("shedy") is also frequently used although less frequently than chedy. And if we know that ychy ("ychy") only occurs four times it is possible to predict that a glyph group ochy ("ochy") should also exist and that the groups ychdy ("ychdy"), yshy ("yshy") and osheedy ("osheedy") probably occur less than four times.

[14] To consider the peculiarities of the VMS script, the edit distance is defined as follows: If a glyph is deleted, added or replaced by a similar glyph, this is counted as one change. Also, the change from into or from or into is counted as one change. If a glyph is replaced by a non-similar glyph, this is treated as deleting one glyph and adding another glyph. This is counted as two changes.

The grid reveals that the words of the VMS are connected to each other. It is possible to generate another word from the word pool by replacing a glyph by a similar one, or by adding or deleting a glyph. How was it possible to construct a language with "generated" words and to write a text containing over 37,000 words with determinable word frequencies? Was the scribe counting the words he was writing? This seems very unlikely. A better explanation would be the assumption that it is an unintended side effect of the manufacturing or encoding process that similarly spelled words occur with predictable frequencies.

## 4 Labels

Graph 1 shows the proportion of words related to daiin ("daiin"), ol ("ol") and chedy ("chedy") for all pages of the manuscript.[15] The proportion of words related to daiin is around 15%. The mean value for words related to ol ("ol") and chedy ("chedy") is 33% and 31%, respectively. This means that words of the ol- and for the chedy-series are used twice as frequently as words of the daiin-series. A possible explanation for this distribution is that there are also twice as many spelling variations for the ol- and for the chedy-series.[16]

There are pages where words of a particular series are frequently used and pages where they are rare.[17] Graph 1 also shows that the number of words which are part of the grid decreases for the pages from <f67r1> to <f73v>. These pages belong to the Cosmological section and to the section with Zodiac illustrations. Since the grid contains all words used at least four times, this means that more unique and rare words occur on these pages.

[15] The fact that a word is not part of the grid only means that this word occurs less than four times. For words occurring three times or less, transcription problems become more important. The number of samples that could be checked for related words close by is limited for them. Since it is very time-consuming to add rare words to the grid, the author has limited his research to words that occur four times and more.

[16] For the daiin–series, 384 spelling variants (4,7%) were found. For the ol–series, 997 spelling variants (12,3%) and for the chedy–series, 958 variants (11,8%) were found.

[17] For instance the maximum proportion for words of the daiin–series is 37% on page <f37v>. The minimum is on page <f27v>. On this page and for this series only the glyph group daidy ("daidy") occurs in line <f27v.P.2>. The maximum use of words of the ol–series is on page <f54r> and is 62%, and for the chedy–series on page <f41r> it is 61%.

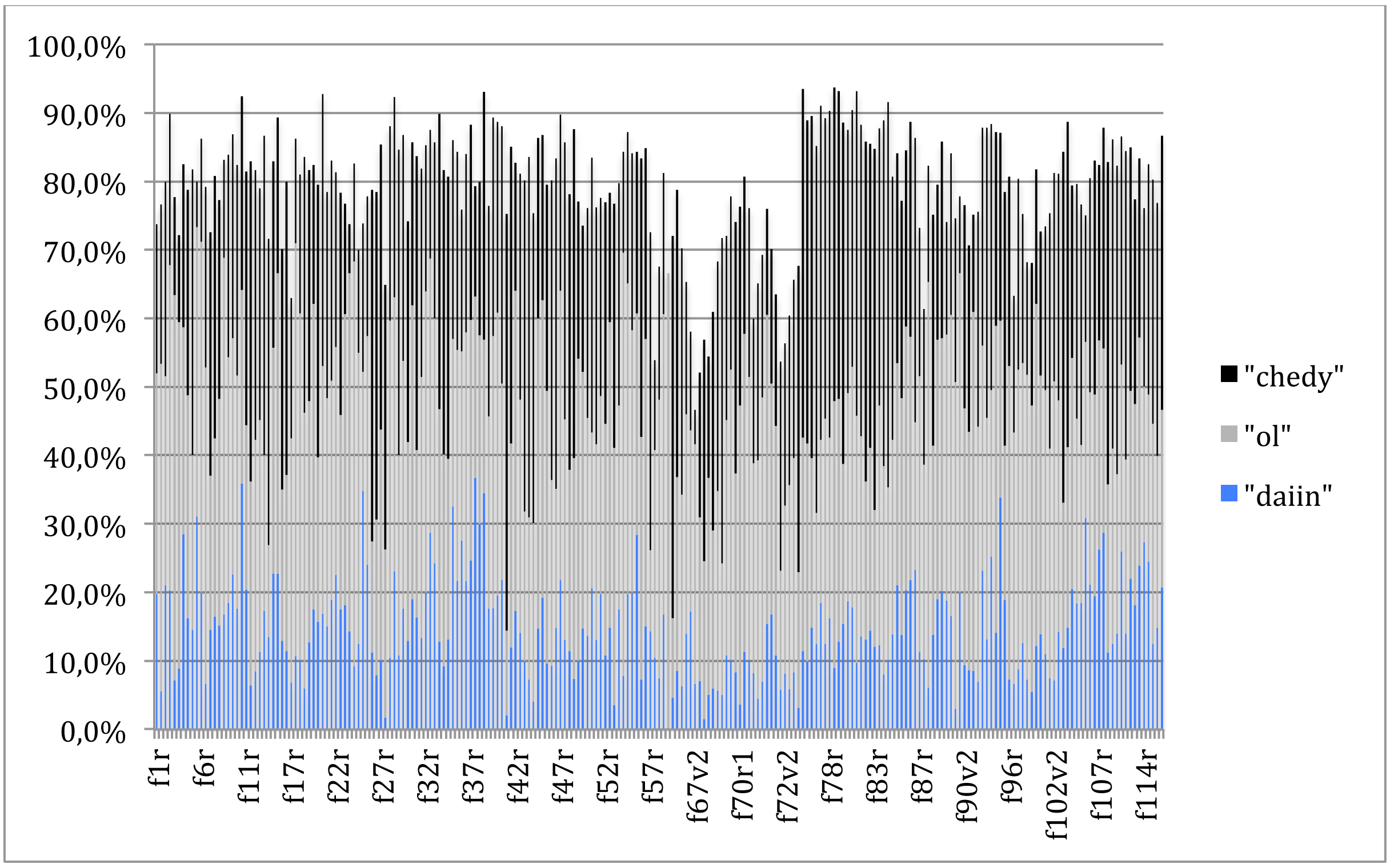

Graph 1: proportion of words of the ("daiin"), ("ol") and ("chedy") series in %

The figures used in the Cosmological and Zodiac section are named by using labels. It is interesting that on some pages it seems that some consecutive labels are "generated" by replacing glyphs with similarly shaped ones. Page <f70v2> provides an example of this:

<f70v2.S1.1> ("otaral")

<f70v2.S1.2> ("otalar")

<f70v2.S1.3> ("otalam")

<f70v2.S1.4> ("dolaram")

<f70v2.S1.5> ("okaram")

<f70v2.S1.8> ("okaldal")

There is another interesting observation for page <f70v2>. In line <f70v2.R3.1> a word ("oteotey") and a word ("oteoteotsho") occur. The repetition of within both words is probably not a coincidence since the first word occurs only four times and the second word is unique.

<f70v2.R3.1> otey ... oteotey ... oteoteotsheo

<f71r.R1.1> oky ... okeoky ... okeokeokeody

Furthermore, on page <f71r> a similar duplication pattern occurred using oke. In this case, the unique words okeoky ("okeoky") and okeokeokeody ("okeokeokeody") were used.[18] These observations raise the question as to whether the generating mechanism for labels is less complex and therefore easier to determine. A possible explanation is that the scribe was focused on arranging the text in circular form on these pages.

An indication that labels were generated during the writing process is the use of some identical labels in the astronomical and in the herbal section. These labels are okary ("okary"), oky ("oky"), otalam ("otalam"), okeoly ("okeoly"), otaly ("otaly"), otoky ("otoky"), otaldy ("otaldy"), otal ("otal"), otol ("otal"), ykeody ("ykeody"), okeody ("okeody"), okeos ("okeos"), otory ("otory"), okody ("okody") and oran ("oran").[19] One would at least not expect several stars or star constellations to be named after plants or parts of plants. Moreover, some similarly spelled labels occur together in both sections.[20] A possible explanation for this observation is that these labels depend on each other.

## 5 Context dependency

In the VMS it is noteworthy that many words look similar to a word written one or two lines above. To quantify this effect it is possible to calculate the number of times a word occurs twice within the same line and within previous lines.[21] Graph 2

---

[18] The repetition pattern for oke and ote occurs on four pages within the Astronomical and Zodiac section. In line <f68v1.C.2> otey ... oteotey, in line <f70v2.R3.1> otey ... oteotey ... oteoteotsheo, on page <f71r> in line <f71r.R1.1> oky ... okeoky okeody ... okeokeokeody ... okeody together with oteotey in line <f71r.S1.9> and in line <f72v2.R2.1> otey ... oteotey oteoldy. Only seven glyph groups contain three gallow glyphs and only three of these groups contain the same gallow glyph three times. Therefore, the occurrence of oteoteotsheo and okeokeokeody at almost exactly the same place on two consecutive zodiac pages is probably not a coincidence.

[19] See addendum: II. Labels (p. 48).

[20] For instance the labels ykeody ("ykeody") and okeody ("okeody") are used together on page <f69v>, <f73v> and <f102v>. ykeody is used in line <f69v.L.16>, <f73v.S1.7> and <f102v1.L1.1> and okeody occurs on <f69v.L.23>, <f73v.S1.8> and <f102v1.L1.2>. Also okary ("okary") - occurring in line <f72v3.S2.8>, <f73r.S2.2> and <f99r.L1.1> - and oky ("oky") - occurring in line <f72v3.S2.3>, <f73r.S2.5> and <f99r.L1.3> - are used together on three pages.

[21] See addendum: VI. Statistics for graph 2 (p. 83). Single letter words are eliminated from this calculation. Additionally, the whole book is treated as a unit. Therefore, the first line of a page is compared with the last line of the previous page and so on. Even if a word occurs multiple times

shows that there is a peak of repeated words for three consecutive lines. In other words, if a word appears in one line there is an increased chance that this word will also appear within one of the next three lines.

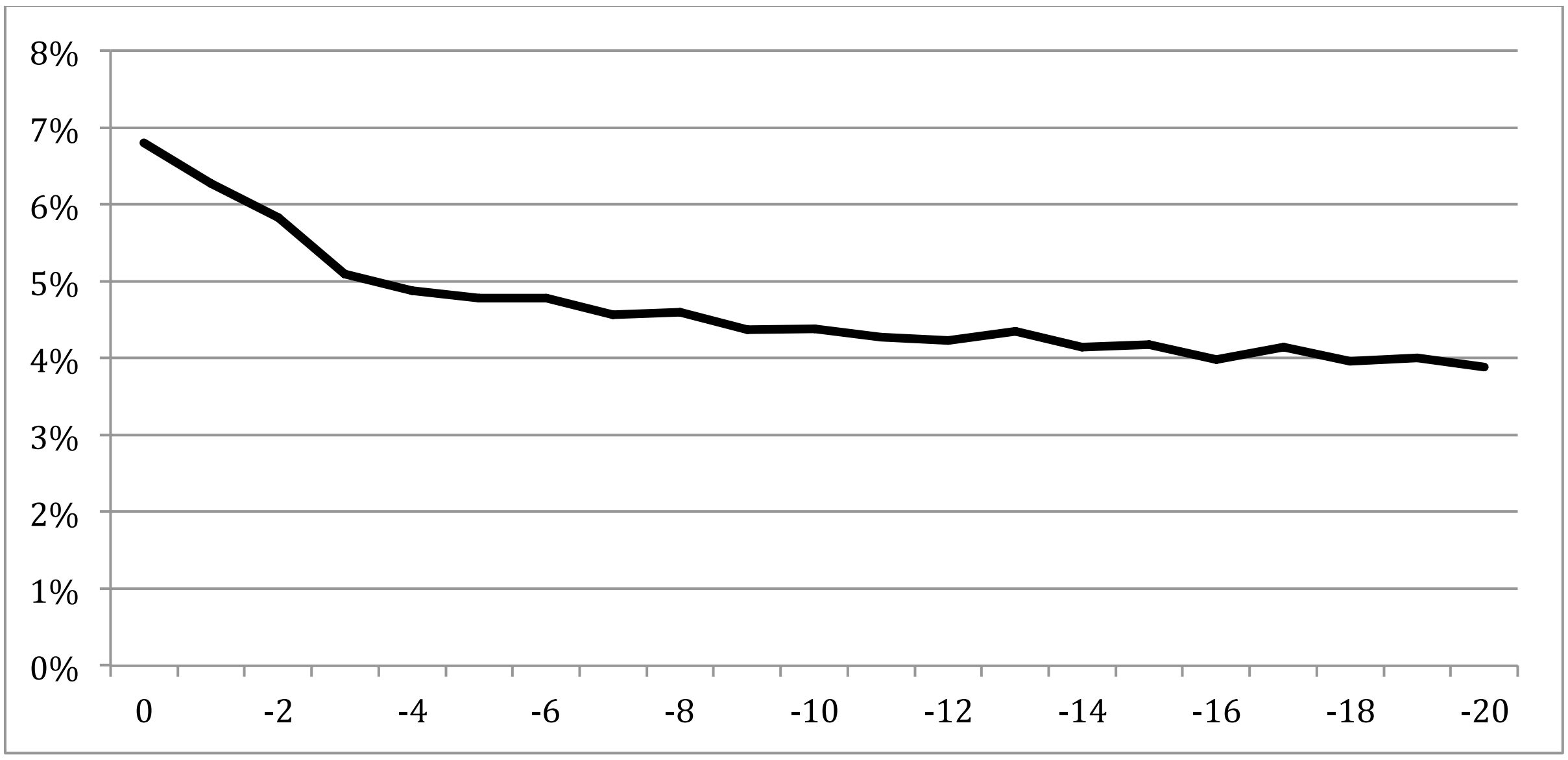

Graph 2: repeated words within the previous 20 lines for each line of the VMS

This effect increases if similar words are compared (see graph 3). For the following calculation, glyph groups with an edit distance of one are regarded as similar.[22]

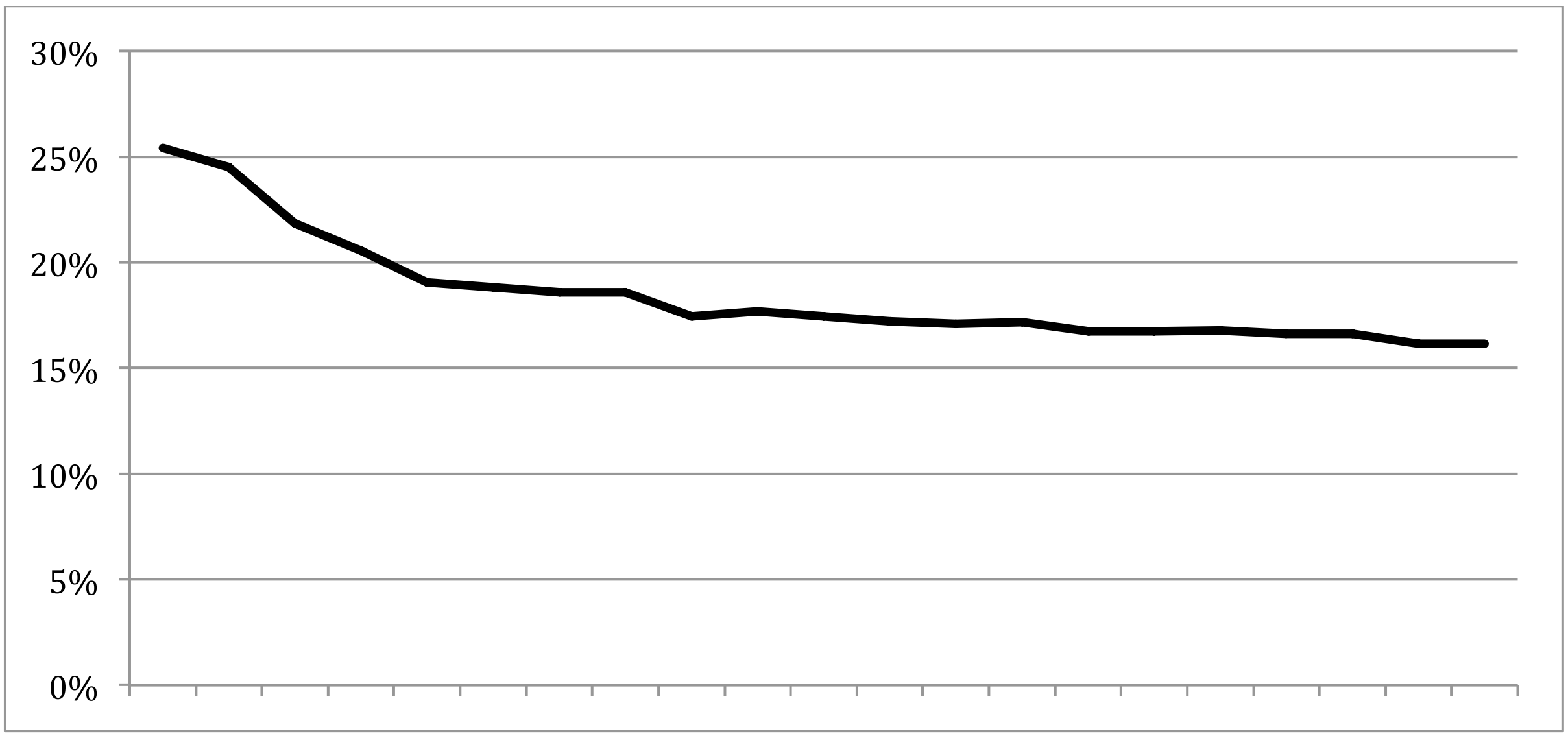

Graph 3: similar words within the previous 20 lines for each line of the VMS

---

within the start-line it is only compared once with previous lines. For the first line, on average only 50% of that line remains to search for a word pair. To obtain a comparable result the number of occurrences for repeated words is counted. Example: If, in consecutive lines, a word occurs multiple times within each line, the repeated word count is two for every line in which a word occurs twice, three for every line in which a word occurs three times and so on.

[22] See addendum: VII. Statistics for graph 3 (p. 82). To use an elementary algorithm to calculate the edit distance it is accepted that the edit distance between cc ("ee") and cτ ("ch") is calculated as two.

As shown in graph 4, the level of repeated words for a Latin text[23] is lower than it is for the VMS.

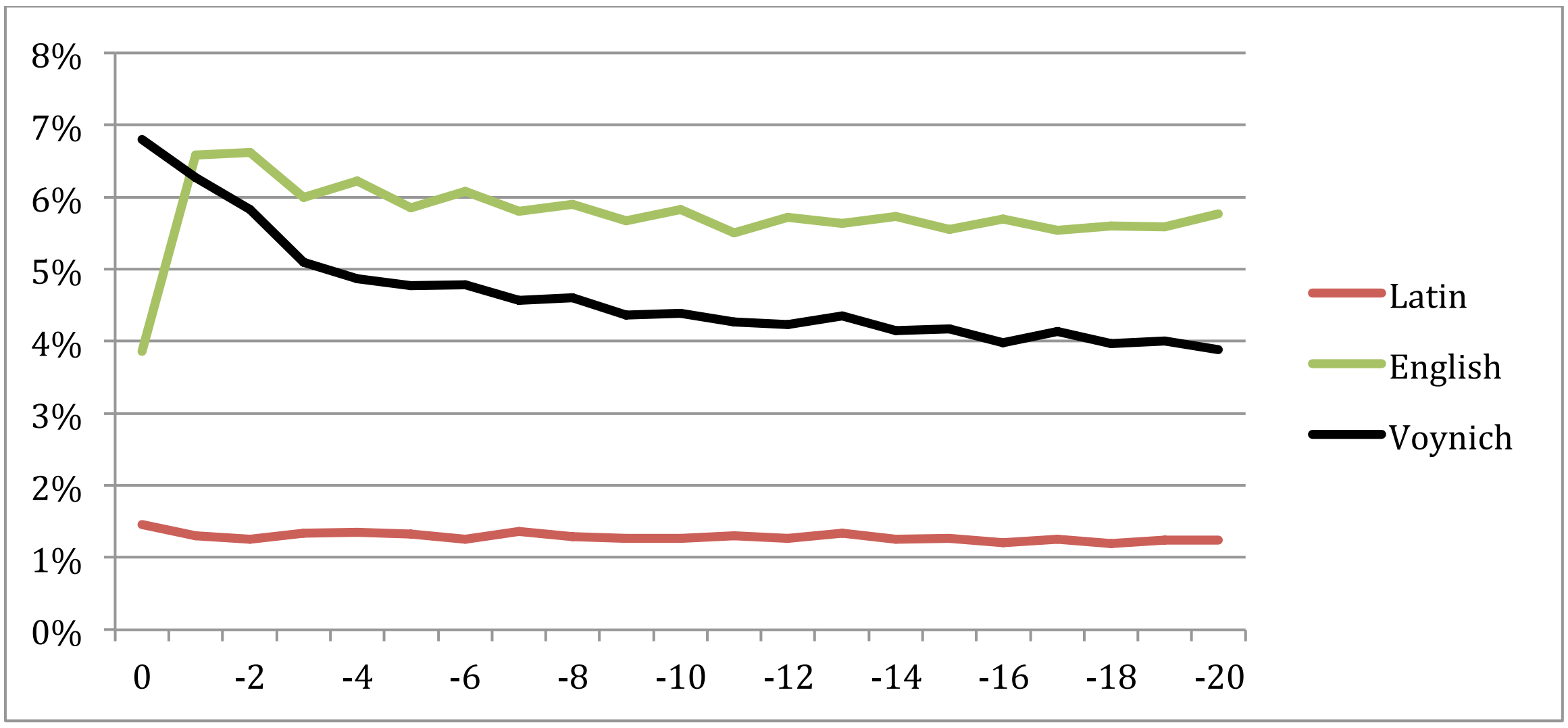

**Graph 4: repeated words within the 20 lines for the VMS an Latin and an English text**

Latin as an inflecting language uses markers like suffixes to express different grammatical functions. Therefore, the number of repeated words is very small for Latin. The level of repeated words in an English text[24] is comparable to the VMS. The increase in repeated words for consecutive lines is a feature which cannot be found in a Latin or an English text.[25] The examined English text shows only a small increase in repeated words for consecutive lines. After this small peak, the number of repeated words remains the same, whereas for the VMS the number of repeats permanently decreases. It seems that the high level of context-dependency is specific for the VMS and therefore needs explaining.

Schinner found a similar result in 2007 when using the random walk method to analyze long-distance relations within the VMS. He demonstrated that the probability of the occurrence of a similar word decreases with distance:

> "Interpreting normal texts as bit sequences yields deviations of little significance from a true (uncorrelated) random walk. For the VMS, this only holds on a small scale of approximately the average line length; beyond positive correlation build up: the presence/absence of a symbol appears to increase/decrease the tendency towards another occurrence." [Schinner: p. 105]

---

[23] The Aeneid by Virgil 80-19 BC [Virgil]

[24] The Aeneid by Virgil translated by Dryden in 1697 [Dryden]

[25] For the English text the number of repeated words within a line decreases. This effect is probably a result of the rhyme scheme used in the poem "The Aeneid". This rhyme scheme leads to words which always occur in the same position within a line and which therefore can only occur once in a line.

The calculation reveals a pattern for the usage of similar words. They are not randomly distributed within the VMS but are used on the same pages next to each other.

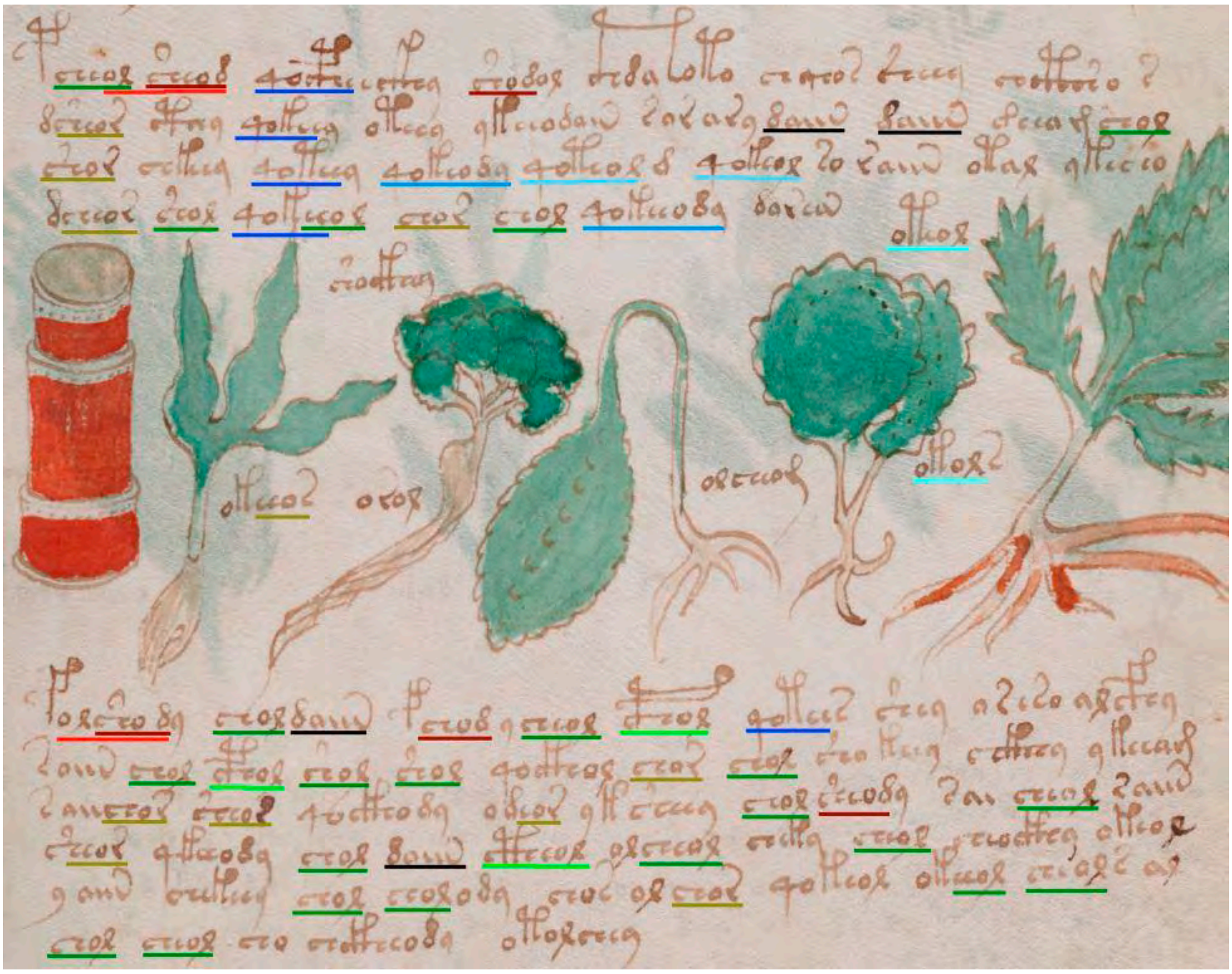

Figure 2: part of page <f100r>

For an example see figure 2. Similarly spelled glyph groups are marked in this figure with different colored lines. For instance, the initial glyph group for the second paragraph in line <f100r.P2.5> is ("folshody"). The first paragraph on this page has a similar sequence ("ol sheod") in line <f100r.P1.1> (both sequences are marked red).[26]

## 6 Evidence

On page <f100r> similarly spelled glyph groups like ("qokeey") and ("qoteey") or ("chol") and ("cheol") occur near to each other. Also for glyph groups sharing a rare feature it is typical that they occur near to each other. One

[26] A good way to verify the described pattern is by checking rare words. For instance, ("dolcheedy") and ("rolchedy") occur only once. But both words can be found next to each other in line <f77v.P.33>.

example is the subgroup ("on"). For the whole VMS only four groups contain such a sequence:

<f29v.P.5> ("tochon")

<f37v.P.3> … ("qokchon … chon")

<f37v.P.5> ("ykchon")

On page <f29v> the word ("tochon") did occur together with similar spelled glyph groups like ("chol"), ("shor") and ("shan"). And on page <f37v> the words ("qokchon"), ("chon") and ("ykchon") occur near to similar glyph groups like ("qotchor"), ("dchor") and ("ytchor").

Another example is the rarely used glyph ("x").[27] This glyph is part of several similar spelled groups appearing together on the same pages:

<f46r.P.2> ("xor")

<f46r.P.7> ("xar")

<f66r.R.13> ("axor")

<f66r.R.16> ("chxar")

<f111r.P.21> ("oxor")

<f111r.P.25> ("oxar")

Are all this examples exceptions or do similar groups typically appear together on the same pages? This hypothesis can be confirmed by checking a control sample using all glyph groups appearing a certain number of times within the VMS.

In this paper all words occurring seven times and all words occurring eight times are used as two separate control samples. To limit the samples, all words also appearing as subgroups of other words are excluded. The result for the remaining 20 words occurring only seven times is that in 120 out of 140 cases (85%) similar words can be found near to each other.[28] For the group of words used eight times, the result is that in 66 out of 72 cases (91%) at least one similar word was found within a maximum distance of three lines and a maximum edit distance of three.[29] Furthermore, if near is defined as both glyph groups must be used one after another or in two consecutive lines one above the other, the result is still interesting. In 62 out of 140 cases (44%) and in 25 out of 72 cases (35%) a similar glyph group can be found for both samples. In other words, similar glyph groups can be found above each other twice as often as they can be found side by side.[30]

[27] There are 24 groups containing a -glyph. They occur on 18 pages. For example three times on page <f46r> in line <P.1> ("chepchx"), <P.2> ("xor") and <P.7> ("xol"), on page <f55r> in line <P.8> … ("xar … xaloeees"), on page <f66r> in line <R.13> ("axor") and <R.16> ("chxar"), <f85r1.P.12> ("dxar"), <f86v6.P.10> ("oxy"), <f94r.P.6> ("oxockhey"), <f104r.P.34> ("cholxy"), <f105r.P2.30> ("shx"), on page <f105v> in line <P.19> ("rokaix") and <P.27> ("arxor") etc.

[28] Near to each other means within a range of three lines before and after a word, and similar means that it is possible to transform them into each other by changing three or fewer glyphs (edit distance <=3).

A feature of the VMS is that similarly spelled glyph groups are used together on the same pages near to each other. This means, the reason why similarly written words have similar frequencies is that they appear together on the same pages. In other words, the scribe was writing similarly spelled glyph groups near to each other because they depend in some way on each other. This result also explains the observation that repeated phrases typically consist of glyph groups spelled the same or similarly. Such phrases occur multiple times because of the connection between similarly spelled glyph groups.

## 7 The text generation method

Taking the glyph groups used seven and eight times as an example it is possible to describe the changes for similar glyph groups occurring near to each other:

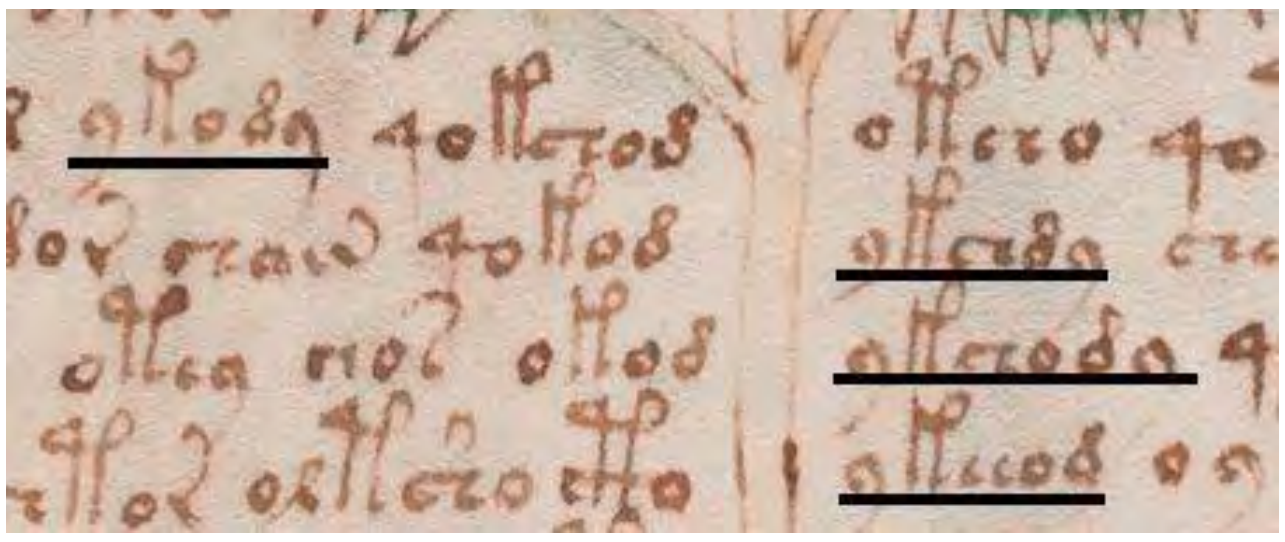

Figure 3: part of page <f53r> with glyph groups similar to [Voynich glyph group] underlined

<f53r.P.5> [Voynich glyph group]

<f53r.P.6> [Voynich glyph group] (o removed and [Voynich glyph] added)

<f53r.P.7> [Voynich glyph group] (the previously removed o was added)

<f53r.P.8> [Voynich glyph group] ([Voynich glyph] replaced with [Voynich glyph] and the final [Voynich glyph] removed)

---

[29] See addendum: III. Glyph groups occurring seven times (p. 49) and IV. Glyph groups occurring eight times (p. 61).

[30] The ratio is 39:22 for the words occurring seven times and 16:11 for the words occurring eight times.

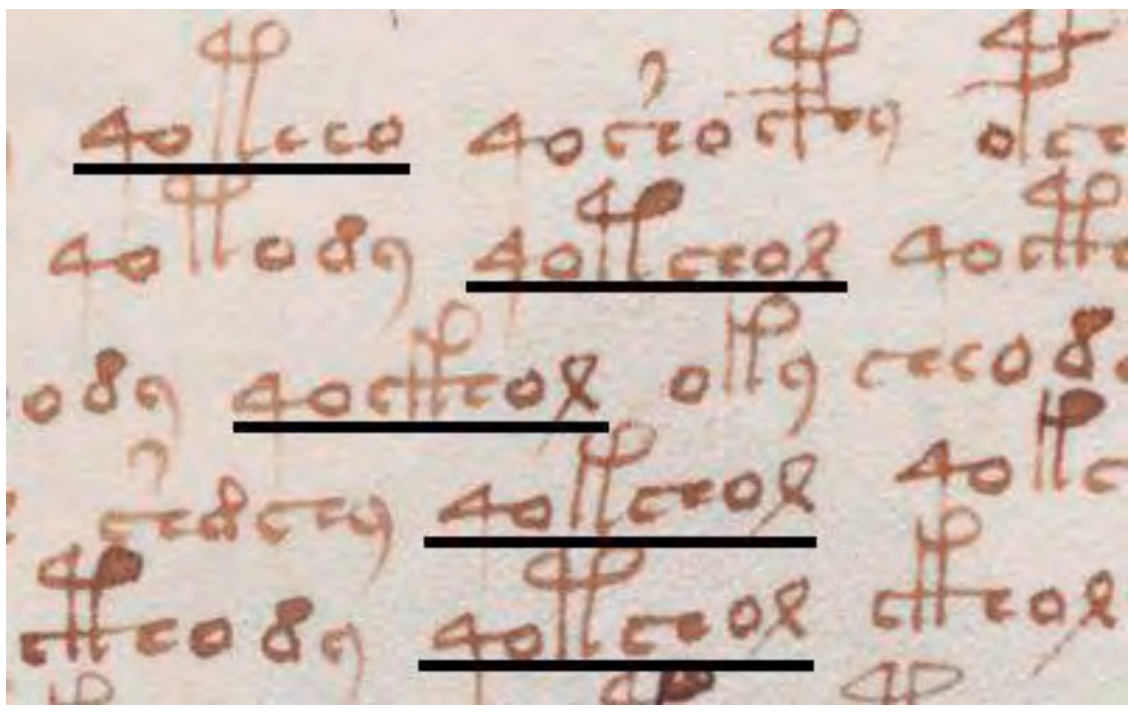

Figure 4: part of page <f93v> with glyph groups similar to [glyphs] underlined

<f93v.P.1> [glyphs]

<f93v.P.2> [glyphs] ([glyphs] replaced with [glyphs] and a final [glyphs] added)

<f93v.P.3> [glyphs] ([glyphs] replaced with [glyphs])

<f93v.P.4> [glyphs] ([glyphs] replaced with [glyphs])

<f93v.P.5> [glyphs] ([glyphs] replaced with [glyphs])

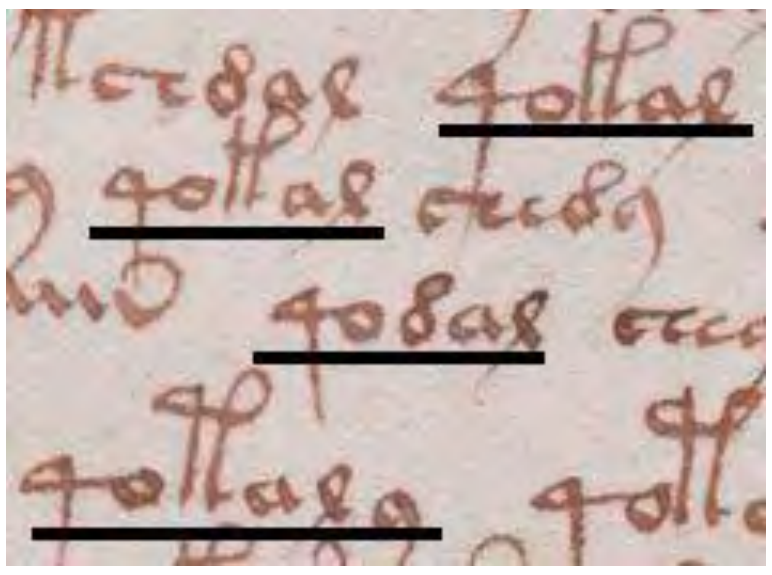

Figure 5: part of page <104r> with glyph groups similar to [glyphs] underlined

<f104r.P.33> [glyphs]

<f104r.P.34> [glyphs] (nothing changed)

<f104r.P.35> [glyphs] ([glyphs] removed and [glyphs] added)

<f104r.P.36> [glyphs] ([glyphs] removed and [glyphs] and the final [glyphs] added)

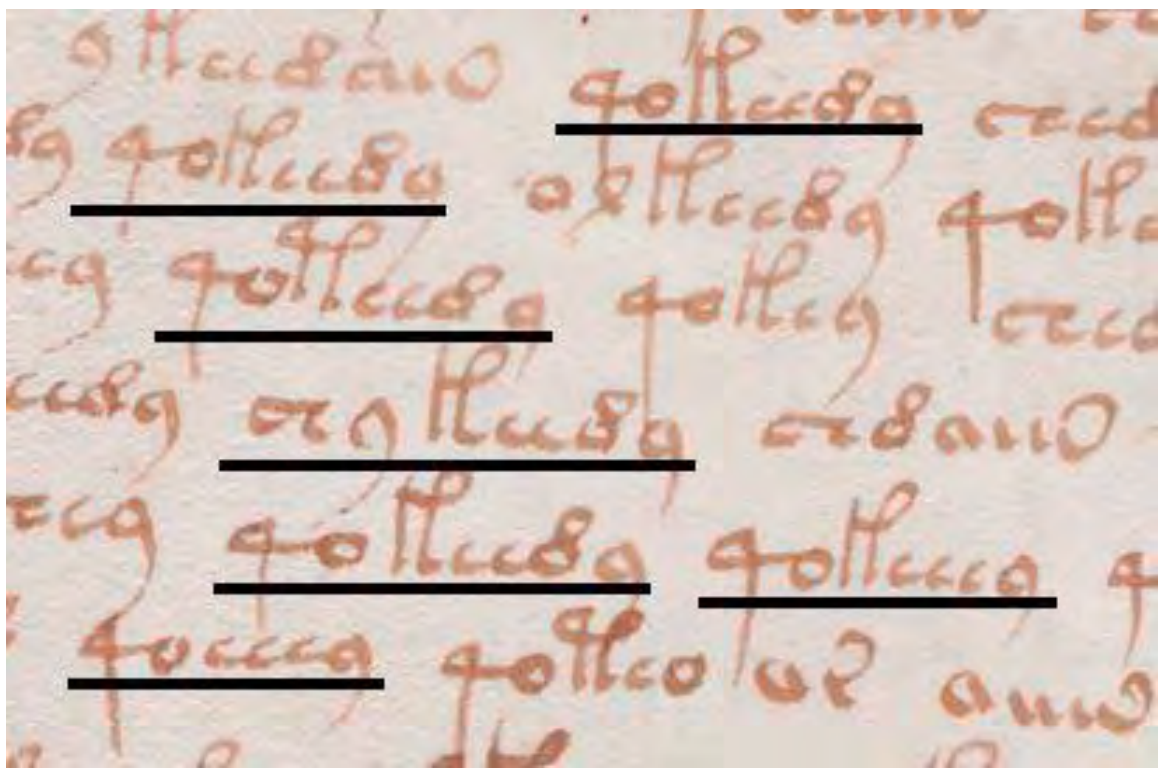

Figure 6: part of page <112v> with glyph groups similar to [glyphs] underlined

<f112v.P.14> *qokeedy*

<f112v.P.15> *qokeedy* (nothing changed)

<f112v.P.16> *qoteedy* (*k* replaced with *t*)

<f112v.P.17> *chykeedy* (*t* replaced with *k* and *qo* with *chy*)

<f112v.P.18> *qokeedy* (*chy* replaced with *qo*)

<f112v.P.18> *qokeeey* (*ee* replaced with *eee* and *d* removed)

<f112v.P.19> *qoeeey* (*k* removed)

Now the following rules can be used to describe the way the scribe was generating the text:

I) Copy an already written glyph group and replace one or more glyphs with similarly shaped glyphs. An example is *dain dail dair* ("dain dail dair") in line <f45v.P.4>.[31]

II) Copy a glyph group and add one or more glyphs. An example is *okedy qokedy qokeedy* ("okedy qokedy qokeedy") in line <f31r.P.10>.

III) Copy a glyph group and delete one or more glyphs. See, for instance, line <f27v.P.1> *fochof chof cho* ("fochof chof cho") or *lkeeedy* ("lkeeedy"), *lkeedy* ("lkeedy") and *lkedy* ("lkedy") in line <f111r.P.37>.

IV) Create a new glyph group by combining two groups. Example: *ol chedy* ("ol chedy") and *olchedy* ("olchedy") in line <f75v.P2.18> and line <f75v.P2.19>.

V) Create two words by splitting a glyph group created by rule IV). Example: *olchedy* ("olchedy") and *ol chedy* ("ol chedy") in line <f80r.P.5> and <f80r.P.6>.

VI) Create a new glyph group by duplicating a group or subgroup. Examples: line <f75v.P2.18> *or aror … olol ol* ("or aror … olol ol") or *chechey* ("chechey") and *chey* ("chey") in line <f111v.P.30> and <f111v.P.31>.

VII) From time to time copy a glyph group without changing anything. See for instance *qokain* ("qokain") in line <f116r.P.7>, <f116r.P.8> and <f116r.P.9>.

VIII) For the first word of a paragraph a gallow glyph like *p* ("p") or *f* ("f") was commonly added. Furthermore, both gallow glyphs were preferred within the first line of a paragraph.

IX) Use a combination of the rules I-VI and VIII.

In order to use the described rules for an empty page, it was probably useful to use a second page as a source. There are indications that the scribe preferred the last completed sheet for this purpose. One peculiarity for the words used seven or eight times is that they often appear on subsequent pages. For instance *qodal* ("qodal") appears on four consecutive sheets:

[31] Glyph groups like *daid* ("daid") are rare. It seems that a glyph was normally not replaced by a glyph which was already part of the group.

<f51v>, <f52r>, <f53v> and <f54v>. For the glyph groups which occur seven or eight times, this happens more often on subsequent sheets (37 times) than on the front- and back of a page (5 times).[32]

The connection between consecutive lines and between similar glyph groups exists because the text is a copy of itself. The statistical features of the text can be explained by the hypothesis that the author of the VMS was using the described self-referencing system to generate the text. This text generation mechanism also explains the observation that for common glyph groups almost all spelling variations occur. The use of different spelling variations is no coincidence, because the scribe was generating the text by varying glyph groups already written.

Many rare or unique words were used as labels for the pages <f67r1> to <f73v> in the Cosmological and Zodiac section (see graph 1 on p. 7). A possible explanation for this is that the glyph groups written in circular form were harder to copy. The scribe probably used these groups less frequently as a source for the generation of new text.

By copying and manipulating the glyph groups again and again, the character of the text would change over time. One way to avoid this effect would be to invert the modification rules from time to time. And indeed this type of repetition did occur. For instance, on page <f77v> two similar word sequences occur in lines 5 and 15 (see figure 7: both sequences are marked with black rectangles).

<f77v.P5> ("sal shedy qokal chedy qokedy")

<f77v.P15> ("sal chedy qokedy qokal shed")

There are two explanations for the similarity of these two sequences. On this page, many glyph groups are copied by inverting a formerly used modification rule. Additionally, there are three chains of glyph groups connected to ("chedy") on this page (they are marked with orange, red and dark red). An observation for this page is that in most cases similar glyph groups are arranged one above the other, whereas the glyph group below is shifted a step to the right, in the direction of writing. A possible explanation for this observation is that the scribe was generating the text while writing by copying glyph groups he was able to see.

32 Further examples can be found by checking rare or unique glyph groups. For instance, the unique word ("osaro") appears on page <f100r>. A possible source for this word is the unique glyph group ("osary") on page <f99v>. In a similar way, a possible source for the first word on page <f79v> ("poldshedy") is the rare word ("olshedy") on the previous sheet in line <f78v.P.2>.

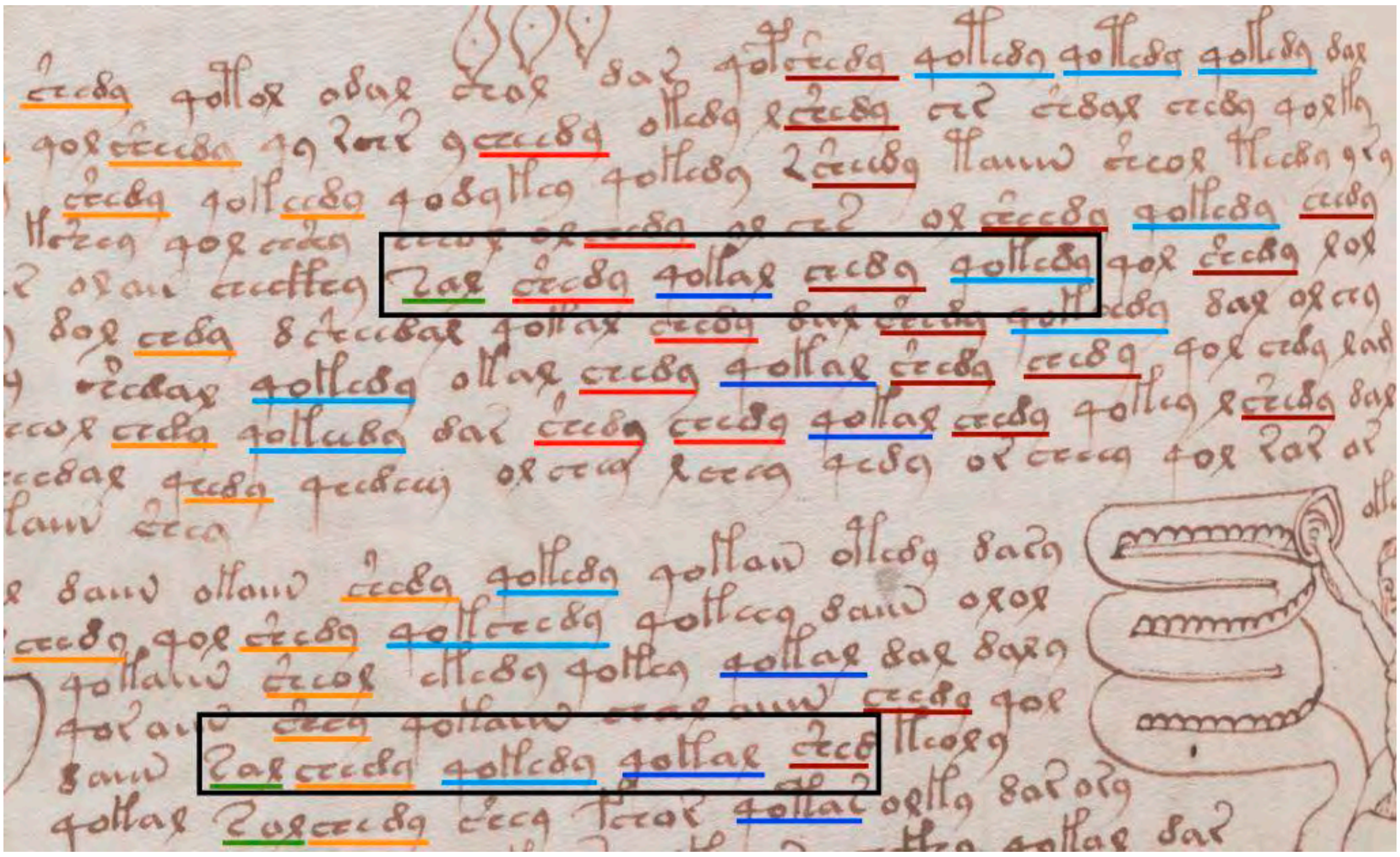

Figure 7: part of page <f77v>

## 8 The line as a functional entity

One result of the text generation mechanism is that similar elements can be found in the same position for subsequent lines.

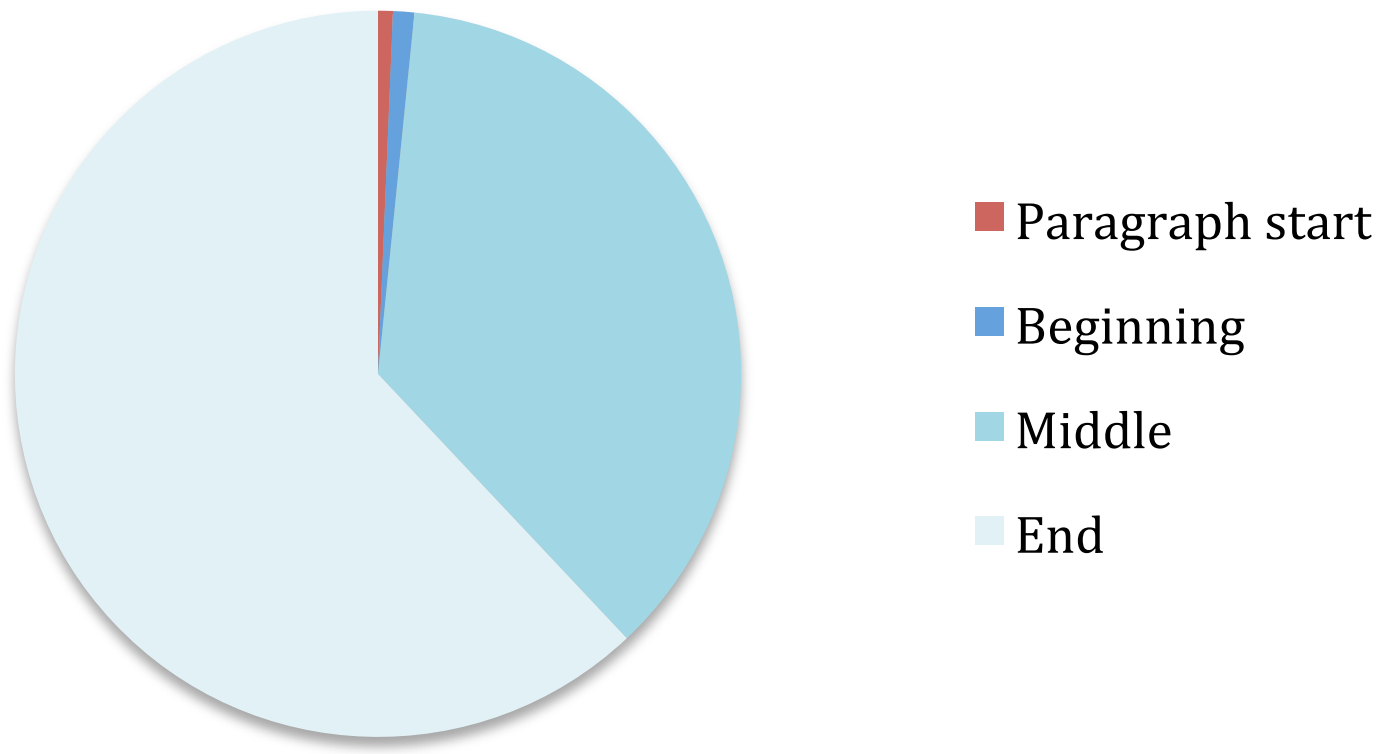

Graph 5: Usage of glyph groups with ꝯ ("m") as the last glyph for different positions within a line

It is practical to copy a glyph group from the same position some lines above. One observation for the VMS is, therefore, that some elements are typical for a specific position within a line. For instance, 62% of the occurrences of the m-glyph ("m") are at the end of a line (see graph 5).[33]

Another example is that the glyph group at the beginning of a paragraph is usually highlighted by an additional gallow glyph (f, p, k, t) as the first sign (see graph 6). This is the case in 617 out of 716 paragraphs (86%). Within a paragraph the words at the beginning of a line frequently start with a glyph 9, o, 8 or 2. This occurs more frequently at the beginning of a line (68%) than this happens within a line (50%).[34]

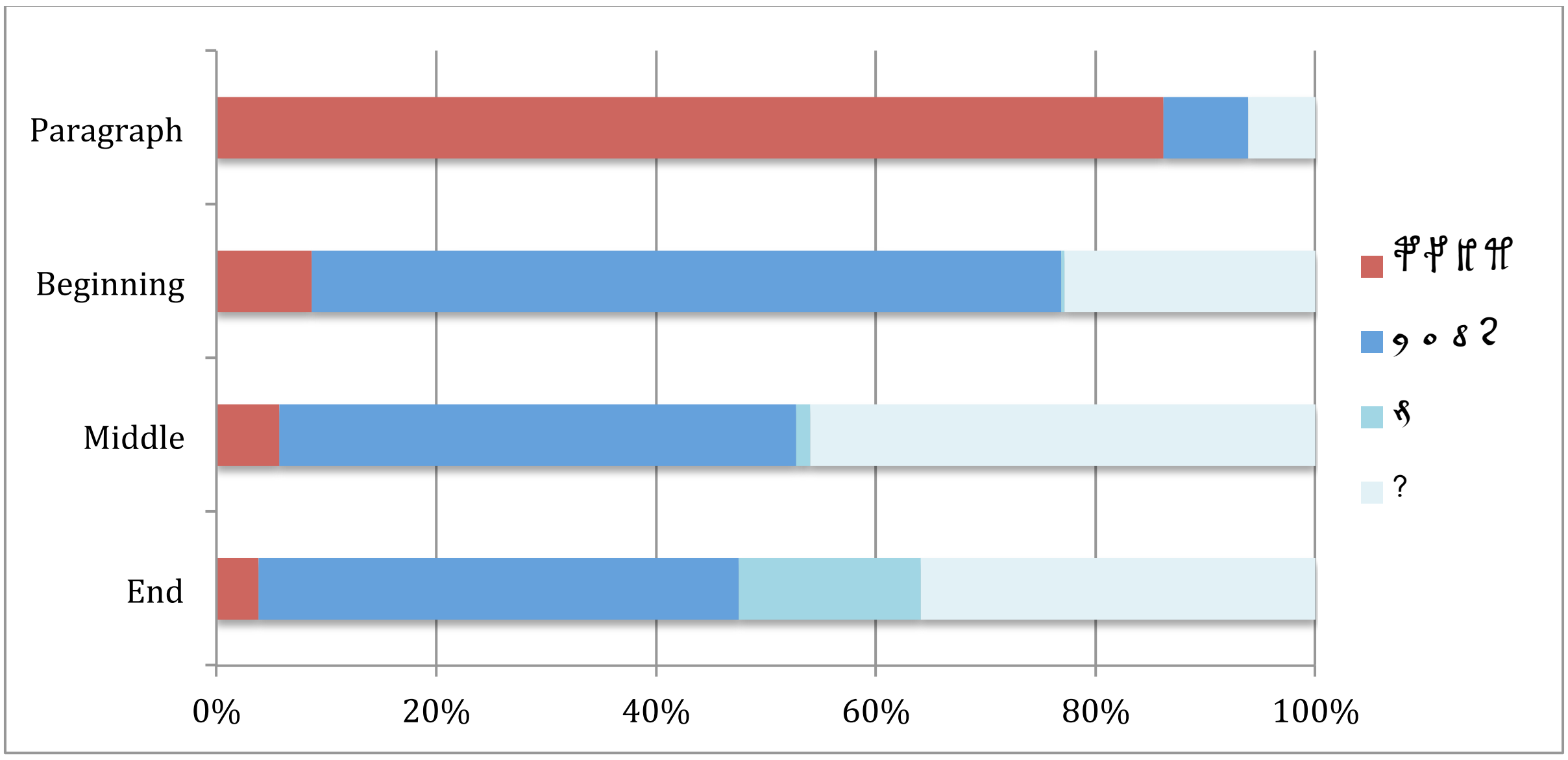

**Graph 6: Usage of glyph groups with typical start or end glyphs at certain positions within a line**

Since 9, o, 8 or 2 were added frequently to the first glyph group within a line and since the gallow glyphs f, p, k or t were added regularly to the initial glyph group within a paragraph the average length of the first words within a line increases. In a statistical analysis Vogt thus comes to the conclusion: "1. The first word with $i$ = 1 of a line is *longer* than average, $\bar{l}_1 > \bar{l}$, 2. The second word with $i$ = 2 is shorter, $\bar{l}_2 < \bar{l}$" [Vogt: p. 4]. In fact, the second glyph group is shorter than the first group in 48% of the lines and longer in only 32%.[35]

Both observations can be explained as an unintended side effect of the text generation method. The source for the first word in each line could only be found within the previous lines. Since the first and the last word in each line are easy to spot, the most obvious way is to pick them as a source for the generation of a group at the beginning or at the end of a line.

33 An example is given in addendum: XVII. Statistics for dam ("dam") (p. 88)
34 See addendum: IX. Glyph group statistics for all pages (p. 85).
35 See addendum: XXI. Similar consecutive glyph groups / table 16 (p. 89).

For the second glyph group it is also possible to select the first group as a source. Since the first group in a line usually has a prefix (see o and 9 underlined in red in figure 8) the simplest change is to remove this prefix. And it is indeed possible to find examples of such changes. For instance for the first paragraph on page <f3r> there are two occurrences in which the leading 9 ("y") is also removed for the second glyph group:

<f3r.P.2> ("ycheor chor")

<f3r.P.8> ("ysheor chor")

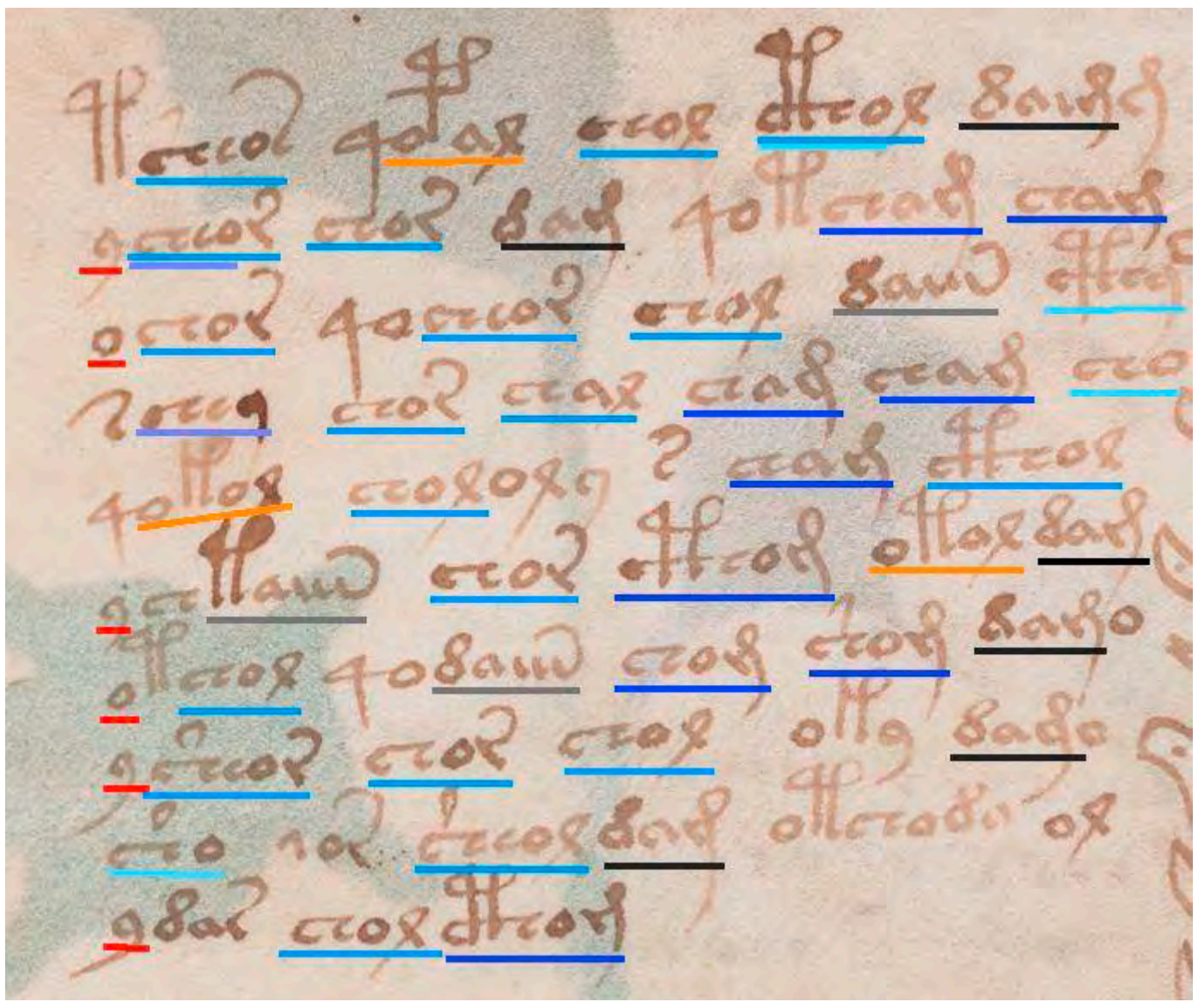

Figure 8: part of page <f3r>

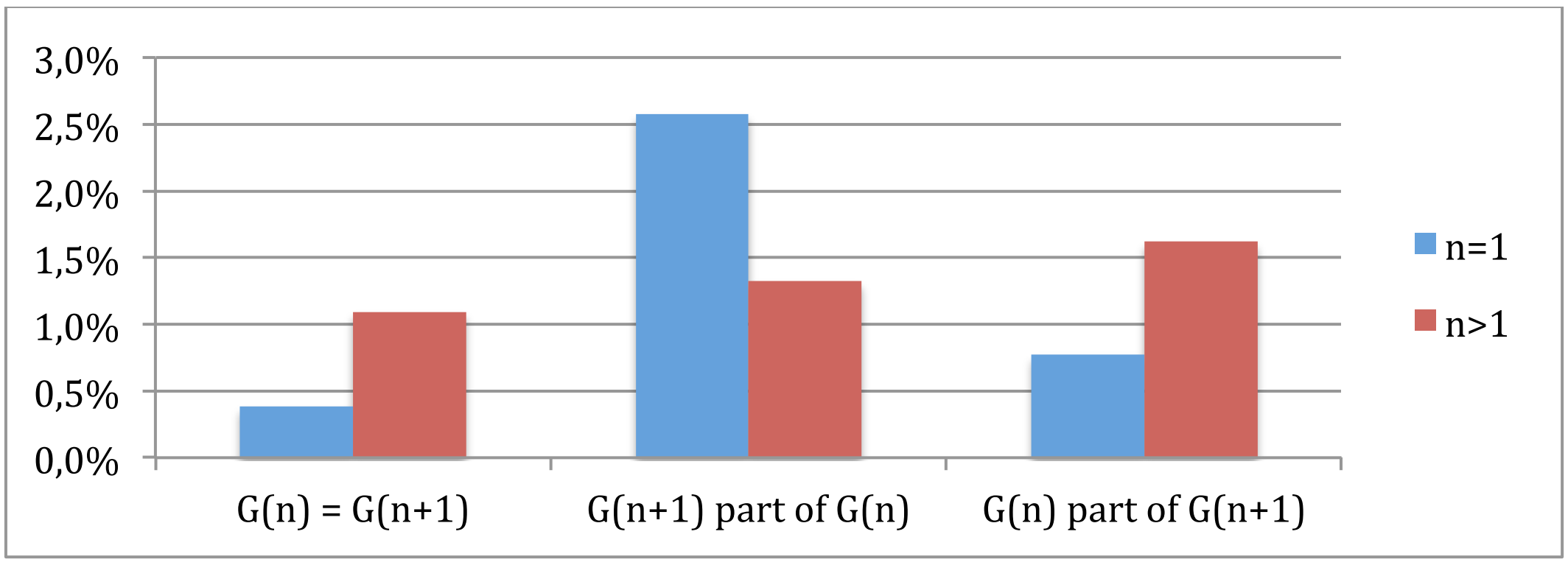

Graph 7: Percentage of consecutive groups which are spelled the same or similar

Statistically, the second glyph group in a line occurs twice as often as a subgroup of the first group (2.6%) than this is the case for any other groups in a given line (1.3%) (see blue bar vs. red bar "$G_{(n+1)}$ part of $G_{(n)}$" in graph 7).[36] In contrast, cases in which the first glyph group is also part of the second group are only half as frequent (see "$G_{(n)}$ part of $G_{(n+1)}$" in graph 7). As expected, the most common change for the

36 See addendum: XXI. Similar consecutive glyph groups / table 17 (p. 89)

second word is that a prefix y (8 times), d (6 times) or s (6 times) is removed. Within a line the most common change for similarly spelled consecutive groups is the removal of a prefix such as q (42 times), ch (20 times) or o (18 times).[37]

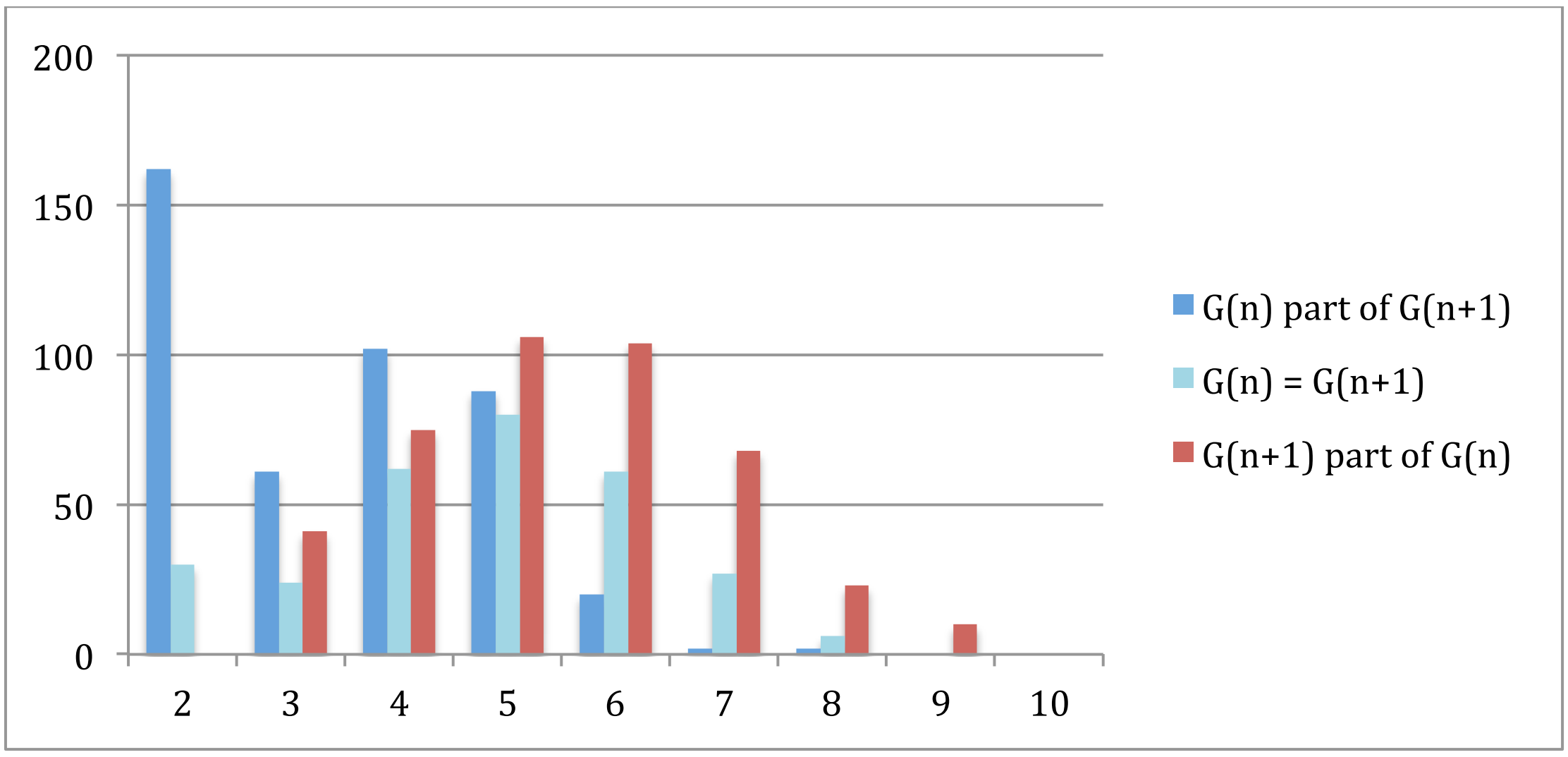

**Graph 8: Sequence count grouped by the length of the first glyph group and by the type of similarity[38]**

Graph 8 shows that glyph groups which also occur as part of the next group are mostly shorter than the average word length of 5.5 (see blue bars in graph 8). For groups longer than average the next group is more frequently part of this group (see red bars in graph 8).

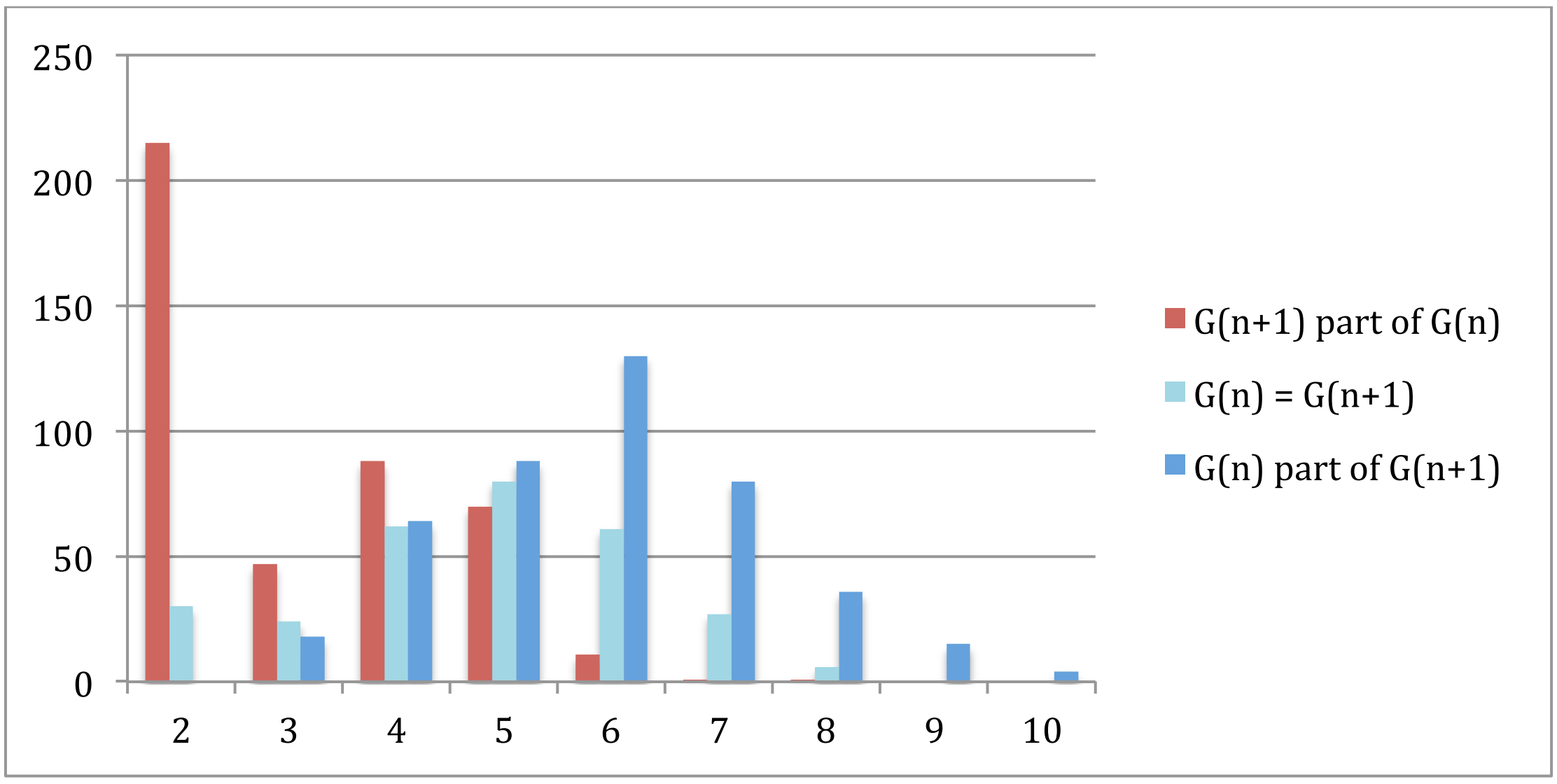

**Graph 9: Sequence count grouped by the length of the second glyph group and by the type of similarity[39]**

---

[37] For a more complete list see addendum: XXI. Similar consecutive glyph groups / table 22 (p. 91).

[38] Glyph sequences such as ol ("ol"), al ("al"), or ("or"), ar ("ar") and dy ("dy") can occur as prefixes, suffixes or standalones. These groups explain the high number of 162 for glyph groups with two glyphs.

[39] See addendum: XXI. Similar consecutive glyph groups / table 20 (p. 90).

Is there a connection between similarly spelled consecutive glyph groups? If this is not the case, the order in which the glyph groups occur should not make any difference. But there is a significant difference. Graph 8 uses the length of the first group for the x-axis and graph 9 the length of the second glyph group. Since the order of the glyph groups is inverted for the second graph, the bars for "$G_{(n+1)}$ part of $G_{(n)}$" in graph 9 (red bars) should show the same behavior as the bars for "$G_{(n)}$ part of $G_{(n+1)}$" in graph 8 (blue bars). However, this is not the case. There is a higher value for groups with a length of two for "$G_{(n+1)}$ part of $G_{(n)}$" in graph 9 (red bars). Also, more groups with six or more glyphs for "$G_{(n)}$ part of $G_{(n+1)}$" are counted in graph 9 (blue bars).

It is also possible to find an example for a change in which the order of the groups matters. There are 17 cases in which a group with a prefix ("qok") or ("qot") is followed by a similar group with the prefix removed.[40] But a group without a prefix is never followed by a similar group to which a prefix ("qok") or ("qot") was added.

This means that in the case of two similar consecutive glyph groups something is frequently added to short groups and something is frequently removed from longer groups in order to generate the next group (see graph 10). This finding confirms that the second glyph group in a line is indeed shorter on average, since the first glyph group is longer on average. Furthermore, it confirms a connection between similarly spelled consecutive glyph groups.

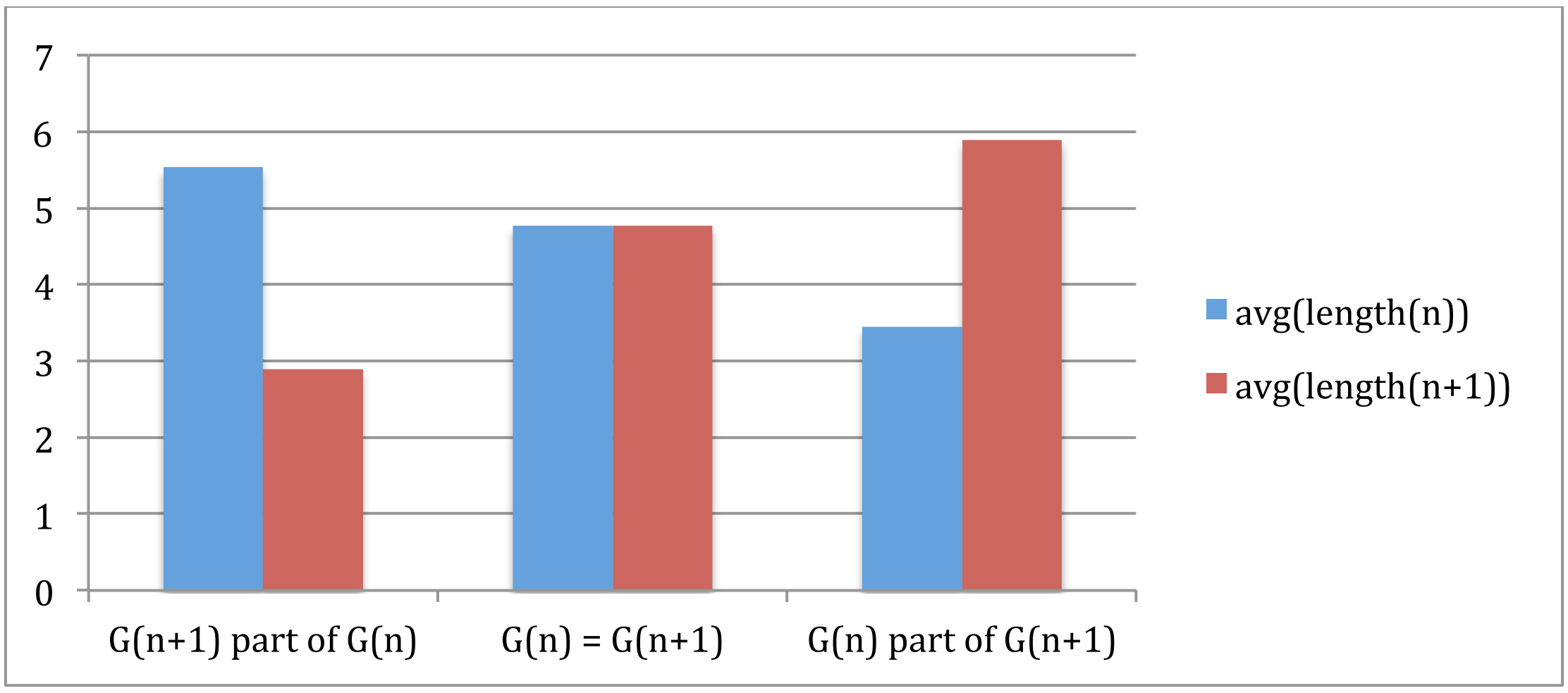

**Graph 10: Average glyph group length for consecutive groups which are similar to each other**

Since there are as many different glyph groups with a length of five as there are with a length of six glyphs,[41] it is an unexpected finding that something is frequently added to groups with five glyphs but that this type of change is rare for groups with six or more glyphs (see blue bars in graph 8).

[40] See <f3r.P.2> ("qotcham cham"), <f7r.P.4> ("qotcho cho"), <f10r.P.5> ("qotcho chol") etc.

[41] See the word length distribution by [Stolfi].

The first reason for the fact that this occurs is that the usage of prefixes is restricted by a number of rules. For instance, normally no other prefix is added in front of ("q") and normally no prefix is added twice. Typical prefix groups for the VMS are, for example, glyph sequences such as ("l"), ("ch"), ("ok") or ("qok"). Therefore, prefixes added by text generation rule II usually contain between one and three glyphs. This means that the maximum length of a prefix group is limited.

The second reason is that longer glyph groups are often a result of text generation rule IV, in which short groups are concatenated to a longer group.[42] Since the most common basic glyph groups such as ("daiin"), ("ol") and ("chedy") consist of five or fewer glyphs, the length of the source groups to which this rule was applied is also limited.

Both circumstances prefer glyph groups with five or fewer glyphs as a source for the generation of longer groups. The fact that something is frequently added to shorter-than-average source words and something is frequently removed from longer-than-average words also explains the binomial distribution of the word length.[43] It is a side effect of the permanent copying of the glyph groups that they are related to each other and that their length distribution is balanced.

## 9 Relations between similar glyph groups

In 1976 Currier described at least two different hands and two different languages within the manuscript [see Currier].

> "There are two different series of agglomerations of symbols or letters, so that there are in fact two statistically distinguishable languages." [Currier]

According to Currier two or more scribes created the manuscript, and each scribe only wrote a part of the manuscript. A good way to distinguish between these two "languages" is the usage of ("chedy"). is the third most frequent word. This word is missing for the pages using the language named Currier A and frequent on pages using Currier B. Instead of

---

[42] For instance ("olchedy") combines ("ol") with ("chedy"), ("chedol") combines ("ched") with ("ol") and ("cheodaiin") combines ("cheod") with ("daiin").

[43] Stolfi plotted "the relative number of distinct VMS words of each length, ignoring their frequencies in the text" [Stolfi].

("chedy"), the words ("chody") and ("shody") occur for pages using language A (see graph 11). And also the interim word ("cheody") appears on pages using language A, whereas ("chody") is also used on pages using language B.[44]

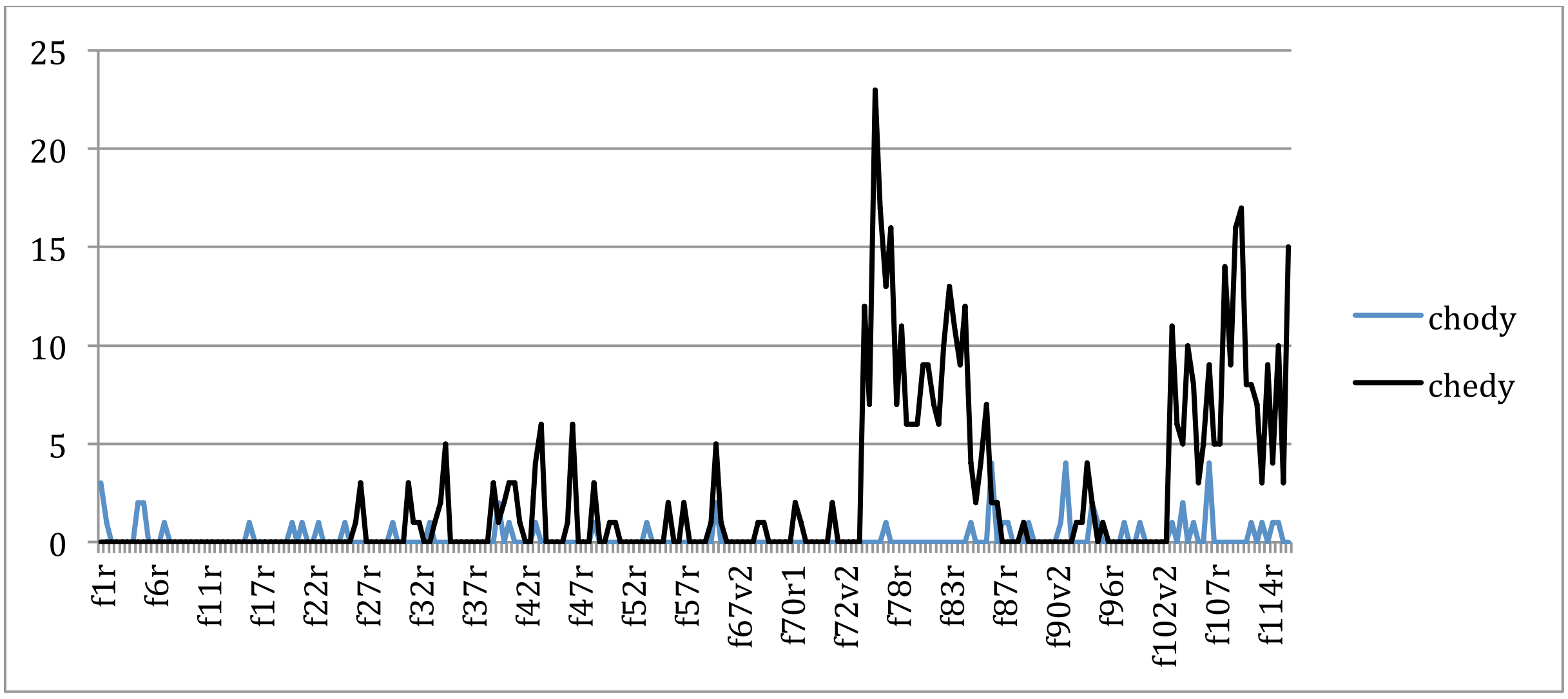

Graph 11: appearance of ("chody") and ("chedy")

A similar distribution also occurred for other word pairs connected by the usage of ("od") and ("ed").[45] In all cases, the variants using occurred for the whole VMS, whereas the variants using instead of are missing on the pages Currier has identified as written in language A (see graph 12).[46]

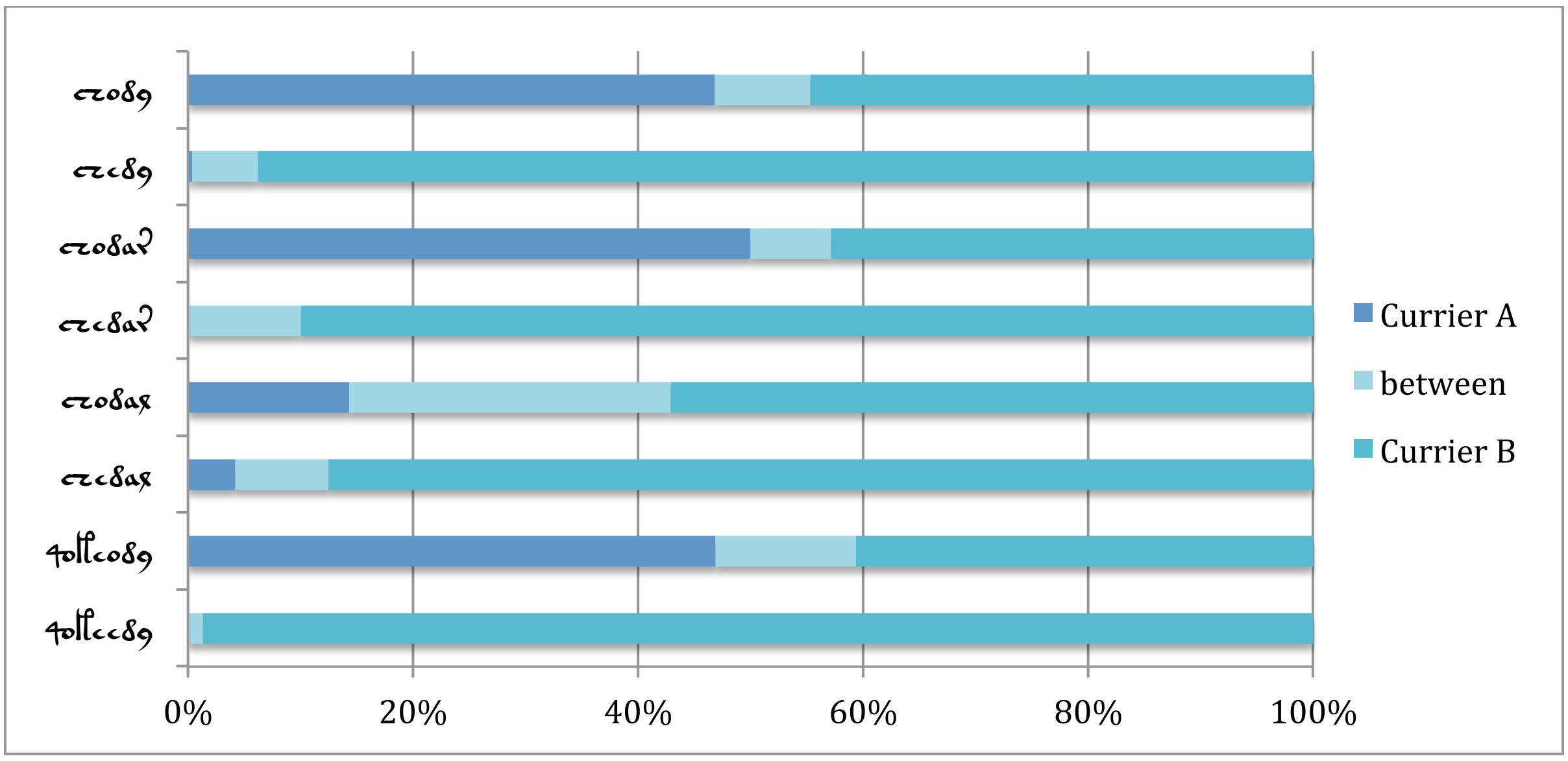

Graph 12: distribution of glyph groups using ("od") and ("ed") on pages using Currier A and Currier B

[44] See addendum: XII. Statistics for ("chody") and ("chedy") (p. 86).
[45] See addendum: XIII. Statistics for ("chodar") and ("chedar") (p. 86), XIV. Statistics for ("chodal") and ("chedal") (p. 87) and XV. Statistics for ("qokeody") and ("qokeedy") (p. 87).
[46] See addendum: IX. Glyph group statistics for all pages (p. 85), X. Glyph group statistics for pages in Currier A (p. 85) and XI. Glyph group statistics for pages in Currier B (p. 85).

Another important observation is that some typical spelling variants exist for certain positions within a paragraph or line. There is no difference between Currier A and B regarding the usage of these spelling variants. In both "languages" an additional gallow glyph , , or was added to the initial glyph group of a paragraph, and there is the same preference to add , , or to the first group in a line for both "languages".[47] The method for creating the text is the same for the whole manuscript. The difference between both "languages" is only that the scribe may have changed his preferences while writing the manuscript. A possible explanation for this observation is that spelling variants invented while writing did not appear on pages already completed. This leads to the conclusion that the pages using Currier B were written after the pages in Currier A.

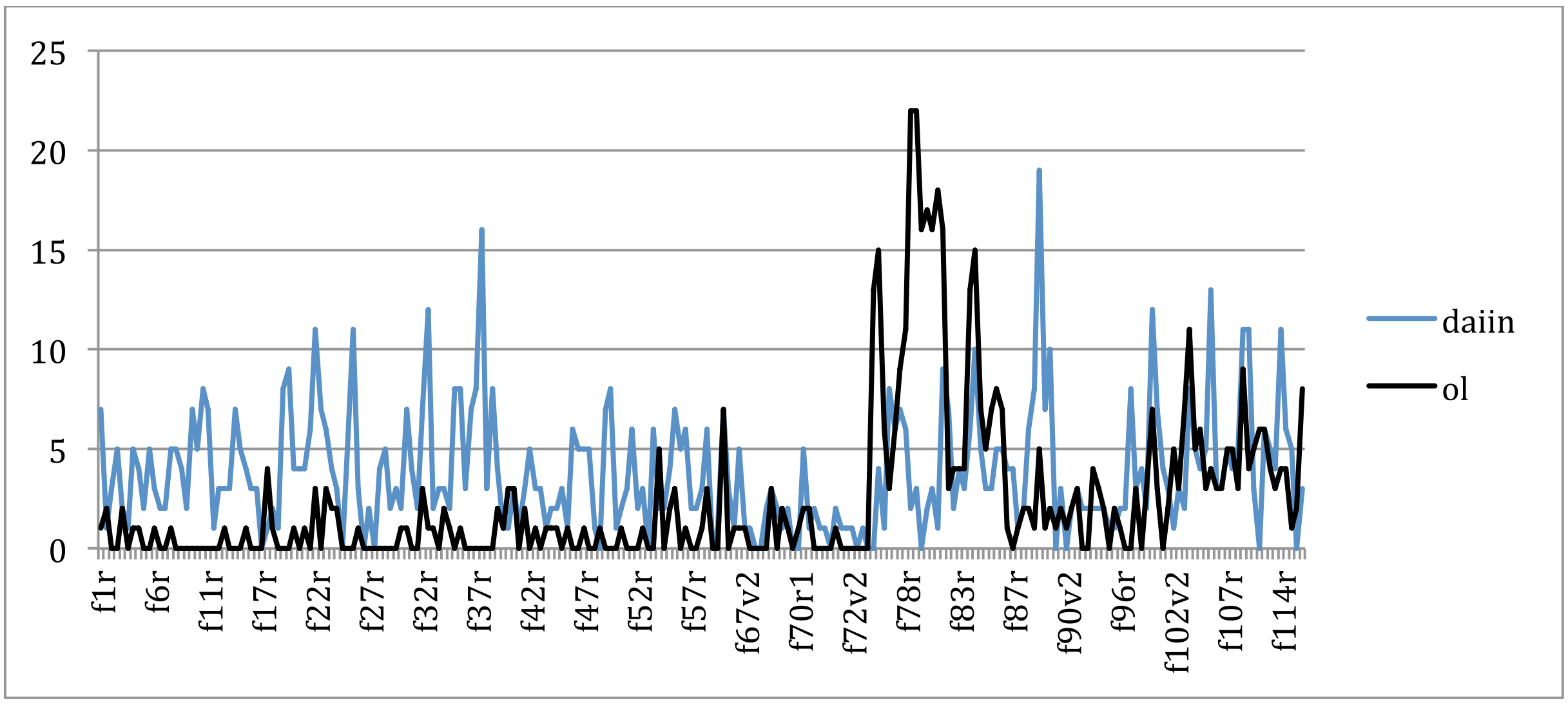

**Graph 13: appearance of ("daiin") and ("ol")**

Not only ("chedy") but also ("ol") is unequally distributed across the pages of the VMS (see graph 13). It seems that only the most frequent word ("daiin") occurs with similar frequencies throughout the manuscript.[48] It is interesting that the frequency of does not increase in the sections with pages containing more text.[49] This means that even the distribution of frequently used words is inhomogeneous for the VMS.

The examination of the relation between similarly spelled glyph groups for the whole manuscript only showed a weak correlation. For instance, in the case of ("dal") and ("dar") it is hard to detect a relation between both frequency graphs (see graph 14). The correlation coefficient between the frequency of occurrence for ("dal") and the frequency of

47 See addendum: XVI. Statistics for ("od") and ("ed") (p. 87).
48 See addendum: XVIII. Statistics for ("daiin") (p. 88)
49 Such parts are the biological section (pages <f75r> – <f84v>) and the stars section (pages <f103r> – <f116r>).

occurrence for ("dar") is only 0.2.[50] The text length of a page better explains the frequency of occurrence for both words, since the correlation coefficient between text length and frequency is 0.4 in both cases. This means that one word is not related to a specific similarly spelled word. In fact, each word is related to all other similarly spelled words. In the case of ("dal") and ("dar") this means that they are also related to glyph groups like ("ar"), ("dor") and ("sar") etc. This explains why similarly spelled glyph groups have similar frequencies. The more often a glyph group appears, the more often it was used as a source for generating other glyph groups.

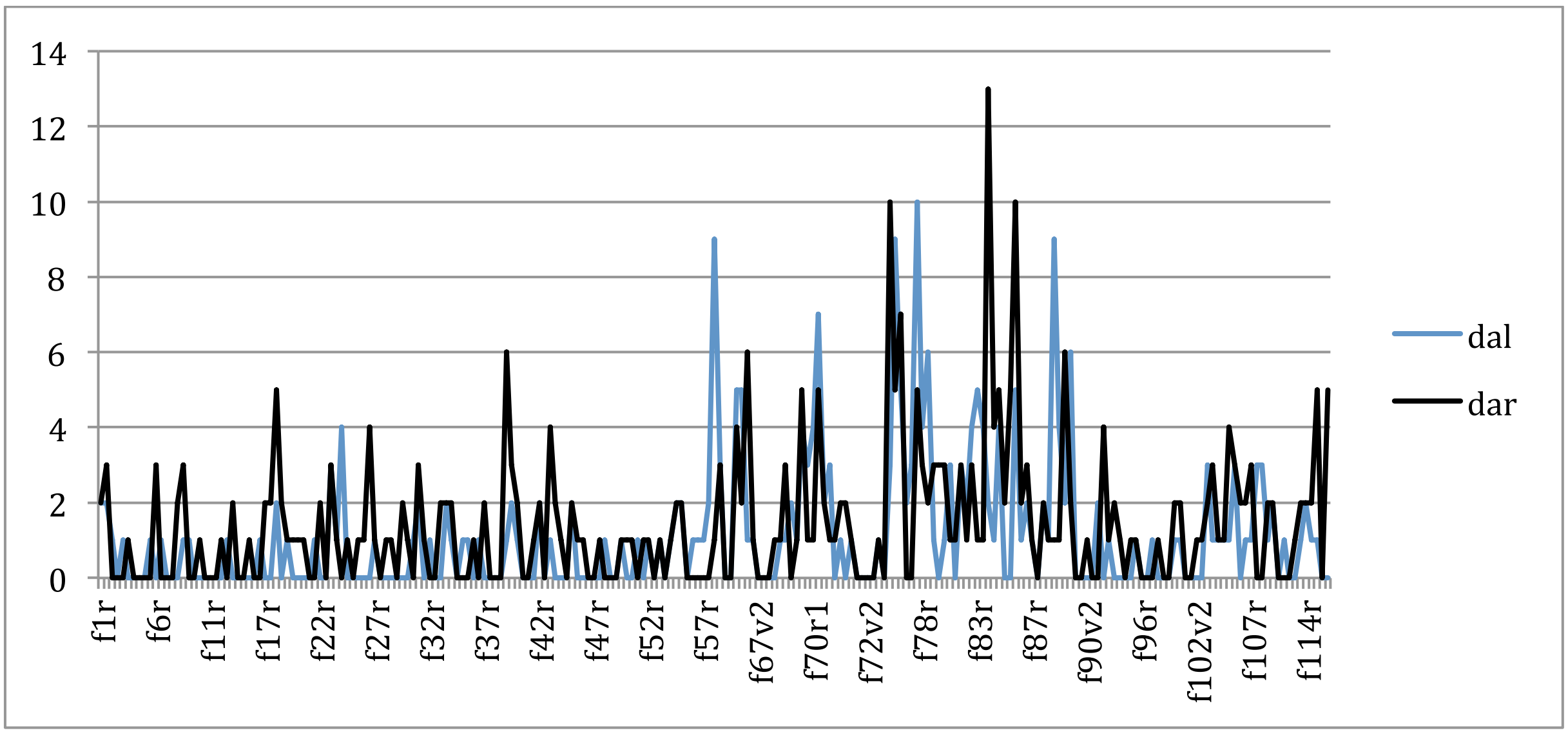

Graph 14: occurence of ("dal") and ("dar")

This also explains the exceptions in the relationship between edit distance and frequency for the grid. Since ("dair") is similar to ("dain") and ("dar") it is more frequent than would be expected from the similarity to alone. In other words, the increased frequency for ("dair") and ("dar") could be explained by a path of similarities connecting ("daiin") with ("ol"): - - - - - - .[51]

## 10 The paragraph as a functional entity

The most interesting glyphs in the VMS are the gallow glyphs , , and . As described in text generation rule VIII they play an important role for the first line of a paragraph. In 86% of cases the paragraphs start with a gallow glyph, and the glyphs ("p") and ("f") are frequent for the first line of a paragraph but rare anywhere else.[52]

[50] For calculating the correlation coefficient the frequencies were normalized with respect to the text length of each page.

[51] This similar glyph groups occur together at least once: <f10v.P.7> , <f10r.P.7> , <f69r.R.17> , <f39r.P.7> , <f81v.P.2>

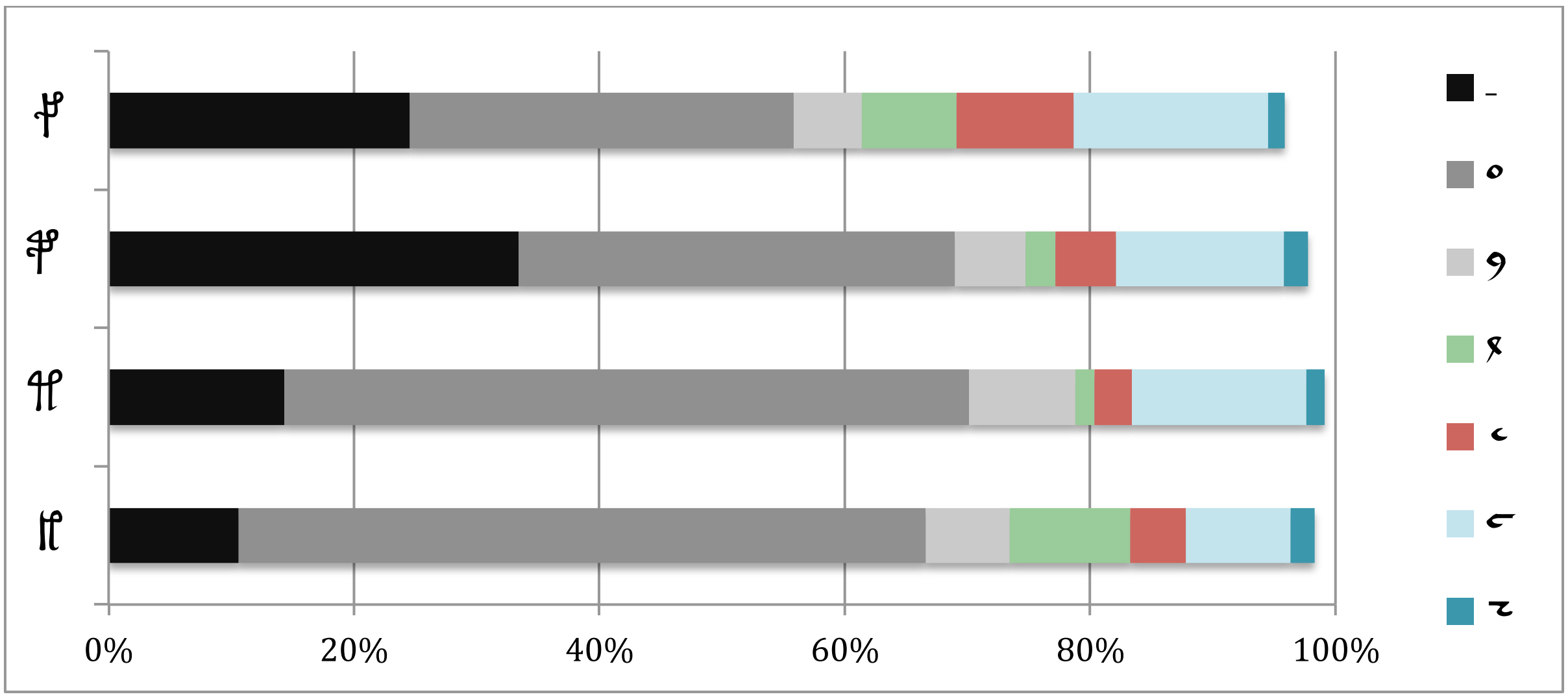

Graph 15: distribution of glyphs in front of , , and

There are two interesting statistical patterns for the gallow glyphs. The first pattern is that a gallow glyph following ("l") is most likely a glyph ("k") or ("f") with only one loop (see green bar in graph 15). The second pattern is that there are only a few cases in which ("e") occurs after ("p") or ("f") (see red bar in graph 16). The reason for this is that after and the -glyph ("e") transforms into ("ch") or ("che") (see dark blue bar in graph 16).[53]

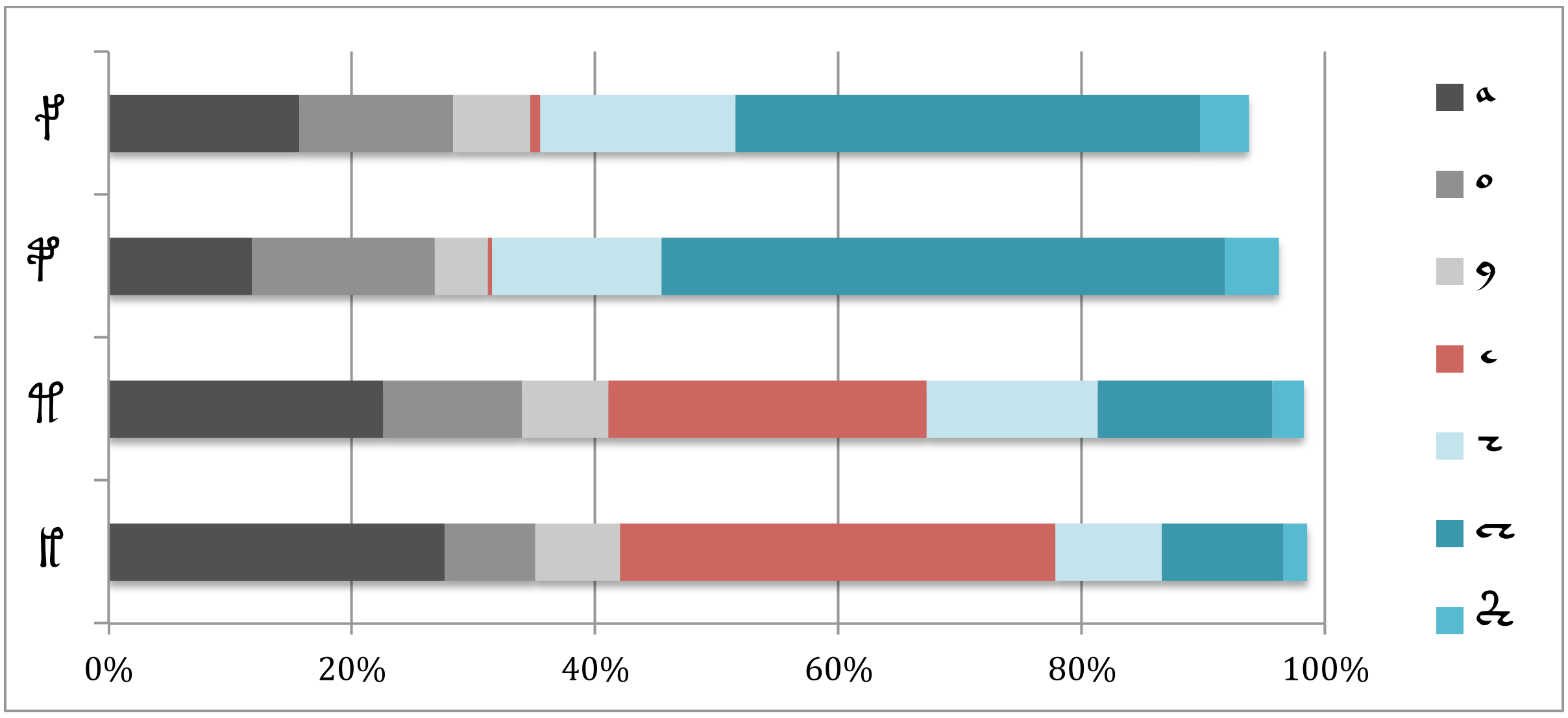

Graph 16: distribution of glyphs following to , , and

---

[52] See addendum: XXII. Distribution of glyph groups before and after , , and (p. 92).

[53] It seems that this happens for instance in <f76r.R.5> ("qokeedy qopchedy"), <f78r.P.32> ("qokeedy qopchedy"), <f112v.P.30> ("opchedy qokedy opchedy") etc.

Except of the change between ϲ ("e") and ϲϲ ("ch") the statistics for all four gallow glyphs behave similar. This indicates that they are used in a similar way. In addition, the patterns indicate that a glyph can depend on the preceding glyph.[54]

Another important observation is that the average word length for the first line of a paragraph is longer than in other lines. For the first line the average word length is 5.45, whereas for the following lines it is 5.07. As graph 17 shows, this is a result of the circumstance that more glyph groups with six or more glyphs occur in the first lines of the paragraphs. As a result the average number of glyph groups within the first line is lower.[55]

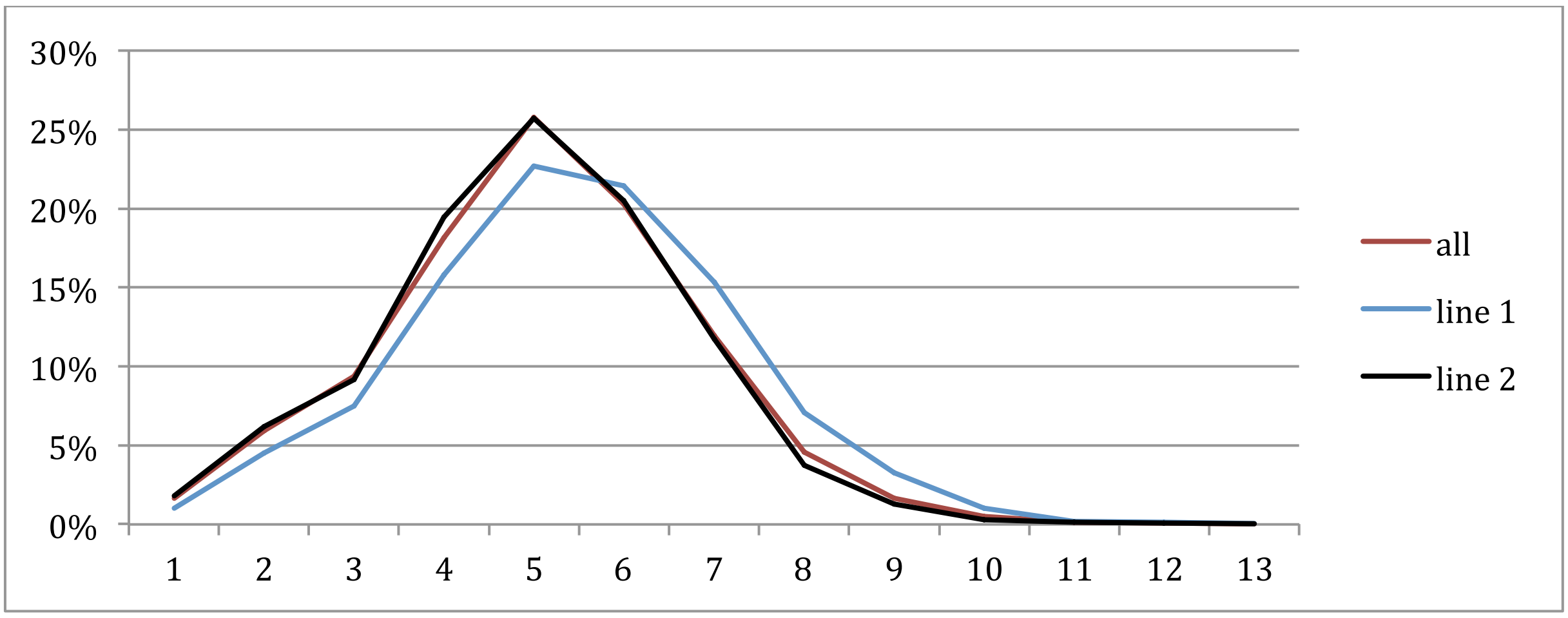

Graph 17: glyph group length distribution for the first and the second line in a paragraph

What is the reason for this difference? To answer this question it is interesting to compare the word length distribution for pages in Currier A and B. There is a difference between both "languages".

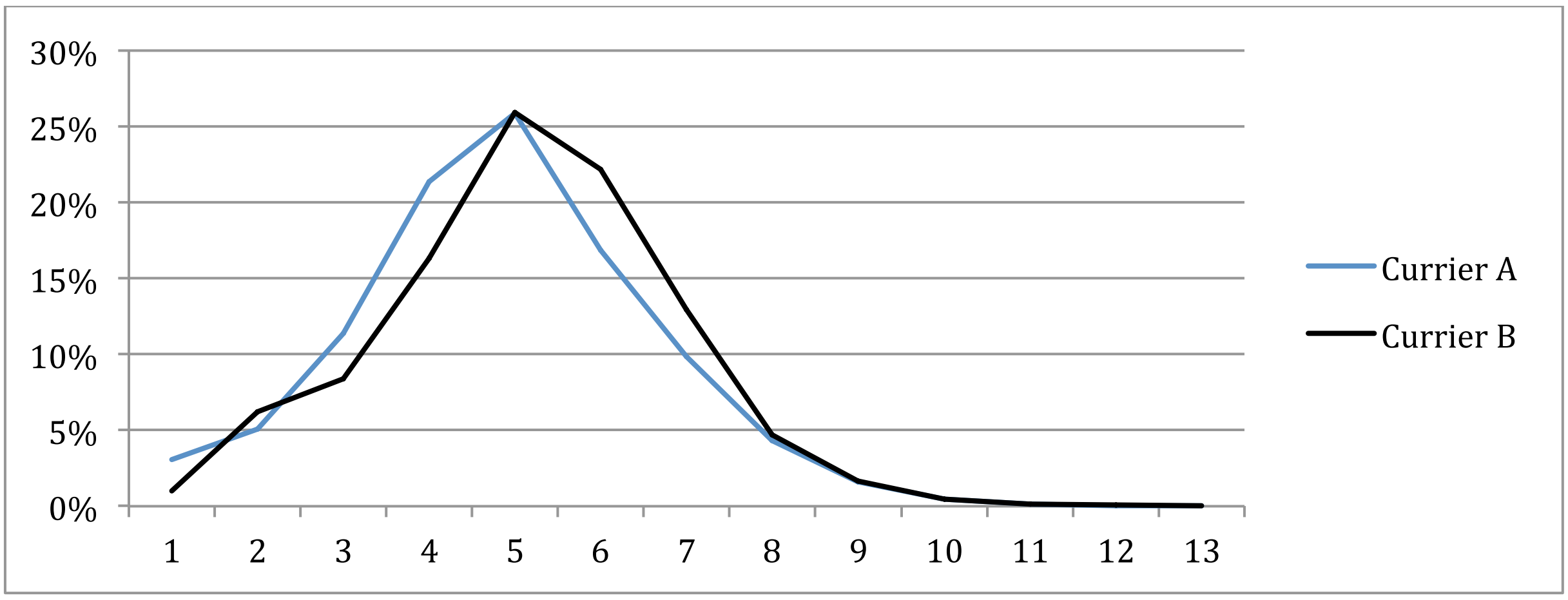

Graph 18: glyph group length distribution for Currier A and B

---

54 Another example is that a group ending with 9 ("y") increases the chance that the next group will start with 4 ("q"). Whereas 40% of the groups end with a 9-glyph, a group which ends with a 9-glyph precedes 64% of the groups starting with a 4-glyph.

55 The first line of a paragraph contains, on average, 9.4 glyph groups, whereas the second line contains, on average, 9.8 groups.

The average length of a glyph group is higher in pages written in Currier B. For paragraphs on pages written in Currier A the average word length is 4.9, whereas for pages in Currier B it is 5.2. As graph 18 shows, this is a result of the circumstance that groups containing six or seven glyphs are more frequent in Currier B.

Other than this difference, the behavior for the first line is similar for both "languages" (see graphs 19 and 20). For both "languages", groups with more than five glyphs are more frequent for the first line of a paragraph.

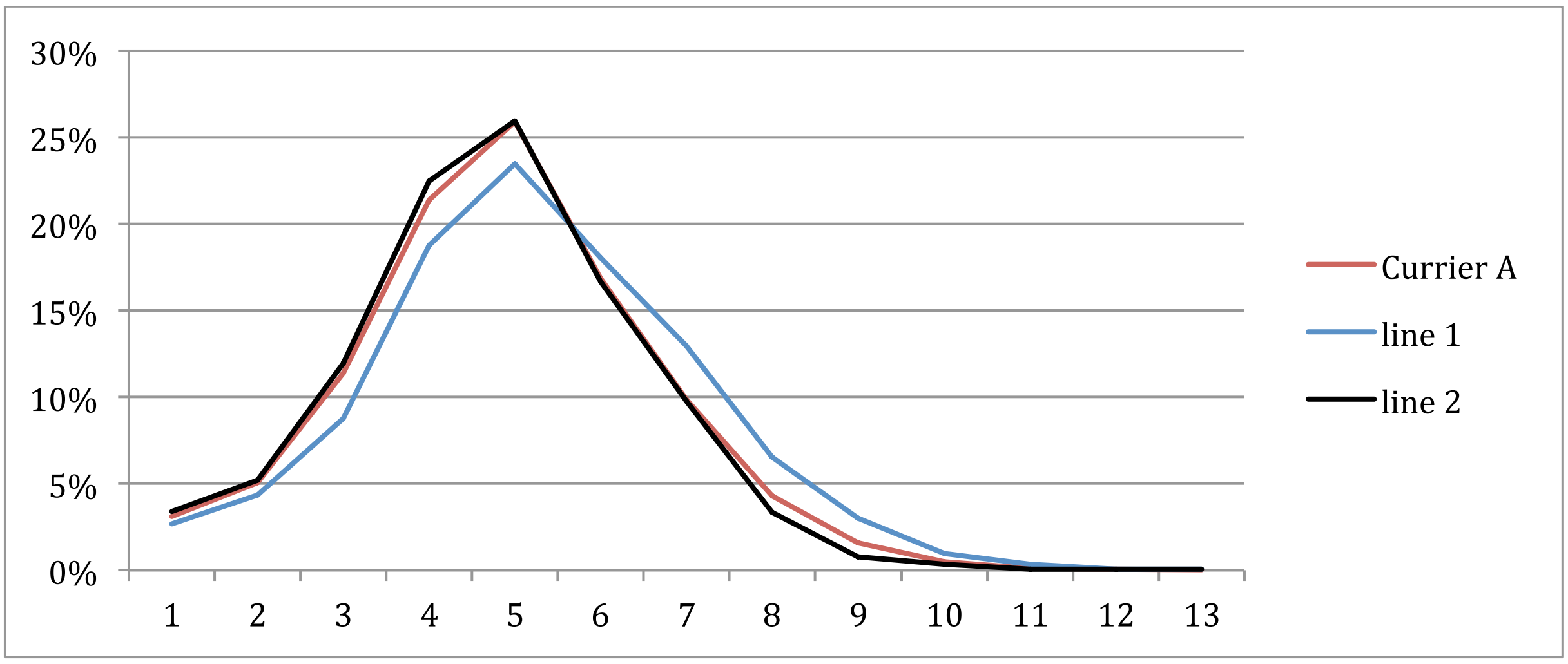

Graph 19: glyph group length distribution for Currier A

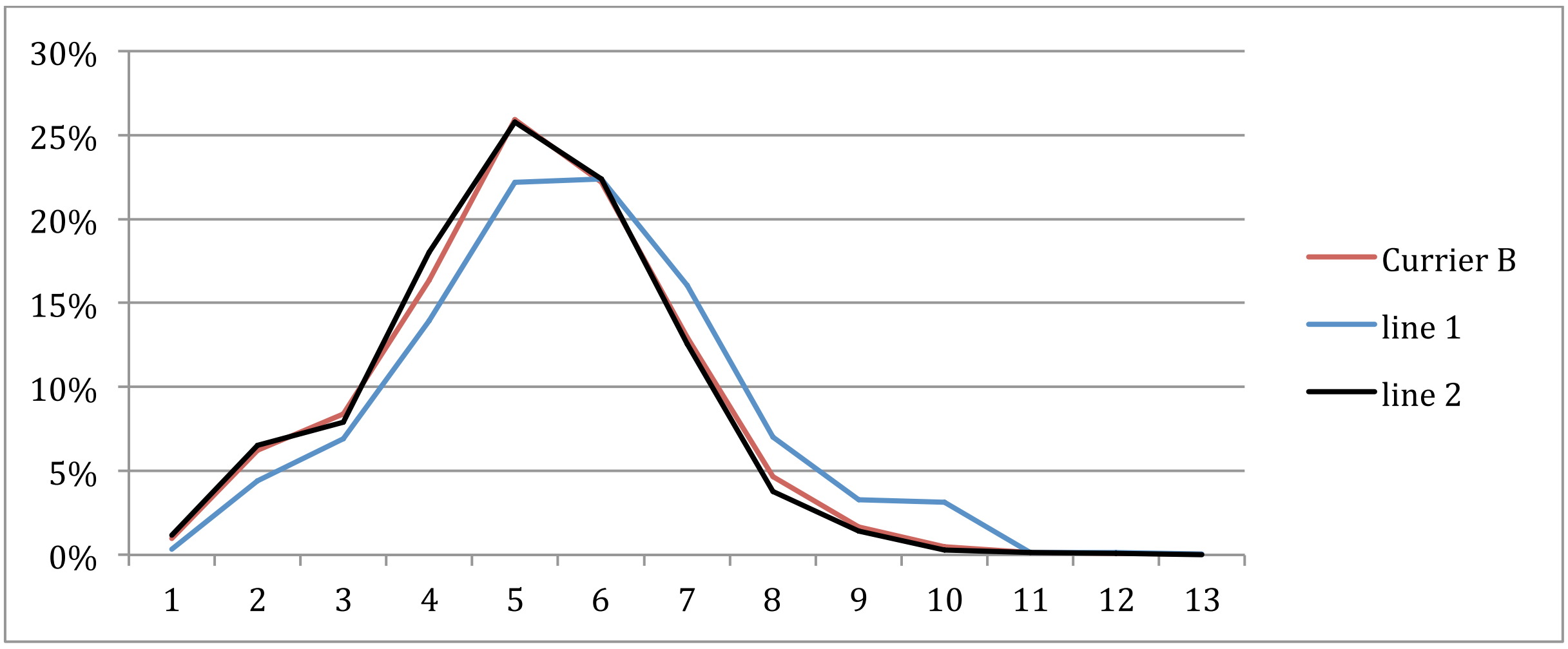

Graph 20: glyph group length distribution for Currier B

The first reason for the higher average word length in the first lines of a paragraph is that the paragraph initial glyph groups are, on average, longer. Their average word length is 6.3 as opposed to 5.0. The second reason is that because of the change from ("e") into ("ch") after and , groups including these glyphs are also longer on average. The third reason is that more groups using a gallow glyph occur in the initial lines of a paragraph (see graph 21).

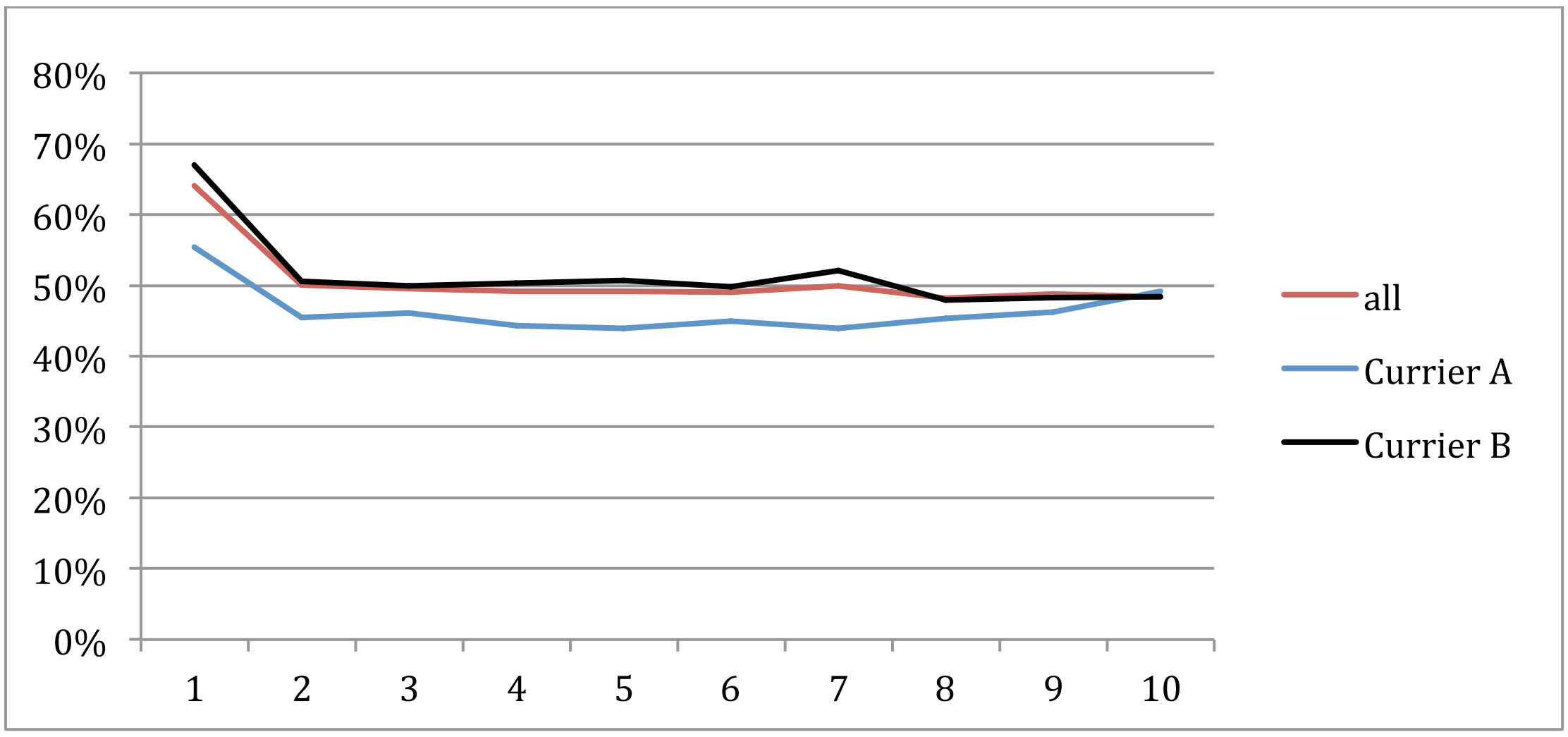

Graph 21: percentage of glyph groups with a gallow glyph in relation to the position of the line in a paragraph

In general, groups including a gallow glyph are more complex than groups without a gallow glyph. This is because the gallow glyph works as a separator. Therefore, gallow glyphs are often used in groups with two parts, such as ("chopchy"), ("chopchol") or ("chopchal").[56] For this reason, groups using a gallow glyph usually contain more glyphs on average (see graph 22).[57] Therefore, the fact that the average word length is higher on pages in Currier B can be explained by the fact that Currier B uses more gallow glyphs than Currier A (see graph 21).

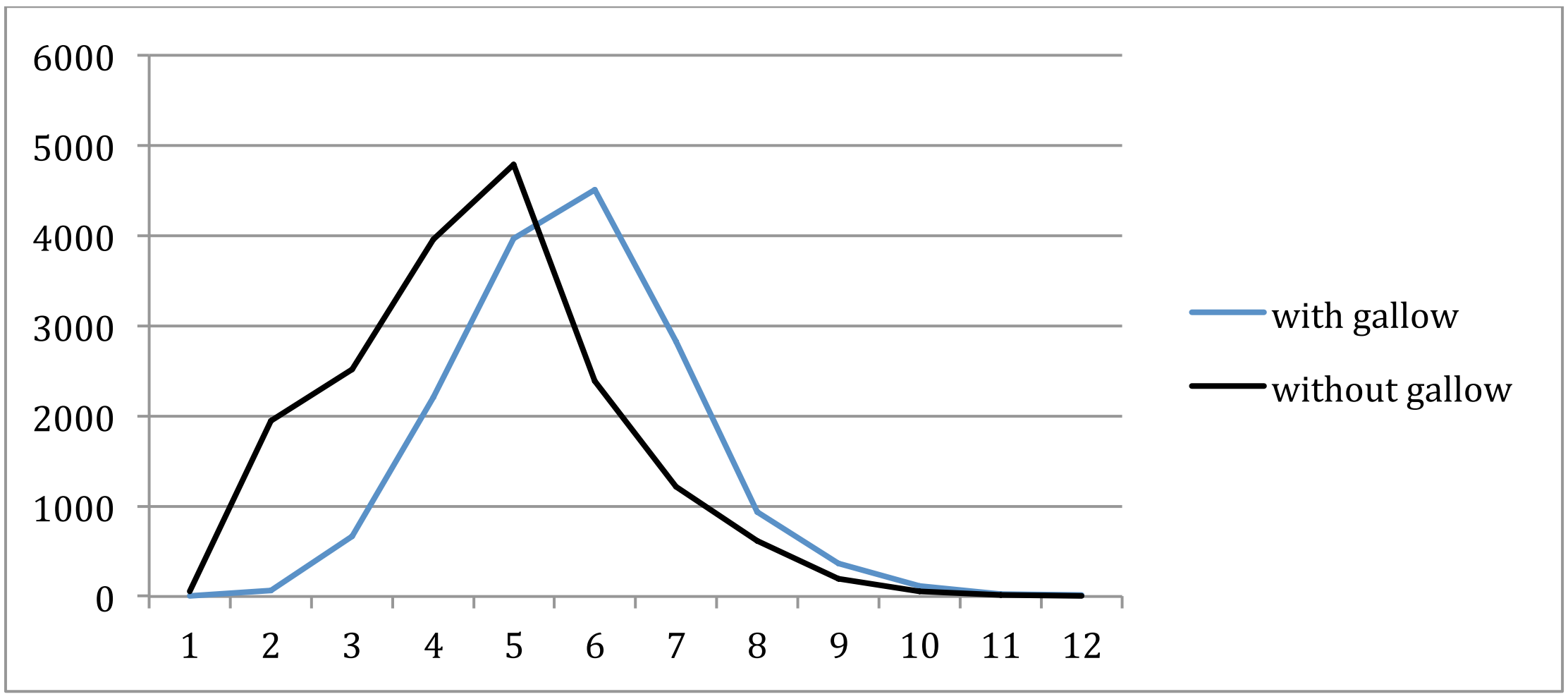

Graph 22: glyph group length distribution for groups with and without gallow glyphs

To use the described text generation method for the initial line of a page it was necessary to use another page as a source. The fact that the first lines of a paragraph behave like the first lines of a page can only mean that they are generated in a similar way. In this context it is interesting to note that on some pages several paragraphs begin with similarly spelled glyph groups.[58] For instance, on page <f3r> two paragraph initial groups ending with a final -glyph occur.

[56] A group ("chopchy") occurs in line <f4v.P.1>, ("chopchol") occurs in line <f6r.P.1> and ("chopchal") occurs in line <f10r.P.1>.
[57] The average word length for glyph groups with a gallow is 5.74 and for groups without a gallow it is 4.5.

<f3r.P.15> ("tsheoarom")

<f3r.P.18> ("pcheoldom")

In the whole of the VMS, there are only seven paragraph initial groups with a final -glyph. Therefore, it is remarkable that two of these occur on page <f3r>. Moreover, both groups are slightly similar to each other and it is possible to identify a possible source group for them on the same page in line <f3r.P.9>.

<f3r.P.9> ... ("sheoldam")

Also on page <f100r> the second paragraph starts with a group containing elements of the start groups of the first paragraph on the same page (see figure 2 p.12).

<f100r.P1.1> ("pcheol sheod")

<f100r.P2.5> ("folshody")

Furthermore, paragraph <f100r.P1> contains many groups similar to ("qokeedy"), whereas the second paragraph contains many groups similar to ("chol"). The fact that both paragraphs behave so differently is probably also a result of the larger gap between them. Because of this gap, more effort was needed to choose a group from the first paragraph as a source while writing the second paragraph. For consecutive paragraphs without a larger gap there are usually more similarities.[59]

It seems that the first line of a paragraph initializes the text generation mechanism of the VMS. For this purpose, the initial group and the gallow glyphs play an important role. A source word from a different context was used at least for the initial group of a paragraph. Additionally, glyph groups containing the gallow glyphs ("p") and ("f") followed by ("ch") or the corresponding ligatures ("cph") and ("cfh") play an important role in paragraph initial lines.

## 11 Lack of corrections

One important observation for the VMS is the lack of corrections [see Reddy: p. 79]. Did the scribe not make any mistakes? If the intention was to avoid repetitions, it is a mis-

58 See addendum: XXIV. Similarities for paragraph initial glyph groups (p. 95).

59 See for instance page <f77v> in figure 7 (p. 17).

take to repeat something. The easiest way to remove something repeated is to change it. One feature of the script used for the VMS is that in many cases one additional quill stroke is enough to change a glyph into another one. For instance, it would easily be possible to change ch into sh or e into s.

In this context it is interesting to note that at least two variants of the s glyph exist. The first variant consists of two quill strokes. In this case, it seems as if e was changed into s. In the second variant, the s was written with one quill stroke. Did this mean that the s is a ligature of e and ɔ or did this mean that the scribe was frequently changing e into s?[60]

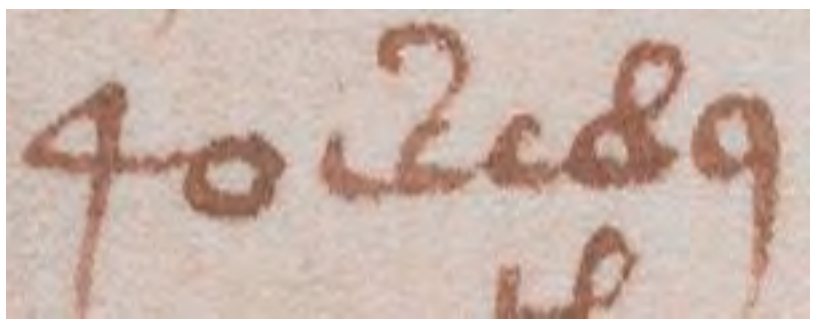

Figure 9: qoesedy ("qoesedy") in line <f82r.P2.24>

One example for the variant using two quill strokes is the unique glyph group qoesedy ("qoesedy") in line <f82r.P2.24> (see figure 9). The fact that the corresponding spelling variation qoeeedy ("qoeeedy") using e instead of s, appears three times in the VMS is an argument in favor of the hypothesis that the scribe was only changing e into s.[61]

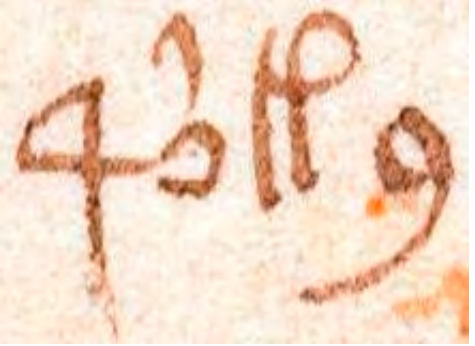

Figure 10: qoky ("qoky") in line <f2r.P10>

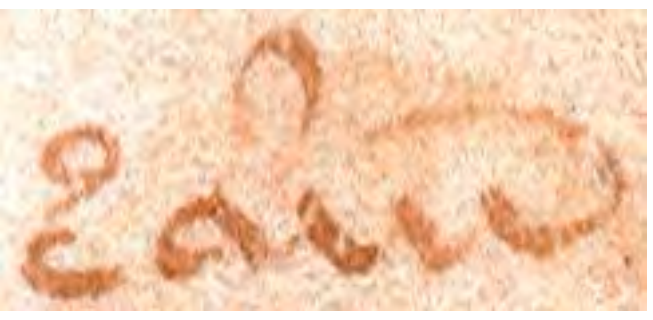

Figure 11: saiin (saiin") in line <f2r.P10>

---

[60] One hypothesis could be that the glyph r ("r") is also a ligature of i ("i") and ɔ. A similar hypotheses is that of Currier: "We have the fact that you can make up almost any of the other letters out of these two symbols i and e; it doesn't mean anything, but it's interesting." [Currier]

[61] qoeeedy ("qoeeedy") occurs in lines <f76.P.6>, <f76v.P.17> and <f105r.P2.13>.

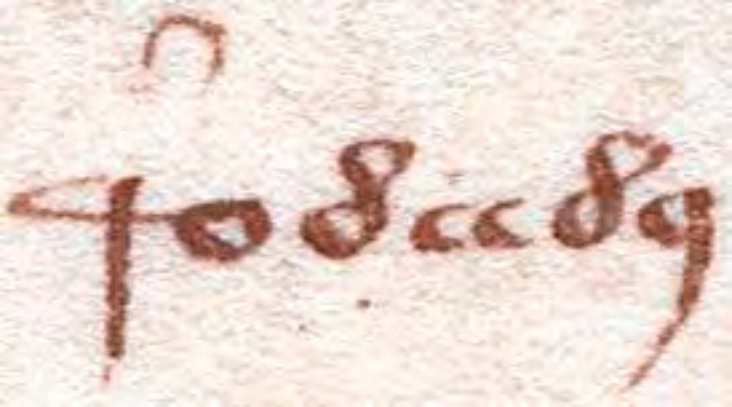

Figure 12: qodeedy ("qodeedy") in line <f105r.P2.25>

Normally, an additional quill stroke was used to change e into s or ch into sh. In some cases, such a plume was also added to other glyphs. This is at least the case for qoky ("qoky") and saiin ("saiin") in line <f2r.P.10>, for qo ("qo") in <f49r.P.14>, for qocho ("qocho") in line <f49r.P.16>, for ol ("ol") in line <f84r.P1.4> and for qodeedy ("qodeedy") in line <f105r.P2.25> (for examples see figures 10, 11 and 12). Since these exceptions exist, the question arises as to whether ch and sh are indeed two different glyphs or whether the additional '-stroke has its own meaning? The second hypthesis is also supported by the occurrence of the glyph group 's in line <f4r.P.2>.

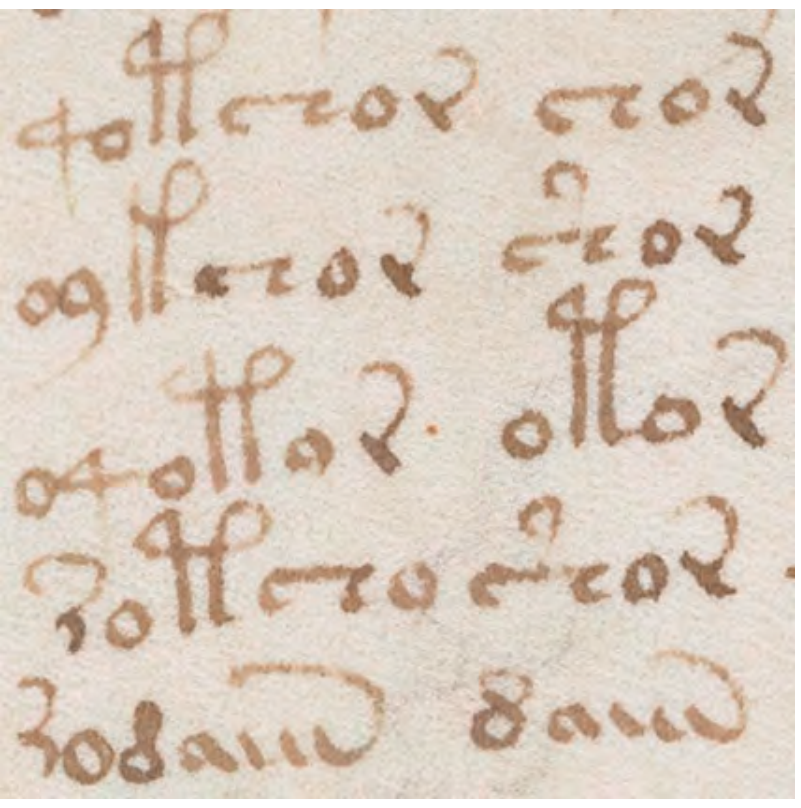

Figure 13: line <f10r.P.8> - <f10r.P.12>

Furthermore, on page <f10r> a glyph 3 occurs twice (see figure 13). For this glyph the question is whether it is a variant of r or a ligature of two ' glyphs?
The lines <f10r.P.8 - 12> start with qo, oy, oq, 3o and 3o (see figure 13). This is interesting since only 20 words starting with oq can be found in the VMS. It seems as if the scribe was developing new spelling variants at this place. First, glyph groups starting with oq and, second, a new glyph 3.

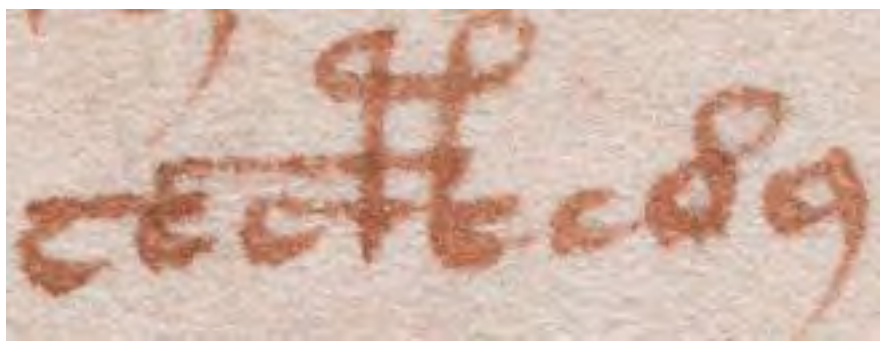

Figure 14: chcthedy ("chcthedy") in line <f115v.P.32>

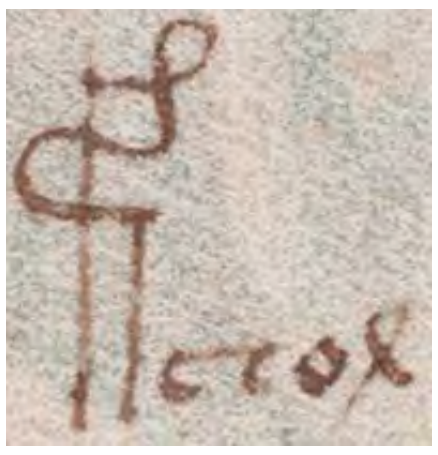

Figure 15: ("pchol") in line <f14r.P.8>

For gallow glyphs some irregular variants exist.[62] For instance, ("chcthedy") is written six times as expected. But on page <f115v> the group is written with an additional quill stroke connecting with (see figure 14). Another example of an ambiguous glyph is the in line <f14r.P.8> (see figure 15). It is uncertain if this glyph is used as or as or as a new character. It seems as if the scribe was sometimes testing different design variants for the glyphs used.

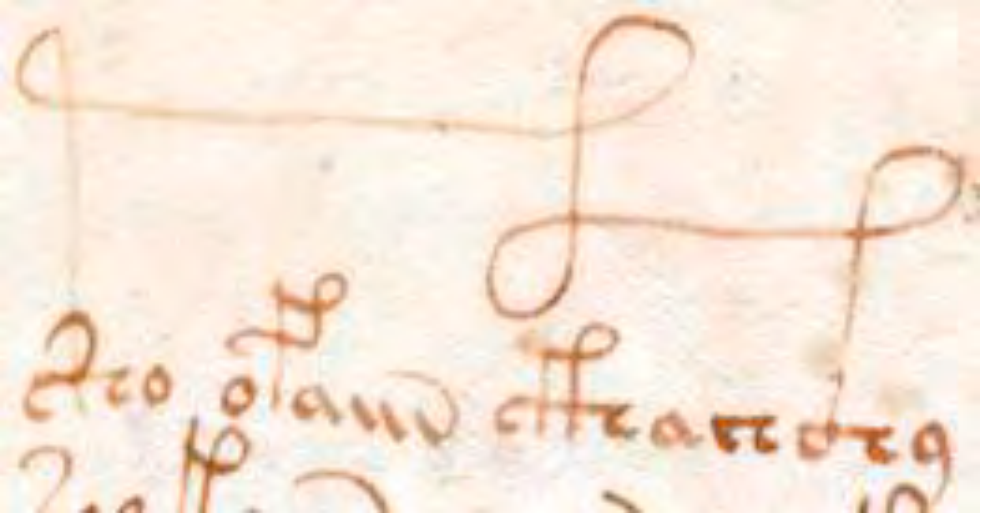

Figure 16: the first word on page <f42r>

Other examples are curlicue gallow glyphs used as initial markers for some pages. One example is on page <f42r> (see figure 16). In this case, the question arises as to whether this glyph is only used as decoration or has some meaning.[63]

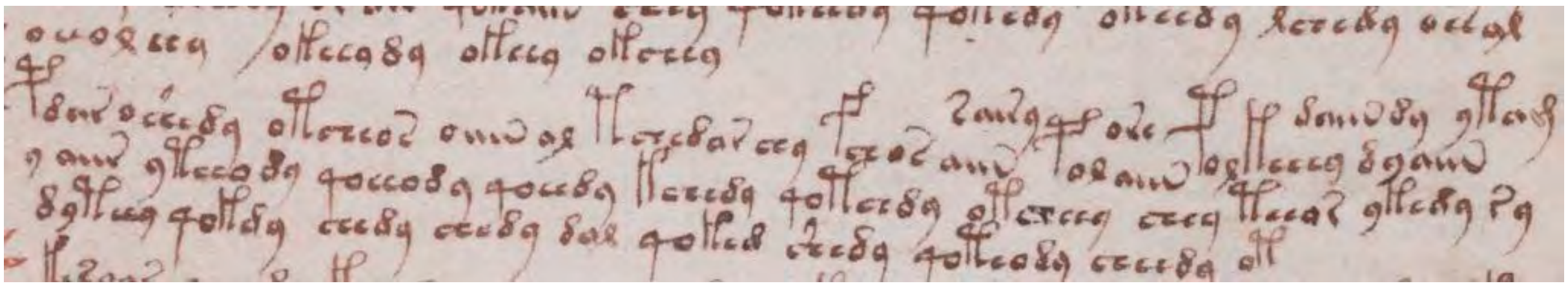

Figure 17: part of page <f105r>

Another example of ambiguousness in the VMS can be found on page <f105r> in lines 9a and 10. When writing line 10, the scribe left a larger gap (see figure 17). It seems that the scribe was not satisfied with the resulting layout. To make the gap less obvious, he used larger gallow glyphs and filled the gaps between them with additional glyph groups.[64] This raises the question whether the additional glyph groups in line 9a constitute a separate line or whether they belong to line 10.

---

62 Irregular gallow glyphs can be found in lines <f8r.P3.14>, <f9r.P.6>, <f14r.P.8>, <f24v.P.1>, <f25v.P.3>, <f36r.P.1>, <f42.P.2>, <f76v.P.25>, <f79r.P.1>, <f84r.P.15>, <f86v6.P.1>, <f87v.P.1>, <f90v2.P.1>, <f95v1.P.1>, <f114r.P1.1> and <f115v.P.32>.

63 In the case of page <f42r> different transcriptions are conceivable. Currier leaves the additional glyph unaccounted for. He reads ("sho ofaiin cthachchy"). The transcription of the first study group is ("tsho ofaiin cthachchy"). And Takhahasi reads ("sho ofaiin cthachcthy").

It is to be expected that the scribe would run out of space at the end of a line. Therefore it is remarkable that the end of the lines nearly always fit into the available space. In addition, on many pages the text is placed around illustrations (see for instance figure 1 on p. 1). In such cases, the available space was also limited. Therefore it would be no surprise if sometimes the last glyphs in a line were squeezed into the available space. However, there are no such crowded places in the VMS. This behavior was described by Currier in 1976 as follows:

> "The ends of the lines contain what seem to be, in many cases, meaningless symbols: little groups of letters which don't occur anywhere else, and just look as if they were added to fill out the line to the margin. Although this isn't always true, it frequently happens." [Currier]

One possible explanation for this feature is that it was possible for the scribe to select glyph groups which fitted into the available space. With other words the lines fits into their margins because the text layout was made during writing.

## 12 Discussion

What conclusions can be drawn from the observations made for the text of the VMS? First, it is possible to describe a text generation method for the VMS. The text was generated by copying and modifying already written text. This method explains the fact that similarly spelled glyph groups are usually near to each other. The text generation method found also explains why the set of used words is constantly changing and why repeated sequences with more than five words are missing in the VMS. The hypothesis that the text was generated by copying already written text leads to the conclusion that the glyph groups only behave like words, but have a different meaning.

Writing the text of the VMS was clearly a monotonous and laborious process and required considerable time and motivation. However, only knowledge available in the 15th century was needed to use the described text generation mechanism. Therefore there is no indication that the text belongs to a later

64 These words fit into the gaps between the gallow glyphs or the gallow glyphs fit into the gaps between the words. That the gallow glyphs were written after line <f105r.T1.9a> is less likely because in this case it would be necessary to predict the gallow glyphs while writing line 9a.

period than the parchment, which was carbon-dated to the $15^{th}$ century.

The script used to write the VMS uses similarly shaped glyphs. The design of these glyphs is based on some simple forms starting with  and . By combining these simple forms with some additional ink strokes, complex forms like 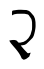 or  were also generated. In the same way as the scribe combined some ink strokes to design similarly shaped glyphs he also combined these glyphs to build similarly spelled glyph groups and these groups to generate a text.[65]

If everything is possible, nothing can be excluded. It is not possible to detect a falsely identified glyph from its context, because for nearly every group multiple similar spelling variations exist. For "reading" the VMS this means that it would be necessary to identify every single glyph. But since similar glyphs can replace each other, the identification of glyphs is even more difficult than necessary. This means that it is far from being easy to transcribe the VMS. For this reason it is no surprise that all published transcriptions of the VMS frequently differ.

Do similar spelled glyph groups share similar meaning? For the possibility that a relation between similar spelled groups exists, speaks that such groups occur with similar frequency and that it is typical for such groups that they occur near to each other. Therefore it seems unreasonable to assume that every added or replaced glyph would imply a complete change of the meaning of a glyph group. But if this is the case how is it possible to explain the ambiguousness of the VMS? The first thinkable explanation is that it is not necessary to identify each glyph in order to read the VMS because similar glyphs share the same meaning. The second explanation is that only some glyphs have meaning, and the third explanation is that it is not necessary to read the glyphs because they have no meaning.

The assumption that similar glyphs share the same meaning would result in a monotonous text repeating the same words again and again. Even without such an assumption the VMS contains monotonous sequences[66] such as the following two:

---

[65] Similarly, the plant illustrations in the herbal section contain known elements combined with fantasy elements.

[66] Pages written in Currier B, especially the pages in the biological section (pages <f75r> – <f84v>), are more monotonous than pages written in Currier A. The result of the copying mechanism depends on how often a new source word is selected and how often a source word is selected from a previous paragraph or page. On the other hand, the most efficient way is to use a source word multiple times or to select the next source word straight from the last line. A possible explanation is, therefore, that the scribe was more thoughtful when writing the pages which use Currier A and less careful when writing the pages which use Currier B.

<f75r.P.38> qokeedy qokeedy qokedy qokedy qokeedy xdy
<f108v.P.39> qokeedy qokeedy qokeedy qotey qokeey qokeey otedy

If similar glyphs share the same meaning, such sequences would only repeat the same information multiple times.

The text generation method found confirms the conclusion that the glyph groups in the VMS are not used as words. All features of the VMS speak against such a hypothesis. The first important feature is the weak word order. The second feature is that it is possible to group the words into a daiin-, an ol- and a chedy-series. The third feature is that the line behaves as a functional entity. And last but not least the fact that similarly spelled glyph groups occur with similar frequencies also speaks against such a hypothesis. For a natural language or for a constructed language the words should be used because of their meaning, and relations between words should be expressed by grammatical rules. Since the only relation found for the words within the VMS is that similarly spelled glyph groups are used near to each other, an unknown natural language or a constructed language can be ruled out.[67]

More interesting is the hypothesis that the VMS is an encoded book of secrets. One thinkable hypothesis is that only a subset of glyphs has a meaning and the gaps between them are filled up with meaningless content. However, for common glyph groups in the VMS each glyph can be replaced or deleted. Furthermore, repeated word sequences with more than five words are missing for the VMS. Therefore it is impossible that the VMS was encoded using a simple cipher such as: only the first glyph of a glyph group should be read. But if the cipher is more complex, this raises the question why it was necessary to construct a new script and to simulate words using letters for vowels and consonants. Would it not be easier to use a normal text to hide a message? Nobody would assume a hidden message behind a text that can be read. Therefore it is unlikely that only a small group of glyphs for each page have a meaning. Furthermore, the assumption that it is possible to distinguish between meaningful and meaningless glyphs would give a decipherer a degree of freedom, which means anything could be read into the VMS. Therefore, it would not be possible to prove or disprove such a hypothesis.

Another conceivable hypothesis is that one glyph group of the text stands for a letter or a syllable. In this case, the large number of differently spelled glyph groups could be explained if multiple groups stand for one and the same letter or syllable. One argument in favor of such an assumption is that the scribe was generating the same glyph groups numerous times on different pages. In most cases, it is possible to distinguish the glyph groups of the three series by their end characters, which are in most cases ("n"), ("l") or ("y"). In combination with the use of common prefixes such as ("qo"), ("l") and ("ch") it seems possible to build an encoding system for an alphabet using at least 24 letters.[68]
A method of encoding a text in a similar way has been known at least since the 16th century. Johannes Trithemius described such a method in his book Polygraphiae in 1508 [see Hermes: p. 139-152]. Trithemius´ method uses a code table to assign multiple words to each letter.[69] The use of such a method would explain the occurrence of similarly spelled glyph groups, the word length distribution and the occurrence of three different word series. On the other hand, 8133 is a large number of different glyph groups. Even if hapax legomena[70] are excluded, 2486 glyph groups and therefore approximate 100 different encoding variants for each letter of the plaintext alphabet will remain. The use of an encoding table containing such a number of glyph groups would make the encoding process unreasonably laborious. An explanation for these observations could be that the scribe was using synonymous glyphs or was adding meaningless glyphs. In such a case the problem would be to distinguish between meaningful and meaningless spelling variations. To use the encoding method described by Trithemius is by no means an easy procedure. As corrections are missing in the VMS, a method to mark deleted glyph groups or glyphs would be needed. One hypothesis could be that glyphs were marked as deleted by an additional quill stroke, such as the stroke changing into or into .

[67] Montemurro and Zanette also argue with respect to the context dependency: "Words that are related by their semantic contents tend to co-occur along the text." [Montemurro]. They come to the opposite conclusion: that there is a genuine linguistic structure behind the text. In fact, the context dependency only points to a self-referencing system, but this alone is not enough to allow the conclusion that this must be a linguistic system.

If the text was encoded using the method described by Trithemius the observed weak word order would lead to the conclusion that a plaintext word, which occurred multiple times was encoded differently each time. How could it be possible? One explanation could be that a local element was used for encoding. One observation for the VMS was that similarly spelled glyph groups are used near to each other. An explanation found for this feature is that the scribe was using glyph groups in his sight field to generate new text. In this way it would be possible to use a local element for the encoding process. On the other hand, the process of searching for an encoded glyph group and changing this group would make the encoding process more complicated than necessary. It would seem to be easier to select the next code word from a code

68 Since all three word series are used frequently and in similar proportion (1:2:2) for the pages of the VMS one hypothesis could be that in such a case the -series was used to encode fewer letters than the - and the -series.

69 See addendum: XXV. Code table as described by Trithemius (p. 96).

70 Hapax legomena (lat.): words which occur only once tablet than to search for a suitable glyph group within the text already written.

Sometimes it is possible to expand repetitive phrases under the assumption that similarly spelled glyph groups are related to each other. For the hypothesis of an encoded text using the method described by Trithemius such phrases would be a good starting point for searching for glyph groups standing for the same letter or syllable. From the following two examples on pages <f84r> and <f83v> – <f84v> it could be concluded that for 4olfc89 it is possible to replace c with cc and for 8o9 that it is possible to delete the line initial 8 and to replace o with a.

<f84r.P.3> 2cc89 4olfc89 4olfcc89 4olfc89 ccc89 olfaw ccc9

<f84r.P.10> 2cc89 4olfc89 4olfcc89 4olfcc89 ccc89 2aw ccc9

<f83v.P1.2> 8o9 2cc89 4olfc89 2cc89

<f84r.P.6> o9 2cc89 4olfc89 2cc89

<f84v.P.26> 8a9 2cc89 4olfc89 2cc89

Both phrases occur on the same or on repetitive pages. Therefore, it is possible to explain them by the assumption that the scribe generated similar phrases several times by copying from the same source or by copying them from one another. It is thus possible to explain all features of the VMS by the hypothesis that only permutations of the same glyph groups are repeated over and over again. In other words, it is not necessary to assume that the text of the VMS contains a message. Moreover, the observed weak word order is a strong argument against the assumption that the sequence of glyph groups is influenced in some way by a message behind the text.

Another simple method of generating distinguishable glyph groups while writing could be to use only some basic geometrical elements such as linear strokes and curved strokes to carry meaning. It is, for instance, conceivable that it was only necessary to count the number of c or ι strokes to encode a glyph group. But even for such a method, numerous repetitive phrases, standing for repeated plain text words, could be expected. Thus, the fact that longer repetitive phrases are missing speaks against such a method. Another conceivable hypothesis is that the paragraph initial gallow glyphs are used as markers for a change in the encoding procedure. In such a case, one and the same glyph group would stand for a different meaning if the paragraph were marked by a different initial glyph. But such a hypothesis would not explain that similar groups tend to occur in the same positions within the lines. Also, the absence of corrections speaks against such a hypothesis. It is unreasonable to assume that an ambiguous encoding method can be used without making any errors. Even under the assumption that an already encoded text was copied from another source, numerous copy errors would be expected. Moreover the fact that the end of the text lines nearly always fit into the available space indicates that the text was generated during writing. Therefore the scribe was probably using a rather simple method.

The grid demonstrates that all glyph groups in the VMS are connected to each other. Most glyph groups contain common elements such as aiin, ol, ar, dy and common prefixes such as qo, ch or l. Even hapax legomena, such as daiinol ("daiinol"), ollain ("ollain") or qockhedy ("qockhedy") mainly consist of such elements. Moreover, the word series are also related to each other. It is for instance possible to demonstrate multiple paths between daiin ("daiin") and ol ("ol"). One example is daiin - dain - dair - dar - ar - al - ol.[71] A path connecting ol ("ol") with chedy ("chedy") can also be found: ol - chol - cheol - cheod - cheody - chedy.[72] If similar groups have similar meaning, the fact that all groups in the VMS are connected to each other can only mean that they share similar meaning or have no meaning at all.

The most plausible explanation for the text generation method described in this paper is that the glyphs have no meaning. To use a glyph group already written as a source for generating another group is only efficient if it is possible to select any group. However, this is only possible under the assumption that the glyph groups have no meaning. In this case, it would not matter which source word the scribe chose. By changing meaningless glyph groups it would be possible to generate the text while writing. The scribe only needed to choose a source word, a rule and to select between similar-looking glyphs. And he could be sure that someone not familiar with his idea would only see different words. The reason for this is that we are used to reading similar glyphs like o and a or q and y differently. It is reasonable to assume that with some training it is very efficient to copy a text using the text itself as a source. For doing so it is not necessary to switch between an external source and the text being written. Nor is it necessary to check or to correct copy errors if the generated text has no meaning.

Such a method would also explain why similarly spelled glyph groups occur frequently one above the other and have a comparable frequency. The most plausible hypothesis is therefore that it is not possible to read the VMS because the glyph groups are meaningless. With this hypothesis it is possible to explain the features of the VMS described in this paper. Par-ticularly, the fact that similarly spelled glyph groups occur with similar frequencies could be explained. If the scribe preferred words he was familiar with, he would unconsciously generate such words more often. Since a more frequently occurring glyph group would be used more often as a source for other glyph groups, the spelling variations of this group would also occur more frequently.

[71] <f10v.P.7> daiin dain, <f10r.P.7> dain dair, <f69r.R.17> dair dar, <f39r.P.7> dar ar al, <f81v.P.2> al ol

[72] <f17v.P.18> ol chol, <f89r2.P2.5> chol cheol, <f31r.P.8> cheol cheod, (<f38v.P.5> cheod chody, <f48v.P.3> chody chedy), <f68r3.C1.1> cheody chedy

An observation for the paragraph initial groups was that an element from another context was used as a source word for them. If the scribe was generating a pseudo text it makes sense to disrupt the monotonous copying process from time to time. It is therefore possible to explain the paragraph initial lines by the hypothesis that the scribe was adding periodically some new elements to the copying process to prevent the text becoming too monotonous.

Another indication which points to a pseudo text is the observation that there are characteristic properties for glyph groups in a certain position within the VMS-lines. This occurs because the scribe preferred glyph groups in the same position some lines above as a source for generating new text. A side effect of this text generation method is that the position is encoded within the glyph groups. For a meaningless pseudo text the existence of such patterns is expected, since it is not possible to write something meaningless off the top of one’s head without repeating the same patterns again and again.

Furthermore, it makes sense to optimize the writing process during writing a pseudo text. An indication that this was the case for the VMS is that the text on the pages in Currier B is more repetitive than that on the pages in Currier A. Last but not least the hypothesis of a pseudo text also explains the absence of corrections in the VMS, since there would be no need to correct something meaningless.

By using similar glyphs and similarly spelled glyph groups it would be possible to write the same glyph groups again and again in different ways. Using this procedure, it would be possible to generate a meaningless text efficiently. Even ambiguous glyphs do not do any harm if these glyphs have no meaning. Finally, it makes sense to illustrate a manuscript nobody can read. The illustrations would attract attention and everyone would assume that the text might be explaining the secret he or she could see within the strange illustrations. Or, in other words, if the text cannot be read or decoded then a background story or some fanciful illustrations are needed to make a manuscript interesting.

In the end, the most plausible hypothesis for the Voynich manuscript is that the text generation method described in this paper was used to generate a meaningless pseudo text.

Some questions still remain. For what reason did the scribe invent a text generation mechanism to write the VMS? Was the manuscript some type of art? Was the VMS a test to prove that it is possible to fill a whole manuscript with such a method? Was the purpose of the manuscript to create a secret to impress someone? Or was the purpose to sell it for money?

**Acknowledgements**

The author wishes to thank Jürgen Hermes and Nick Pelling for stimulating discussions and helpful suggestions.

**Bibliography**

Prescott H. Currier, 1976: New Research on the Voynich Manuscript: Proceedings of a Seminar. Unpublished communication, retrieved July 19, 2014 from http://www.voynich.nu/extra/curr_main.html

John Dryden, 1697: The Aeneid of Virgil
Retrieved July 19, 2014 from
http://www.sacred-texts.com/cla/virgil/aen/index.htm

Jürgen Hermes, 2012: Textprozessierung - Design und Applikation. Dissertation, Universität zu Köln.

Gunther Ipsen, 1954: Zur Theorie der Entzifferung. Studium Generale, Jhg. 7, Heft 7, p. 416-423.

M.E. D'Imperio, 1980: The Voynich Manuscript: An Elegant Enigma, Aegean Park Press.

Marcelo A. Montemurro, Damián H. Zanette, 2013: Keywords and Co-Occurrence Patterns in the Voynich Manuscript: An Information-Theoretic Analysis. PLoS One, 8(6), e66344.

Sean B. Palmer, 2014: Which Voynich MS glyphs are related?
Retrieved July 19, 2014 from
http://inamidst.com/voynich/related

Sravana Reddy / Kevin Knight, 2011: What We Know About The Voynich Manuscript, Proc. ACL Workshop on Language Technology for Cultural Heritage, Social Sciences, and Humanities, p. 78-86.

Andreas Schinner, 2007: The Voynich Manuscript: Evidence of the Hoax Hypothesis, Cryptologia, Vol. 31, Iss. 2, p. 97-107.

Jorge Stolfi, 2000: On the VM Word Length Distribution
Retrieved July 19, 2014 from
http://www.ic.unicamp.br/~stolfi/voynich/00-12-21-word-length-distr/

Daniel Stolte, 2011: UA Experts Determine Age of Book 'Nobody Can Read', Retrieved July 19, 2014 from
http://uanews.org/story/ua-experts-determine-age-book-nobody-can-read

Takeshi Takahashi
Retrieved July 19, 2014 from
http://voynich.freie-literatur.de/index.php?show=extractor

Virgil (P. Vergilius Maro 70-19 BC), The Aeneid
Retrieved July 19, 2014 from
http://www.sacred-texts.com/cla/virgil/aen/index.htm

Elmar Vogt, 2012: The Line as a Functional Unit in the Voynich Manuscript: Some Statistical Observations
Retrieved August 10, 2014 from
http://voynichthoughts.files.wordpress.com/2012/11/the_voynich_line.pdf

René Zandbergen / Gabriel Landini, 2000: EVA Alphabet,
Retrieved July 19, 2014 from
http://www.voynich.nu/extra/eva.html

## Addendum

### I. Repeated sequences using the same words

There are 35 repeated sequences using the same three words which occur at least three times within the VMS (transcription after Takahashi). In five cases, the order is uniform. In all other 30 cases, the word order is changed.[73]

It is notable that in most cases the repeated sequences consist of words with similar or the same spelling, as in chol shol cthol ("chol.shol.cthol") or qokeey qokeedy qokeey ("**qokeey**.qokeedy.**qokeey**"). 10 sequences contain the same word **twice**. 14 sequences contain at least two similar spelled words. Two words are treated as similar if they can be transformed into each other by adding removing or replacing one glyph (edit distance=1).

#### a) Uniform word order

5 | 5 ol shedy qokedy ("ol.shedy.qokedy") <f75v.P2.21> <f81v.P.18> <f84r.P.6> <f84r.P.10> <f84v.P.14>

4 | 4 chey qol chedy ("chey.qol.chedy") <f81v.P.25> <f82r.P2.20> <f104v.P.4> <f111v.P.32>

4 | 4 ol s aiin ("ol.s.aiin") <f55v.P.10> <f78r.P.35> <f85r1.P.13> <f94v.P.9>

3 | 3 r ol dain ("r.ol.dain") <f32r.P.13> <f80r.P.14> <f111v.P.18>

3 | 3 shedy qokedy shedy ("**shedy**.qokedy.**shedy**") <f83v.P1.2> <f84r.P.6> <f84v.P.26>

Note: In four out of five cases there are sequences such as s aiin ("s.aiin"), ol shedy ("ol.shedy"), r ol ("r.ol") and qol chedy ("qol.chedy"). For this sequences similar words like saiin ("saiin") (144 times), olshedy ("olshedy") (23 times), rol ("rol") (20 times) and qolchedy ("qolchedy") (10 times) exist. This raises the question whether there is a transcription problem for sequences such as s aiin ("s.aiin"), ol shedy ("ol.shedy"), r ol ("r.ol") and qol chedy ("qol.chedy")? A check reveals that this may only be the case for r ol ("r.ol") in line <f32r.P.13> and for ol shedy ("ol.shedy") in line <f84r.P.6>. For otchorol dain in line <f32r.P.13> transcriptions like "otcho.rol.dain" (First study group), "otchor.ol.dain" or "otchorol.dain" are also conceivable. For olshedy qokedy in <f84r.P.6> a transcription "olshedy.qokedy" (Currier, First study group) was also proposed.

[73] When two-word sequences are included there are 270 repeated sequences using the same words. In 92 out of 270 cases the word order is uniform.

## b) Changed word order

7 | 3 ("shedy.qokedy.qokeedy") <f75v.P2.21> <f84r.P.3> <f84r.P.10>
2 ("shedy.qokeedy.qokedy") <f77r.P.13> <f78v.P.7>
1 ("qokeedy.qokedy.shedy") <f78r.P.2>
1 ("qokedy.shedy.qokeedy") <f84v.P.18>

6 | 3 ("ol.chedy.qokain") <f78r.P.16> <f80r.P.6> <f80v.P.31>
2 ("qokain.ol.chedy") <f75v.P2.18> <f84v.P.32>
1 ("chedy.qokain.ol") <f77v.P.22>

5 | 2 ("ol.qokar.shedy") <f78v.P.21> <f84v.P.24>
1 ("shedy.qokar.ol") <f75r.P.35>
1 ("qokar.ol.shedy") <f80r.P.31>
1 ("ol.shedy.qokar") <f81v.P.19>

5 | 3 ("or.aiin.ol") <f79r.P.13> <f86v4.P.3> <f86v6.P.7>
1 ("aiin.or.ol") <f54r.P.9>
1 ("or.ol.aiin") <f101v2.P.3>

4 | 3 ("sheedy.qokedy.chedy") <f76r.R.17> <f77v.P.4> <f84r.P.27>
1 ("qokedy.sheedy.chedy") <f81r.P.19>

4 | 3 ("**or.or**.aiin") <f55v.P.9> <f79v.P.39> <f85r2.P.6>
1 ("**or**.aiin.**or**") <f101v2.P.1a>

4 | 2 ("chedy.**qokeey.qokeey**") <f108v.P.45> <f112v.P.16>
1 ("**qokeey.qokeey**.chedy") <f76r.R.14>
1 ("**qokeey**.chedy.**qokeey**") <f103r.P.53>

4 | 2 ("shedy.qokaiin.chedy") <f103v.P.6> <f107v.P.35>
1 ("chedy.qokaiin.shedy") <f77r.P.8>
1 ("qokaiin.shedy.chedy") <f77r.P.26>

3 | 2 ("chol.daiin.cthy") <f3r.P.3> <f15v.P.11>
1 ("daiin.chol.cthy") <f15r.P.3>

3 | 2 ("**chol**.**chol**.daiin") <f56v.P.15> <f56v.P.16>
1 ("**chol**.daiin.**chol**") <f24r.P.18>

3 | 2 ("cthor.chol.chor") <f9v.P.2> <f15v.P.12>
1 ("chol.chor.cthor") <f16v.P.9>

3 | 2 ("**daiin**.**daiin**.dal") <f35v.P.12> <f66r.R.19>
1 ("**daiin**.dal.**daiin**") <f108r.P.16>

3 | 2 ("qokar.**shedy**.**shedy**") <f76r.R.14> <f76r.R.22>
1 ("**shedy**.qokar.**shedy**") <f75r.P.32>
3 | 1 ("qokedy.qokeey.chedy") <f77r.P.13>
1 ("qokeey.qokedy.chedy") <f83r.P.25>
1 ("qokedy.chedy.qokeey") <f111v.P.5>

3 | 2 ("**qokedy**.qokeedy.**qokedy**") <f76r.R.43> <f84r.P.3>
1 ("qokeedy.**qokedy**.**qokedy**") <f75r.P.38>

3 | 2 ("qokedy.**qokeedy**.**qokeedy**") <f84r.P.10> <f108r.P.48>
1 ("**qokeedy**.**qokeedy**.qokedy") <f75r.P.38>

3 | 2 ("**qokeey**.qokeedy.**qokeey**") <f108r.P.42> <f108v.P.13>
1 ("**qokeey**.**qokeey**.qokeedy") <f108r.P.6>

3 | 2 ("qokal.shedy.qokedy") <f77v.P.31> <f83r.P.23>
1 ("shedy.qokedy.qokal") <f103r.P.32>

3 | 2 ("qokal.chedy.qokaiin") <f77v.P.27> <f104r.P.34>
1 ("qokaiin.chedy.qokal") <f83v.P2.28>

3 | 2 ("qol.cheey.chey") <f79r.P.14> <f103r.P.50>
1 ("cheey.qol.chey") <f79r.P.6>

3 | 2 ("**shedy**.ol.**shedy**") <f81r.P.18> <f84v.P.33>
1 ("**shedy.shedy**.ol") <f34r.P.5>

3 | 2 ("shedy.qokain.shey") <f75v.P2.15> <f115r.P.27>
1 ("shey.qokain.shedy") <f103r.P.20>

3 | 2 ("shey.qokar.shedy") <f79r.P.25> <f83v.P2.24>
1 ("qokar.shedy.shey") <f106v.P.22>

3 | 2 ("ol.chedy.qol") <f75v.P2.18> <f84v.P.14>
1 ("chedy.qol.ol") <f81r.P.22>

3 | 2 ("or.aiin.otar") <f85r1.P.32> <f86v5.P.12>
1 ("otar.or.aiin") <f39v.P.5>

3 | 2 ("dar.ar.al") <f39r.P.7> <f115r.P.8>
1 ("al.dar.ar") <f105v.P.6>

3 | 2 ("cheey.chey.qokeey") <f103r.P.50> <f108r.P.15>
1 ("cheey.qokeey.chey") <f111v.P.47>

3 | 1 ("shedy.qokal.chedy") <f77v.P.5>
1 ("chedy.qokal.shedy") <f77v.P.7>
1 ("shedy.chedy.qokal") <f77v.P.8>

3 | 1 ("chol.cthol.shol") <f1v.P.6>
1 ("chol.shol.cthol") <f4r.P.2>
1 ("cthol.chol.shol") <f42r.P2.10>

3 | 1 ("chol.s.cheol") <f3r.P.16>
1 ("cheol.chol.s") <f49v.P.18>
1 ("s.cheol.chol") <f90v1.P.8>

## II. Labels

The following "labels" occur in the Astronomical section, in the Zodiac section and in the Pharmaceutical section. The list contains the findings sorted according to their occurrence on pages <f99>, <f101> and <f102>. Some similarly spelled labels, they are highlighted by underlining, occur together in different sections.

| | Astronomical section <f67r1 - f70r2> | Zodiac section <f70v2 - f73v> | Pharmaceutical section <f88r - f89v> and <f99r - f102v1> |
|---|---|---|---|
| ("okary") | | **f72v3.S2.8** \| **f73r.S2.2** | **f99r.L1.1** \| f99r.L1.9 |
| ("oky") | | **f72v3.S2.3** \| **f73r.S2.5**<br>f73v.S2.2 | **f99r.L1.3** \| f99r.L2.5 |
| ("otalam") | | f70v2.S1.3 | f99r.L1.12 |
| ("okeoly") | | f70v2.S2.14 \| f72v1.S1.4 | f99r.L2.5a |
| ("otaly") | | f70v2.S2.11 \| f72v3.S2.12<br>f73r.S1.2 \| f84r.Y.13 | f88r.t.6 \| f99v.L1.5 |
| ("otoky") | **f67r1.S.2** | | **f88r.t.5** \| f99v.L1.8 |
| ("otaldy") | **f67r1.S.1** | | **f88r.m.1** \| f101v2.R1.2 |
| ("otal") | | f72r2.S2.1 \| f73r.S2.7 | f99v.L3.1 \| f101v2.R1.3 |
| ("okol") | | f73v.S0.1 \| f82v.L3.13 | f88r.m.4 \| f101v2.R2.1 |
| ("ykeody") | **f69v.L.16** | **f73v.S1.7** | **f102v1.L1.1** |
| ("okeody") | **f69v.L.23**<br>f70r1.X.1 | f72v2.S1.14 \| f73r.S1.3<br>f73v.S1.1 \| **f73v.S1.8** | **f102v1.L1.2** |
| ("okeos") | | f73r.S2.1 \| f73r.S1.16<br>f73v.S1.14 | f102v2.L1.4 |
| ("otol") | f68r1.S.21 | f77r.X.4 | f102v2.L1.5 |
| ("otory") | f68r3.X.4 | | f88v.m.2 \| f102v2.L1.6 |
| ("okody") | f69v.L.5 | f70v2.S2.8 | f102v2.L2.1 |
| ("oran") | f67r2.Z.8 | | f102v2.L3.1 |

## III. Glyph groups occurring seven times

As a control sample glyph groups occuring seven times were used. To obtain a limited sample, glyph groups, which also appear as subgroups of other groups, are excluded. For example, chekain ("chekain") occurs seven times. But also the following similar words exists: kchekain ("kchekain"), chekaiin ("chekaiin") and chekaiiin ("chekaiiin"). In order to obtain a valid definition and to limit the number of words to be analyzed, this type of words is excluded from the sample.
The following 20 words remain: okeodar, olkeeody, qoain, qodal, qoeeey, qocthy, qocthol, qolkeedy, qoly, rshedy, sheedy, sham, shcthey, chokeey, chcthey, chcthhy, chcthedy, cthaiin, daiidy, dalam.

The result is that there are 35 consecutive pages (35/((7-1)*20)=29.2%) and 120 similar words 120/(7*20)=85.7%) with a maximum distance of three lines. With other words it is a 80 % rule for the VMS that similar words are used together on the same pages in lines near to each other. With respect to labels, all labels occurring near to each other are treated as in one line. In 62 cases (62/(7*20)=44.3%) the similar words appear together in the same line or in two consecutive lines one above the other.

The first word is ("**oteodar**"). occurs seven times: <f41v.P.4> <f70r2.C.2> <f71v.R1.1> <f72r1.R2.1> <f72r3.R2.1> <f72v2.S1.10> <f105r.P2.35>. One additional observation is that occurs on subsequent sheets (>). On page <f71v> and <f105r> two similar sequences occur and .

Consecutive pages: f70r2.C.2 > f71v.R1.1 > f72r1.R2.1 > f72r3.R2.1 > f72v2.S1.10

Similar sequences: <f71v.R1.1> ("otar.otam.**oteodar**")
<f105r.P2.35> ("(q)otar.**oteodar**.otam")

Similarities: <f41v.P.4> ("**oteodar**") | <f41v.P.5> ("oteody")
<f70r2.C.2> ("oteeodar") ... ("**oteodar**") | <f70r2.C.3> ("oteedy")
<f71v.R1.1> ("**oteodar**") ... ("okeodaly")[74]
<f72r1.R2.1> ("**oteodar**") ... ("oteorar")
<f72r3.R1.1> ("okeo*ar")[75] | <f72r3.R2.1> ("**oteodar**") ... ("oteody") ... ("okeody")
<f72v2.S1.10> ("**oteodar**") | <f72v2.S1.14> ("okeody")[76]
<f105r.P2.29> ("~~oleedar~~")[77] | <f105r.P2.32> ("okeeodair") | <f105r.P2.34> ("otar")
| <f105r.P2.35> ("otar") ... ("**oteodar**")

Word count: "otar" (F=141 times ED=3)[78]; "oteedy" (F=100 ED=3); "oteody" (F=39 ED=2); "okeody" (F=37 ED=3); "okeodar" (F=4 ED=1); "okeodaly" (F=2 ED=3); "oleedar" (F=1 ED=3); "oteeodar" (F=1 ED=1); "okeeodair" (F=1 ED=2); "oteorar" (F=1 ED=1)

Result: 4 consecutive pages; 7 similarities (max. distance 3 lines); 2 similarities (standing next to each other)

74 In line <f71v.R1.1> also a word "****eodal" occurs in the transcription of Takahashi. From the scans available today this is probably or (see http://www.jasondavies.com/voynich/#f71v_f72r1_f72r2_f72r3/0.269/0.526/6.00).

75 "okeo*ar" in line <f72r3.R1.1> is probably ("okeodar") (see http://www.jasondavies.com/voynich/#f71v_f72r1_f72r2_f72r3/0.958/0.474/6.00).

76 ("oteodar") in <f72v2.S1.10> and ("okeody") in <f72v2.S1.14> are used as labels.

77 The distance between line <f105r.P2.29> and line <f105r.P2.35> is larger then the maximum distance of three. To indicate that one criteria is not fullfiled ("~~oleedar~~") is crossed out.

78 F=frequency, ED=edit distance

("**olkeeody**") occurs twice on two consecutive sheets (>). occurs 7 times: <f51v.P.8> <f71r.R1.1> <f85r1.P.29> <f86v5.P.35> <f104r.P.17> <f105r.P2.14> <f115v.P.25>.

Consecutive pages: f85r1.P.29 > f86v5.P.35
f104r.P.17 > f105r.P2.14

Similarities: [Not counted: <f51v.P.7> ("~~qokcheody~~") | <f51v.P.8> ("**olkeeody**")]
<f71r.R1.1> ("**olkeeody**.okody") ... ("okeoky.oteody") ... ("okeokeokeody")
<f85r1.P.25> ("~~oteody~~") | <f85r1.P.28> ("okchdy") | <f85r1.P.29> ("**olkeeody**")
<f86v5.P.31> ("~~olkeedy~~") | <f86v5.P.34> ("olkey") | <f86v5.P.35> ("**olkeeody**")
<f104r.P.16> ("ykeeodey") | <f104r.P.17> ("**olkeeody**") | <f104r.P.19> ("olkchedy")
<f105r.P2.13> ("keeody") ... ("oekeody") | <f105r.P2.14> ("okeedy.**olkeeody**")
<f115v.P.22> ("qokeedy") | <f115v.P.25> ("**olkeeody**") | <f115v.P.26> ("qolkeedy") |

Word count: "qokeedy" (F=305 times ED=3); "okeedy" (F=105 times ED=2); "lkeedy" (F=41 ED=2); "oteody" (F=39 ED=3); "okchdy" (F=21 ED=3);"okeeody" (F=16 ED=1); "okody" (F=16 ED=3); "olkey" (F=12 ED=3); "yteeody" (F=9 ED=3); "keeody" (F=8 ED=2); "qolkeedy" (F=7 ED=2); "olkchedy" (F=6 ED=2); "lkeody" (F=4 ED=2); "qokcheody" (F=3 ED=4); "oekeody" (F=2 ED=3); "ykeeodey" (F=1 ED=3)

Result: 2 consecutive pages; 6 similarities (max. distance 3 lines); 3 similarities (standing next to each other)

---

("**qoain**") occurs 7 times: <f17v.P.17> <f50v.P.4> <f103v.P.38> <f111r.P.22> <f112r.P.23> <f114r.P1.30> <f116r.P.12>
Consecutive pages: f111r.P.22 > f112r.P.23

Similarities: <f17v.P.12> ("~~koaiin~~") | <f17v.P.17> ("**qoain**") | <f17v.P.19> ("qoaiin")
<f50v.P.3> ("okain") | <f50v.P.4> ("**qoain**.olaiin") | <f50v.P.5> ("qodaiin")
<f103v.P.38> ("**qoain**") ... ("qokain") | <f103v.P.39> ("qoar")
<f111r.P.19> ("oaiin") | <f111r.P.22> ("**qoain**") | <f111r.P.23> ("okain")
<f112r.P.20> ("qoaiin") | <f112r.P.23> ("**qoain**.qoiin")
<f114r.P1.30> ("qoaiin") ... ("**qoain**.ain") | <f114r.P1.31> ("okaiin") | <f114r.P1.32> ("qokaiin")
<f116r.P.11> ("aiin") ... ("qotain") | <f116r.P.12> ("**qoain**") | <f116r.P.13> ("qokain")

Word count: "aiin" (F=469 ED=3); "qokain" (F=279 times ED=1); "qokaiin" (F=262 ED=2); "okaiin" (F=212 ED=3); "otaiin" (F=154 ED=3); "okain" (F=144 ED=2); "ain" (F=89 ED=2); "qotain" (F=64 ED=1); "olaiin" (F=52 ED=3); "qodaiin" (F=42 ED=2); "oaiin" (F=26 ED=2); "qoaiin" (F=23 ED=1); "qoar" (F=12 ED=1); "qodain" (F=11 ED=1); "oain" (F=11 ED=1); "koaiin" (F=3 ED=3); "qoiin" (F=3 ED=2)

Result: 1 consecutive page; 7 similarities (max. distance 3 lines); 5 similarities (standing next to each other)

---

qodal ("**qodal**") occurs for four consecutive sheets (>) and twice on page <f53v> (=). qodal occurs seven times: <51v.P.4> <f52r.P.6> <f53v.P.9> <f53v.P.10> <f54v.P.9> <f85r1.P.8> <f104r.P.35>.
Consecutive pages: f51v.P.4 > f52r.P.6 > f53v.P.9 = f53v.P.10 > f54v.P.9

Similarities: <f51v.P.1> qokodar ("qokodar") | <f51v.P.2> qokol ("qokol") | <f51v.P.4> qodal ("**qodal**")
<f52r.P.2> koldal ("koldal") ... dal "dal" | <f52r.P.3> qotam ("qotam") | <f52r.P.4> dar ("dar") | <f52r.P.6> qodal ("**qodal**")
<f53v.P.4> qokol ("~~qokol~~") | <53v.P.9> qodal ("**qodal**")
| <53v.P.10> qodal ("**qodal**")
<f54v.P.8> okal ("okal") | <f54v.P.9> qodal ("**qodal**") ... okal ("okal") | <f54v.P.10> qokal ("qokal")
<f85r1.P.7> qopar ("qopar") | <f85r1.P.8> qodal ("**qodal**") ... dal ("dal") | <f85r1.P.9> qoar ("qoar")
<f104r.P.33> qokal ("qokal") | <f104r.P.34> qokal ("qokal") | <f104r.P.35> qodal ("**qodal**") | <f104r.P.36> qotaly ("qotaly")

Word count: "dar" (F=318 times ED=3); "dal" (F=215 ED=2); "qokal" (F=191 ED=2); "qokar" (F=152 ED=3); "okal" (F=138 ED=3); "qokol" (F=104 ED=3); "qotam" (F=12 ED=3); "qoar" (F=12 ED=2); "qopar" (F=5 ED=3); "qotaly" (F=5 ED=3); "koldal" (F=2 ED=3); "qokodar" (F=1 ED=3)

Result: 4 consecutive pages; 7 similarities (max. distance 3 lines); 4 similarities (standing next to each other)

---

qoeeey ("**qoeeey**") occurs seven times: <f5r.P.6> <f68v2.P.3> <f86v3.P1.2> <f102v2.P2.16> <f104v.P.4> <f106v.P.13> <f112v.P.19>

Similarities: <f5r.P.6> qoeeey qoykeeey ("**qoeeey**.qoykeeey") | <f5r.P.7> qotoeey ("qotoeey")
<f68v2.P.2> qokeey ("qokeey") | <f68v2.P.3> qoeeey ("**qoeeey**")
<f86v3.P1.1> qokeedy ("qokeedy") | <f86v3.P1.2> qoeeey ("**qoeeey**") | <f86v3.P1.4> qoeedy ("qoeedy")
<f102v2.P2.15> okeey ("okeey") | <f102v2.P2.16> qoeeey qokeey ("**qoeeey**.qokeey")
<f104v.P.3> qokeedy ("qokeedy") | <f104v.P.4> chey ("chey") ... qoeeey qokeedy ("**qoeeey**.qokeedy")
<f106v.P.10> oteey ("oteey") | <f106v.P.12> qoeedy ("qoeedy") | <f106v.P.13> qoeeey ("**qoeeey**")
<f112v.P.18> chey qokeedy qokeeey ("chey.qokeedy.qokeeey") | <f112v.P.19> qoeeey ("**qoeeey**")

Word count: "chey" (F=344 times ED=3); "qokeey" (F=308 ED=2); "qokeedy" (F=305 ED=3); "okeey" (F=177 ED=3); "oteey" (F=140 ED=3); "qokeeey" (F=26 ED=1); "qoeedy" (F=20 ED=2); "oeey" (F=6 ED=2); "qoekeey" (F=2 ED=1); "qotoeey" (F=2 ED=3); "qoykeeey" (F=1 ED=2)

Result: 0 consecutive pages; 7 similarities (max. distance 3 lines); 6 similarities (standing next to each other)

---

("**qocthy**") occurs seven times: <f4v.P.4> <f13v.P.9> <f37v.P.1> <f54v.P.10> <f89r1.P1.2> <f90r1.P.7> <f107r.P.32>
Consecutive pages: f89r1.P1.2 > f90r1.P.7

Similarities: <f4v.P.2> ("cthy") | <f4v.P.4> ("qokshy.**qocthy**") ... ("cthey")
<f13v.P.8> ("otchy") | <f13v.P.9> ("**qocthy**") | <f14r.P.2> ("~~qokchy~~")
[Not counted: <f37v.P.1> ("**qocthy**") | <f37v.P.3> ("~~qokchon~~")]
<f54v.P.9> ("qockhey") | <f54v.P.10> ("**qocthy**") | <f54v.P.11> ("qockhy")
<f89r1.P1.2> ("**qocthy**") | <f89r1.P2.3> ("qokechy")
<f90r1.P.5> ("ckhy") | <f90r1.P.7> ("**qocthy**.qokcho")
<f107r.P.32> ("qockhedy") ... ("**qocthy**")

Word count: "cthy" (F=111 times ED=2); "qokchy" (F=69 ED=2); "cthey" (F=50 ED=3); "ckhy" (F=39 ED=3); "qockhy" (F=20 ED=1); "qokechy" (F=13 ED=3); "qokshy" (F=10 ED=3); "qokcho" (F=10 ED=3); "qockhedy" (F=4 ED=3); "qocphey" (F=1 ED=2)

Result: 1 consecutive pages; 6 similarity (max. distance 3 lines); 2 similarities (standing next to each other)

---

For ("**qockhol**") there is a sequence of four sheets. There are similar sequences, this time in reverse order vs. . occurs seven times: <f83r.P.12> <f88v.P2.8> <f93v.P.3> <f99v.P3.13> <f100r.P2.6> <f101r1.P.1> <f101v2.P.6>
Consecutive pages: f99v.P3.13 > f100r.P2.6 > f101r1.P.1 > f101v2.P.6

Similar sequences: <f101r1.P.1> ("qockhol.(s)hor.yteol")
<f101v2.P.6> ("yteol.chor.qockhol")

Similarities: <f83r.P.12> ("**qockhol**") ... ("qokeal")
<f88v.P1.3> ("~~qoekeol~~") | <f88v.P2.8> ("**qockhol**.okol")
<f93v.P.2> ("qotchol") | <f93v.P.3> ("**qockhol**") | <f93v.P.4> ("qokchol")
<f99v.P3.11> ("qokeeor") | <f99v.P3.11c> ("qokeol") | <f99v.P3.13> ("**qockhol**")
<f100r.P1.3> ("qokeol") | <f100r.P1.4> ("qokeeol") | <f100r.P2.5> ("cphol")
| <f100r.P2.6> ("chol.cphol.shol.shol.**qockhol**")
<f101r1.P.1> ("**qockhol**") | <f101r1.P.2> ("qokeol") | <f101r1.P.4> ("qockheol")
<f101v2.P.3> ("keeol") ... ("***opcheol") | <f101v2.P.4> ("qok.ykeol")
| <f101v2.P.5> ("qokeor.qkhol") | <f101v2.P.6> ("**qockhol**")

Word count: "qokeol" (F=52 times ED=2); "qokchol" (F=18 ED=1); "cphol" (F=15 ED=3); "qotchol" (F=13 ED=2); "qokeeol" (F=11 ED=2); "qokeeor" (F=10 ED=3); "qokeor" (F=5 ED=3); "qockheol" (F=4 ED=1); "qokeal" (F=1 ED=3); "qoekeol" (F=1 ED=1); "qkhol" (F=1 ED=2)

Result: 3 consecutive pages; 7 similarities (max. distance 3 lines); 4 similarities (standing next to each other)

---

qolkeedy ("**qolkeedy**") occurs seven times: <f76v.P.18> <f79r.P.24> <f79v.P.5> <f79v.P.11> <f83v.P2.16> <f84v.P.25> <f115v.P.26>
Consecutive pages: f79r.P.24 >= f79v.P.5 = f79v.P.11
f83v.P2.16 > f84v.P.25

Similarities: <f76v.P.17> olkeey ("olkeey") ... qokedy ("qokedy") | <f76v.P.18> qolkeedy qokedy ("**qolkeedy**.qokedy")
<f79r.P.24> qolkeey qolkeedy qokedy ("qolkeey.**qolkeedy**.qokedy") | <f79r.P.26> olteedy qotchey ("olteedy.qotchey")
<f79v.P.4> qokeedy qokeedy ("qokeedy.qokeedy") | <f79v.P.5> qolkeedy qokedy ("**qolkeedy**.qokedy")
| <f79v.P.7> qokeedy qol kedy qokeedy ("qokeedy.qol.kedy.qokeedy") | <f79v.P.9> qokeedy ("qokeedy")
| <f79v.P.10> qokeey ("qokeey") | <f79v.P.11> qolkeedy ("**qolkeedy**")
<f83v.P2.16> solkeedy qokeedy qokeedy ("solkeedy.qokeedy.qokeedy") ... qokeedy qolkeedy ("qokeedy.**qolkeedy**")
| <f83v.P2.17> qolteedy qokeedy ("qolteedy.qokeedy")
<f84v.P.25> qolkeedy ("**qolkeedy**") | <f84v.P.26> qokedy ("qokedy")
<f115v.P.25> olkeeody ("olkeeody") ... qokeedy ("qokeedy") | <f115v.P.26> qolkeedy ("**qolkeedy**")

Word count: "qokeey" (F=308 times ED=2); "qokeedy" (F=305 times ED=1); "qokedy" (F=272 ED=2); "okeedy" (F=105 ED=2);
"qoteedy" (F=74 ED=2); "olkeeody" (F=7 ED=2); "qolkeey" (F=6 ED=1); "solkeedy" (F=5 ED=2); "qol.kedy" (F=2 ED=1);
"qolteedy" (F=1 ED=1)

Result: 3 consecutive pages; 7 similarities (max. distance 3 lines); 7 similarities (standing next to each other)

---

qoly ("**qoly**") occurs seven times: <f75r.P.26> <f75v.P3.26> <f79r.P.16> <f79v.P.20> <f80v.P.4> <f80v.P.34> <f82r.P2.19>
Consecutive pages: f75r.P.26 >= f75v.P3.26
f79r.P.16 >= f79v.P.20 > f80v.P.4 = f80v.P.34

Similarities: <f75r.P.26> qoly ("**qoly**") | <f75r.P.27> qoty ("qoty")
<f75v.P2.23> qol ("qol") | <f75v.P3.26> qoly ("**qoly**") | <f75v.P3.27> qol ("qol") ... oldy oly ("oldy.oly")
<f79r.P.14> qol ("qol") | <f79r.P.15> qokl ("qokl") | <f79r.P.16> qoly ("**qoly**") | <f79r.P.19> oly ("oly")
<f79v.P.16> qol ("qol") ... oly ("oly") | <f79v.P.17> qoky ("qoky") | <f79v.P.20> oty ("oty") ... qoly ("**qoly**")
<f80v.P.1> roly ("roly") | <f80v.P.2> daly "daly" | <f80v.P.3> qol ("qol") | <f80v.P.4> qoly ("**qoly**") | <f80v.P.5> ol ("ol")
| <f80v.P.30> qolky ("qolky") | <f80v.P.32> olky ("olky") | <f80v.P.34> qoly ("**qoly**") | <f80v.P.35> qol ("qol")
<f82r.P1.17> ory ("ory") | <f82r.P2.19> qoly ("**qoly**") | <f82r.P2.20> qol ("qol") | <f82r.P2.21> qol ("qol")

Word count: "ol" (F=537 times ED=2); "qol" (F=151 ED=1); "qoky" (F=147 ED=2); "oty" (F=115 ED=3); "qoty" (F=87 ED=2);
"oly" (F=57 ED=1); "daly" (F=30 ED=3); "oldy" (F=28 ED=2); "olky" (F=22 ED=2); "ory" (F=17 ED=2); "qolky" (F=4 ED=1);
"roly" (F=3)
Result: 4 consecutive pages; 7 similarities (max. distance 3 lines); 3 similarities (standing next to each other)

---

("**rshedy**") occurs seven times: <f75r.P.40> <f78v.P.2> <f78v.P.16> <f79r.P.3> <f82r.P2.26> <f82v.P.16> <f84r.P.11>

Consecutive pages: f78v.P.2 = f78v.P.16 > f79r.P.3
f82r.P2.26 >= f82v.P.16

Similarities: <f75r.P.39> ("yshedy") | <f75r.P.40> ("**rshedy**") | <f75r.P.42> ("dshedy")
<f78v.P.2> ("olshedy") ... ("**rshedy**") | <f78v.P.3> ("qolshedy") | <f78v.P.4> ("lshdy")
| <f78v.P.16> ("lshey.**rshedy**") | <f78v.P.17> ("qol.sheedy") | <f78v.P.19> ("lchedy")
<f79r.P.2> ("lshdy") | <f79r.P.3> ("**rshedy**") ... ("rchedy")
<f82r.P2.26> ("pchedy") ... ("**rshedy**") | <f82r.P2.28> ("lchedy") ... ("lchedy.rchedy")
<f82v.P.11> ("rchedy.pchedy") | <f82v.P.16> ("**rshedy**") | <f82v.P.19> ("yshedy")
<f84r.P.9> ("lshedy") | <f84r.P.11> ("**rshedy**") | <f84r.P.13> ("pchedy")

Word count: "lchedy" (F=119 times ED=2); "sheedy" (F=84 ED=2); "lshedy" (F=42 ED=1); "pchedy" (F=34 ED=2); "dshedy" (F=36 ED=1); "dchedy" (F=27 ED=2); "olshedy" (F=23 ED=2); "lshey" (F=18 ED=2); "rchedy" (F=11 ED=1); "yshedy" (F=10 ED=2); "sshedy" (F=5 ED=1); "lshdy" (F=2 ED=2); "qolshedy" (F=2 ED=3)

Result: 3 consecutive pages; 7 similarities (max. distance 3 lines); 3 similarities (standing next to each other)

---

For ("**schedy**") two sets of consecutive sheets exists. occurs seven times: <f40v.P.11> <f77v.P.32> <f78r.P.7> <f80v.P.3> <f81v.P.7> <f83r.P.6> <f106v.P.14>

Consecutive pages: f77v.P.32 > f78r.P.7
f80v.P.3 > f81v.P.7

Similarities: <f40v.P.10> ("tchedy") | <f40v.P.11> ("**schedy**") ... ("chedy") | <f40v.P.14> ("kchedy")
<f77v.P.31> ("shedy") | <f77v.P.32> ("**schedy**") | <f77v.P.33> ("shedy") | <f77v.P.34> ("kchedy")
<f78r.P.6> ("dshedy") | <f78r.P.7> ("**schedy**") | <f78r.P.8> ("dshedy")
<f80v.P.2> ("tshedy") | <f80v.P.3> ("**schedy**")
<f81v.P.6> ("chedy") | <f81v.P.7> ("**schedy**") | <f81v.P.8> ("cphedy")
<f83r.P.6> ("**schedy**.chedchy")
<f106v.P.11> ("kshedy") | <f106v.P.13> ("kchedy") | <f106v.P.14> ("**schedy**")

Word count: "chedy" (F=501 times ED=1); "shedy" (F=426 ED=2); "dshedy" (F=36 ED=2); "pchedy" (F=34 ED=2); "tchedy" (F=33 ED=2); "kchedy" (F=22 ED=2); "cphedy" (F=8 ED=2); "tshedy" (F=8 ED=2); "kshedy" (F=6 ED=2); "chedchy" (F=1 ED=3)

Result: 2 consecutive pages; 7 similarities (max. distance 3 lines); 5 similarities (standing next to each other)

---

("**sham**") occurs twice on a single page and on two subsequent sheets. Furthermore, in five out of seven cases it is the last group within a line. occurs seven times: <f24r.P.16> <f24r.P.17> <f78v.P.25> <f85r1.P.17> <f106v.P.11> <f115v.P.38> <f116r.Q.37>

Consecutive pages: f24r.P.16 = f24r.P.17
f115v.P.38 > f116r.Q.37

Similarities: <f24r.P.14> ("char") | <f24r.P.16> ("**sham**") ... ("dam") | <f24r.P.17> ("**sham**")
<f78v.P.24> ("shol") | <f78v.P.25> ("**sham**")
<f85r1.P.17> ("shor") ... ("**sham**") | <f85r1.P.18> ("dam") | <f85r1.P.19> ("dam.lam")
<f106v.P.9> ("sheas.am") | <f106v.P.11> ("**sham**")
<f115v.P.35> ("lcham") | <f115v.P.38> ("**sham**")
<f116r.Q.37> ("cheam.**sham**") | <f116r.Q.42> ("~~*cham*~~")

Word count: "shol" (F=186 ED=2); "dam" (F=98 ED=3); "shor" (F=97 ED=2); "char" (F=72 ED=2); "cham" (F=20 ED=1); "am" (F=8 ED=2); "lam" (F=6 ED=3); "cheam" (F=5 ED=2); "lcham" (F=1 ED=2); "sheas" (F=1 ED=2)

Result: 2 consecutive pages; 7 similarities (max. distance 3 lines); 3 similarities (standing next to each other)

---

("**shcthey**") occurs seven times: <f10v.P.4> <f21r.P.9> <f31r.P.9> <f68v2.P.1> <f77r.P.3> <f83r.P.11> <f111v.P.19>

Similarities: <f10v.P.1> ("cthor") | <f10v.P.2> ("cthy") | <f10v.P.4> ("**shcthey**") | <f10v.P.6> ("shcthy")
<f21r.P.9> ("**shcthey**") | <f21r.P.10> ("ctheey") | <f21r.P.11> ("chcthy")
<f31r.P.7> ("checkhey") | <f31r.P.8> ("checthy") | <f31r.P.9> ("**shcthey**")
<f68v2.P.1> ("**shcthey**") ... ("shocthy") | <f68v2.P.2> ("shekeey") ... ("sheetey")
| <f68v2.P.3> ("shekeey")
<f77r.P.2> ("chetey") | <f77r.P.3> ("**shcthey**")
<f83r.P.11> ("**shcthey**") | <f83r.P.12> ("sheckhy") ... ("cseckhdy")
<f111v.P.17> ("chckhy") | <f111v.P.19> ("**shcthey**") | <f111v.P.20> ("shcthy")

Word count: "chckhy" (F=140 times ED=3); "cthy" (F=111 ED=3); "chcthy" (F=79 ED=2); "sheckhy" (F=35 ED=3); "shcthy" (F=31 ED=1); "checthy" (F=28 ED=3); "shocthy" (F=12 ED=2); "shckhey" (F=12 ED=1); "chcphy" (F=11 ED=3); "checkhey" (F=10 ED=2); "shekeey" (F=6 ED=2); "chetey" (F=5 ED=3); "sheetey" (F=2 ED=2); "cseckhdy" (F=1 ED=3)

Result: 0 consecutive pages; 7 similarities (max. distance 3 lines); 2 similarities (standing next to each other)

---

("**choteey**") occurs seven times: <f49v.P.21> <f68r1.P.1> <f68v2.R.2> <f70v2.R3.1> <f72r3.R1.1> <f73r.R1.1> <f108v.P.33>

Consecutive pages: f68r1.P.1 > f68v2.R.2
f72r3.R1.1 > f73r.R1.1

Similarities: [Not counted: <f49v.P.17> ("~~*choty*~~") | <f49v.P.21> ("**choteey**")]
<f68r1.P.1> ("chteey.**choteey**") | <f68r1.P.2> ("cheeteey")
<f68v2.C.1> ("chokeeey") | <f68v2.R.2> ("**choteey**") | <f68v2.R.3> ("otey")
<f70v2.R3.1> ("otey") ... ("**choteey**.choeteedy")
<f72r3.R1.1> ("okeey") ... ("shocthy.**choteey**")
<f73r.S0.2> ("chockhy") | <f73r.R1.1> ("chotchy") ... ("**choteey**.cheteey")
<f108v.P.32> ("oteey") | <f108v.P.33> ("**choteey**")

Word count: "okeey" (F=177 times ED=3); "oteey" (F=140 ED=2); "otey" (F=57 ED=3); "chockhy" (F=21 ED=2); "chokchy" (F=16 ED=2); "chotchy" (F=12 ED=1); "shocthy" (F=12 ED=3); "cheeteey" (F=4 ED=3); "chteey" (F=3 ED=1); "cheteey" (F=3 ED=1); "chokeeey" (F=2 ED=3); "ochokeey" (F=1 ED=2); "tchotchey" (F=1 ED=3); "choeteedy" (F=1 ED=2)

Result: 2 consecutive pages; 6 similarities (max. distance 3 lines); 5 similarities (standing next to each other)

---

("**chcthedy**") occurs seven times: <f34r.P.15> <f66r.R.22> <f75r.P.33> <f76v.P.5> <f104r.P.2> <f111r.P.38> <f115v.P.32>

Consecutive pages: f75r.P.33 > f76v.P.5

Similarities: <f34r.P.13> ("chckhy") | <f34r.P.15> ("**chcthedy**") | <f34r.P.16> ("cheolchcthy")
<f66r.R.22> ("**chcthedy**") | <f66r.R.23> ("~~*sheocthy*~~") | <f66r.R.24> ("chckhedy")
<f75r.P.31> ("chckhy") | <f75r.P.33> ("**chcthedy**")
[Not counted: <f76v.P.4> ("~~*shecthy*~~") | <f76v.P.5> ("**chcthedy**")]
[Not counted: <f104r.P.1> ("~~*qopchedy*~~") | <f104r.P.2> ("**chcthedy**")]
<f111r.P.37> ("chckhy") | <f111r.P.38> ("**chcthedy**") | <f111r.P.39> ("chckhedy")
<f115v.P.32> ("**chcthedy**") | <f115v.P.33> ("chcthy") | <f115v.P.35> ("checthey")

Word count: "chckhy" (F=140 times ED=3); "chcthy" (F=79 ED=2); "*qopchedy*" (F=32 ED=**4**); "*shecthy*" (F=20 ED=**4**); "chckhedy" (F=11 ED=1); "checthey" (F=4 ED=2); "cheolchcthy" (F=1 ED=3)[79]; "*sheocthy*" (F=1 ED=**5**)

Result: 1 consecutive page; 5 similarities (max. distance 3 lines); 1 similarities (standing next to each other)

---

[79] The edit distance for the whole glyph group "cheolchcthy" would be 7. The value 3 is calculated by splitting the word into two parts ("cheol") and ("chcthy"). This results in 2 for the substring "chcthy" and +1 for the splitting.

("**chcthhy**") occurs seven times: <f46r.P.7> <f70r2.P.11a> <f72v3.S2.4> <f78r.P.4> <f79v.P.39> <f104v.P.16> <f111r.P.4>

Consecutive pages: f78r.P.4 > f79v.P.39

Similarities: <f46r.P.7> ("**chcthhy**") | <f46r.P.8> ("chckhy")
<f70r2.P.9> ("shocthhy") | <f70r2.P.11a> ("**chcthhy**")
[No similarities found: <f72v3.S2.4> ("**chcthhy**")]
[Not counted: <f78r.P.3> ("~~qckhedy~~") | <f78r.P.4> ("**chcthhy**")]
[Not counted: <f79v.P.38> ("~~shckhdy~~") | <f79v.P.39> ("**chcthhy**")]
<f104v.P.15> ("chocthy") | <f104v.P.16> ("**chcthhy**")
<f111r.P.2> ("chcthy") | <f111r.P.4> ("**chcthhy**")

Word count: "chckhy" (F=140 times ED=2); "cthy" (F=111 ED=3); "chcthy" (F=79 ED=1); "checkhy" (F=47 ED=3); "chocthy" (F=18 ED=2); "chckhedy" (F=11 ED=3); "shocthhy" (F=1 ED=2)

Result: 1 consecutive page; 4 similarities (max. distance 3 lines); 0 similarities (standing next to each other)

---

("**chcthey**") occurs seven times: <f30r.P.9> <f49v.P.18> <f82r.P2.27> <f86v3.Q.14> <f111v.P.8> <f112r.P.21> <f114v.P.26>

Consecutive pages: f111v.P.8 > f112r.P.21

Similarities: <f30r.P.9> ("**chcthey**") | <f30r.P.10> ("chctho")
[Not counted: <f49v.P.7> ("~~chotchy~~") | <f49v.P.18> ("**chcthey**")]
[Not counted: <f82r.P2.27> ("**chcthey**") | <f82r.P2.31> ("~~chckhey~~")]
[Not counted: <f86v3.Q.14> ("~~qochey~~.**chcthey**")]
<f111v.P.8> ("**chcthey**") ... ("checkhy") | <f111v.P.10> ("shcthy")
<f112r.P.21> ("**chcthey**") | <f112r.P.24> ("cheeteey")
<f114v.P.26> ("**chcthey**") | <f114v.P.27> ("chokeey")

Word count: "checkhy" (F=47 times ED=3); "shcthy" (F=31 ED=2); "chckhey" (F=13 ED=2); "chotchy" (F=12 ED=3); "chokeey" (F=11 ED=3); "shckhey" (F=12 ED=2); "cheeteey" (F=4 ED=3); "chctho" (F=3 ED=2); "cheoctheey" (F=1 ED=3); "tchotchey" (F=1 ED=3)

Result: 1 consecutive page; 4 similarities (max. distance 3 lines); 2 similarities (standing next to each other)

---

("**cphaiin**") occurs seven times: <f4r.P.10> <f8r.P3.18> <f37v.P.14> <f44v.P.5> <f44v.P.10> <f52v.P.1> <f104r.P.27>

Consecutive pages: f44v.P.5 = f44v.P.10

Similarities: <f4r.P.10> ("**cphaiin**") | <f4r.P.12> ("chaiin.chaiin")
<f8r.P3.17> ("chaiin") | <f8r.P3.18> ("**cphaiin**.chaiin")
<f37v.P.12> ("sheaiin") | <f37v.P.14> ("**cphaiin**")
<f44v.P.4> ("shaiin") | <f44v.P.5> ("**cphaiin**") ... ("cthain") | <f44v.P.6> ("ytaiin")
| <f44v.P.7> ("tchaiin") | <f44v.P.8> ("cthaiin") | <f44v.P.10> ("**cphaiin**")
[Not counted: <f52v.P.1> ("**cphaiin**") | <f52v.P.2> ("~~cthodaiis~~")]
[Not counted: <f104r.P.27> ("**cphaiin**.~~daiin~~")]

Word count: "chaiin" (F=45 times ED=1); "ytaiin" (F=44 times ED=3); "shaiin" (F=20 ED=2); "choiin" (F=13 ED=2); "cthaiin" (F=13 ED=1); "sheaiin" (F=9 ED=3); "cthain" (F=4 ED=2); "tchaiin" (F=1 ED=2)

Result: 1 consecutive page; 5 similarities (max. distance 3 lines); 2 similarities (standing next to each other)

---

("**daiidy**")[80] occurs seven times: <f3v.P.2> <f42r.P2.11> <f47v.P.2> <f57r.P.9> <f68v3.P.0.1> <f81v.P.9> <f114r.P1.20>

Similarities: [Not counted: <f3v.P.2> ("**daiidy**") | <f3v.P.10> ("~~daiim~~")]
[Not counted: <f42r.P1.2> ("~~daiiry~~") | <f42r.P2.11> ("**daiidy**")]
<f47v.P.2> ("**daiidy**") ... ("daiiy") | <f47v.P.7> ("~~dair~~")
<f57r.P.7> ("daiis") | <f57r.P.9> ("**daiidy**")
[Not counted: <f68v3.P.0> ("**daiidy**") | <f68v3.O.1> ("~~dais~~")]
<f81v.P.6> ("daldy") | <f81v.P.9> ("**daiidy**")
[Not counted: <f114r.P1.19> ("~~daiin~~") | <f114r.P1.20> ("**daiidy**")]

Word count: "dair" (F=106 times ED=3); "daldy" (F=17 ED=3); "daiim" (F=5 ED=2); "daiis" (F=5 ED=2); "dais" (F=4 ED=3); "daiiry" (F=1 ED=2) | "daiiy" (F=1 ED=1)

Result: 0 consecutive pages; 3 similarities (max. distance 3 lines); 0 similarities (standing next to each other)

---

80 Apparently the glyph group ("daiidy") combines an element of the -series and an element of the -series. This two elements are ("daii") and ("dy"). Such combinations are rarely used. Nevertheless a similar example also exist for the -series and the -series in line <f25v.P.7> ("daiiol").

("**dalam**") occurs seven times: <f58r.P.38> <f65v.P.2>, <f67v2.S.1> <f70r2.P.8> <f86v6.P.4> <f107v.P.13> <f108v.P.36>

Consecutive pages: f107v.P.13 > f108v.P.36

Similarities: <f58r.P.38> ("**dalam**") | <f58r.P.39> ("aram")
[Not counted: <f65r.L.1> ("~~*alam*~~") | <f65v.P.2> ("**dalam**")]
[Not counted: <f67v2.C.2b> ("~~*odam*~~") | <f67v2.S.1> ("**dalam**")]
<f70r2.P.6> ("dal") | <f70r2.P.8> ("alal.**dalam**")
<f86v6.P.2> ("lram") | <f86v6.P.4> ("**dalam**") | <f86v6.P.8> ("dam")
<f107v.P.12> ("~~*otam*~~") | <f107v.P.13> "**dalam**" | <f107v.P.14> ("dlkal")
[Not counted: <f108v.P.34> ("~~*lchedam*~~") | <f108v.P.36> ("**dalam**") | <f108v.P.37> ("~~*otam*~~")]

Word count: "dal" (F=253 ED=2); "dam" (F=98 ED=2); "*otam*" (F=46 ED=**4**); "aram" (F=12 ED=2); "alam" (F=8 ED=1); "odam" (F=6 ED=3); "alal" (F=5 ED=2); "lram" (F=1 ED=3); "araram" (F=1 ED=3), "*lchedam*" (F=2 ED=**5**); "dlkal" (F=1 ED=3),

Result: 1 consecutive page; 4 similarities (max. distance 3 lines); 3 similarities (standing next to each other)

---

## IV. Glyph groups occurring eight times

As a second control sample, glyph groups occuring eight times were used. Glyph groups which also appear as subgroups of other groups are excluded. The sample consist of 9 words: , , , , , , , , . For this sample 19 consecutive pages (19/((8-1)*9)=30%) and 66 similarities (66/(8*9)=91,7%) can be found. In 25 cases (25/(8*9)=34,7%) the similar groups appear together in the same line or in two consecutive lines one above the other.

---

("**qokedar**") occurs 8 times: <f34v.P.4> <f39r.P.7> <f76r.R.31> <f78r.P.4> <f83v.P1.4> <f108v.P.31> <f111v.P.4> <f112v.P.15>

Consecutive pages: f111v.P.4 > f112v.P.15

Similarities: <f34v.P.4> ("**qokedar**") ... ("qoldar") | <f34v.P.5> ("qokar")
[Not counted: <f39r.P.3> ("~~qokeedar~~") | <f39r.P.7> ("**qokedar**")]
<f76r.R.28> ("~~okeyr.ar~~") | <f76r.R.31> ("**qokedar**") | <f76r.R.32> ("qokal")
<f78r.P.4> ("qotal.dol") ... ("**qokedar**") | <f78r.P.5> ("qokedy.dal")
<f83v.P1.2> ("qotedar") | <f83v.P1.4> ("**qokedar**")
<f108v.P.31> ("qokedain") | <f108v.P.31> ("**qokedar**") | <f108v.P.32> ("qokeedar")
<f111v.P.2> ("qokedam") | <f111v.P.4> ("**qokedar**") | <f111v.P.5> ("qotedal")
[Not counted: <f112v.P.15> ("**qokedar**") | <f112v.P.27> ("~~qokeedar~~") ... ("~~okedal~~")]

Word count: "qokedy" (F=271 times ED=2)[81] "qokedy.dal" (ED=3); "qokal" (F=191 ED=3); "qokar" (F=152 ED=2); "okedal" (F=7 ED=2); "qokeedar" (F=6 ED=1); "qotedar" (F=3 ED=1); "qotedal" (F=3 ED=2); "qokedam" (F=2 ED=1); "qoldar" (F=1 ED=3); "okeyr" (F=1) "*okeyr.ar*" (ED=**4**)

Result: 1 consecutive page; 6 similarities (max. distance 3 lines); 2 similarities (standing next to each other)

---

[81] F=frequency, ED=edit distance

("**qokan**") occurs 8 times: <f75r.P.9> <f78r.P.12> <f79v.P.35> <f80r.P.35> <f95v2.P.3> <f103v.P.25> <f111v.P.33> <f116r.Q.42>
Consecutive pages: f78r.P.12 > f79v.P.35 > f80r.P.35

Similarities: <f75r.P.7> ("qokar") | <f75r.P.8> ("qokain") | <f75r.P.9> ("**qokan**") | <f75r.P.10> ("qokain")
<f78r.P.11> ("qokain") | <f78r.P.12> ("**qokan**") | <f78r.P.13> ("olkain")
<f79v.P.33> ("otar") | <f79v.P.34> ("otain") | <f79v.P.35> ("**qokan**") | <f79v.P.37> ("qokal")
<f80r.P.34> ("qokain") | <f80r.P.35> ("**qokan**") ... ("qotain") | <f80r.P.36> ("qokain")
<f95v2.P.2a> ("okar") | <f95v2.P.3> ("**qokan**") | <f95v2.P.4> ("okar")
<f103v.P.24> ("olan.otan.otain.otain") | <f103v.P.25> ("**qokan**") ... ("qotain.otal")
<f111v.P.31> ("qokain") | <f111v.P.33> ("okan") ... ("**qokan**") | <f111v.P.34> ("qokain")
<f116r.Q.42> ("qokain") ... ("**qokan**") | <f116r.Q.43> ("qokam")

Word count: "qokain" (F=279 times ED=1); "qokal" (F=191 ED=1); "qokar" (F=152 ED=1); "otar" (F=141 ED=3); "okar" (F=129 ED=2);
"otain" (F=96 ED=3); "qotain" (F=64 ED=2); "olkain" (F=33 ED=3); "qokam" (F=25 ED=1); "okan" (F=5 ED=1);
"otan" (F=5 ED=2); "qotan" (F=2 ED=1); "olan" (F=1 ED=3)

Result: 2 consecutive pages; 8 similarities (max. distance 3 lines); 4 similarities (standing next to each other)

---

("**qokshey**") occurs 8 times: <f79v.P.7> <f85r1.P.18> <f85r2.P.5> <f86v6.P.33> <f86v6.P.45> <f86v3.Q.10> <f94v.P.9> <f103v.P.33>

Consecutive pages: f85r1.P.18 = f85r2.P.5 > f86v6.P.33 = f86v6.P.45 = f86v3.Q.10

Similarities: <f79v.P.3> ("~~qokshedy~~") | <f79v.P.7> ("**qokshey**") | <f79v.P.9> ("qokechey")
<f85r1.P.17> ("ofchey") | <f85r1.P.18> ("otshey.**qokshey**")
<f85r2.P.2> ("qokshedy") | <f85r2.P.5> "**qokshey**.qoseey"
<f86v6.P.30> ("qopchey") | <f86v6.P.31> ("olkchey") | <f86v6.P.33> ("**qokshey**")
| [(Not counted: <f86v6.P.45> "**qokshey**")]
<f86v3.Q.10> ("**qokshey**") | <f86v3.Q.11> ("qokchey") | <f86v3.Q.12> ("qokchdy")
<f94v.P.7> ("shey") | <f94v.P.8> ("qotchy") | <f94v.P.9> ("**qokshey**")
<f103v.P.33> ("**qokshey**") | <f103v.P.35> ("oshey") | <f103v.P.36> ("shey.qokeshe")

Word count: "qokeey" (F=308 ED=2); "shey" (F=283 ED=3); "qotchy" (F=63 ED=3); "qokchdy" (F=56 ED=3); "qokchey" (F=30 ED=1);
"qokshedy" (F=11 ED=1); "qopchey" (F=10 ED=2); "oshey" (F=7 ED=2); "otshey" (F=7 ED=2); "ofchey" (F=5 ED=3);
"olkchey" (F=4 ED=3); "qokeshe" (F=1 ED=2); "qokechey" (F=1 ED=2); "qoseey" (F=1 ED=2)

Result: 4 consecutive pages; 7 similarities (max. distance 3 lines); 3 similarities (standing next to each other)

---

("**solchedy**") occurs 8 times: <f75v.P4.31> <f76r.R.5> <f77r.P.19> <f79v.P.7> <f80r.P.6> <f83r.P.29> <f83r.P.40> <f113v.P.12>

Consecutive pages: f75v.P4.31 > f76r.R.5 > f77r.P.19
f79v.P.7 > f80r.P.6
f83r.P.29 = f83r.P.40

Similarities: <f75v.P3.34> ("olchedy") | <f75v.P4.31> ("**solchedy**.solkedy")
<f76r.R.3> ("ol.chedy") | <f76r.R.5> ("**solchedy**") ... ("sol.shedy")
<f77r.P.18> ("lchedy") | <f77r.P.19> ("**solchedy**") | <f77r.P.20> ("olchey")
<f79v.P.6> ("olshedy") | <f79v.P.7> ("**solchedy**") | <f79v.P.10> ("opchedy")
<f80r.P.5> ("olchedy") | <f80r.P.6> ("**solchedy**") ... ("ol.chedy")
<f83r.P.29> ("**solchedy**") | <f83r.P.30> ("sokeedy") | <f83r.P.33> ("lchedy")
| <f83r.P.40> ("**solchedy**.olchedy")
<f113v.P.12> ("**solchedy**") ... ("olchey") | <f113v.P.14> ("olchedy")

Word count: "lchedy" (F=119 times ED=2); "opchedy" (F=50 ED=3); "olchedy" (F=38 ED=1); "olchey" (F=29 ED=2); "olshedy" (F=23 ED=2); "ol.chedy" (F=22 ED=2); "sol.shedy" (F=6 ED=2); "solkeedy" (F=5 ED=3); "solkedy" (F=3 ED=3); "sol.chedy" (F=2 ED=1)

Result: 4 consecutive pages; 8 similarities (max. distance 3 lines); 5 similarities (standing next to each other)

---

("**sheety**") occurs 8 times: <f75r.P.35> <f75r.P.37> <f77r.P.7> <f77r.P.9> <f78r.P.9> <f78v.P.30> <f89v2.P2.5> <f103v.P.26>

Consecutive pages: f75r.P.35 = f75r.P.37
f77r.P.7 = f77r.P.9 > f78r.P.9 >= f78v.P.30

Similarities: <f75r.P.35> ("**sheety**") | <f75r.P.37> ("**sheety**")
<f77r.P.6> ("shedy") | <f77r.P.7> ("**sheety**") | <f77r.P.9> ("sheed") ... ("**sheety**")
[Not counted: <f78r.P.8> ("~~*dshedy*~~") | <f78r.P.9> ("**sheety**")]
<f78v.P.27> ("sheey") | <f78v.P.28> ("chcthy") | <f78v.P.30> ("**sheety**")
<f89v2.P2.5> ("**sheety**") | <f89v2.P2.8> ("cthy") ... ("chcthy")
<f103v.P.26> ("**sheety**") | <f103v.P.28> ("shckhy") | <f103v.P.28a> ("chcthy")

Word count: "shedy" (F=426 times ED=3); "sheey" (F=144 ED=1); "cthy" (F=111 ED=3); "chcthy" (F=79 ED=3); "shckhy" (F=60 ED=3); "*dshedy*" (F=36 ED=**4**); "sheed" (F=6 ED=3)

Result: 4 consecutive pages; 7 similarities (max. distance 3 lines); 1 similarity (standing next to each other)

---

("**chkey**") occurs 8 times: <f25v.P.7> <f48v.P.8> <f68v2.C.1> <f70r2.P.1> <f96r.P.7> <f101r1.P.2> <f102r2.P.12> <f112r.P.40>
Consecutive pages: f101r1.P.2 > f102r2.P.12

Similarities: <f25v.P.5> ("chakeey") | <f25v.P.7> ("**chkey**")
<f48v.P.2> ("chety") | <f48v.P.7> ("chey") | <f48v.P.8> ("**chkey**") | <f48v.P.9> ("cheky")
<f68v2.P.3> ("shekeey.shkeey") | <f68v2.C.1> ("**chkey**") ... ("chey")
<f70r2.C.3> ("ctheey") | <f70r2.P.1> ("**chkey**") | <f70r2.P.2> ("chotey")
<f96r.P.5> ("cthey") | <f96r.P.7> ("**chkey**")
<f101r1.P.1> ("shey") | <f101r1.P.2> ("chokeey.**chkey**.cthey")
<f102r2.P.12> ("**chkey**.cheky") | <f102r2.P.13> ("shey")
<f112r.P.40> ("**chkey**") | <f112r.P.41> ("cheey")

Word count: "chey" (F=344 times ED=1); "shey" (F=283 ED=2); "cheey" (F=174 ED=2); "cheky" (F=65 ED=1); "cthey" (F=50 ED=2); "chety" (F=25 ED=2); "ctheey" (F=13 ED=3); "chokeey" (F=11 ED=2); "shekeey" (F=6 ED=3); "shkeey" (F=3 ED=2); "chakeey" (F=1 ED=2)

Result: 1 consecutive page; 8 similarities (max. distance 3 lines); 3 similarities (standing next to each other)

---

("**chekar**") occurs 8 times: <f33r.P.6> <f34r.P.16> <f72r3.R1.1> <f75r.P.43> <f84r.P.29> <f94r.P.5> <f95r1.P.7> <f108r.P.10>
Consecutive pages: f33r.P.6 > f34r.P.16
f94r.P.5 > f95r1.P.7

Similarities: <f33r.P.5> ("shekal") | <f33r.P.6> ("**chekar**")
<f34r.P.13> ("okor") | <f34r.P.15> ("kar") | <f34r.P.16> ("**chekar**.cheky.shek")
<f72r3.R1.1> ("**chekar**.okar")
<f75r.P.43> ("**chekar**") ... ("chkar")
[Not counted: <f84r.X.2> ("~~*or.shekar*~~")[82] | <f84r.P.29> ("**chekar**.or")]
<f94r.P.5> ("**chekar**") | <f94r.P.7> ("pchedar")
<f95r1.P.7> ("**chekar**") ... ("okar") ... ("chkam")
<f108r.P.7> ("okar") ... ("char") | <f108r.P.10> ("**chekar**")

Word count: "okar" (F=129 ED=3); "cheky" (F=63 ED=2); "kar" (F=52 ED=3); "chkar" (F=12 ED=1); "pchedar" (F=11 ED=3);"*cheekaiin*" (F=2 ED=**4**)

Result: 2 consecutive pages; 7 similarities (max. distance 3 lines); 2 similarities (standing next to each other)

---

[82] There is a label ("or.shekar") in <f84r.X.2> and there is a sequence ("chekar.or") in line <f84r.P.29>.

("**ykchdy**") occurs 8 times: <f34r.P.9> <f46v.P.1> <f53r.P.6> <f66r.R.28> <f68r1.S.28> <f89r1.P2.12> <f100v.M.3> <f104r.P.43>

Similarities: <f34r.P.7> ("ykedy.qokchdy") | <f34r.P.9> ("**ykchdy**") | <f34r.P.11> ("otchdy")
<f46v.P.1> ("ytedy") ... ("yfchey") ... ("**ykchdy**") | <f46v.P.2> ("otchdy") ... ("otedy")
<f53r.P.5> ("ykody") | <f53r.P.6> ("**ykchdy**") | <f53r.P.7> ("ykchody") | <f53r.P.8> ("ykeeod")
<f66r.R.25> ("otchedy") | <f66r.R.27> ("otchdy") | <f66r.R.28> ("**ykchdy**")
<f68r1.S.18> ("ytchody")[83] | <f68r1.S.28> ("**ykchdy**")
<f89r1.P2.12> ("qokchy.**ykchdy**") | <f89r1.P2.13> ("qockhedy.yty")
<f100v.M.2> ("ykchochdy") | <f100v.M.3> ("**ykchdy**") | <f100v.M.4> ("dchdy")
<f104r.P.41> ("yteedy") | <f104r.P.43> ("**ykchdy**.pchedy")

Word count: "qokedy" (F=272 times ED=3); "otedy" (F=155 ED=3); "oteedy" (F=100 ED=3); "qokchy" (F=69 ED=3); "qokchdy" (F=56 ED=2); "pchedy" (F=34 ED=3); "ykeedy" (F=33 ED=1); "otchdy" (F=30 ED=2); "yteedy" (F=28 ED=2); "ytedy" (F=24 ED=3); "ykedy" (F=23 ED=2); "ytchey" (F=12 ED=3); "dchdy" (F=8 ED=3); "ytchody" (F=5 ED=2); "ykchody" (F=4 ED=1); "yfchey" (F=3 ED=3); "ykody" (F=2 ED=2); "ykeeod" (F=2 ED=3); "ykchochdy" (F=1 ED=3)

Result: 0 consecutive pages; 8 similarities (max. distance 3 lines); 3 similarities (standing next to each other)

---

("**dsheey**") occurs 8 times: <f8r.P2.9> <f47v.P.2> <f83r.P.5> <f84v.P.16> <f104v.P.25> <f108v.P.49> <f111v.P.39> <f116r.Q.40>

Consecutive pages: f83r.P.5 > f84v.P.16

Similarities: <f8r.P2.9> ("pcheey") ... ("**dsheey**") | <f8r.P2.10> ("cheey")
<f47v.P.2> ("**dsheey**") | <f47v.P.4> ("sheey")
<f83r.P.2> ("cheey") ... ("shey") | <f83r.P.5> ("**dsheey**")
[Not counted: <f84v.P.12> ("~~dshey~~") | <f84v.P.16> ("**dsheey**") | <f84v.P.20> ("~~oshey~~")]
<f104v.P.24> ("cheeey") | <f104v.P.25> ("**dsheey**") ... ("cheey") | <f104v.P.26> ("cheey")
<f108v.P.46> ("ycheey") | <f108v.P.47> ("sheey") | <f108v.P.49> ("**dsheey**")
<f111v.P.38> ("ycheey") | <f111v.P.39> ("**dsheey**") ... ("shey")
<f116r.Q.38> ("lshey") | <f116r.Q.39> ("chey") ... ("shey") | <f116r.Q.40> ("**dsheey**.shey")

Word count: "chey" (F=344 times ED=3); "shey" (F=283 ED=2); "cheey" (F=174 ED=2); "sheey" (F=144 ED=1); "ycheey" (F=24 ED=2); "lshey" (F=18 ED=2); "cheeo" (F=16 ED=3); "dshey" (F=14 ED=1); "dcheey" (F=13 ED=1); "cheeey" (F=9 ED=3); "oshey" (F=7 ED=3); "pcheey" (F=4 ED=3); "dcheedy" (F=4 ED=2); "dcheeoy" (F=1 ED=2)

Result: 1 consecutive page; 7 similarities (max. distance 3 lines); 2 similarities (standing next to each other)

---

[83] ("ytchody") in line <f68r1.S.18> and ("ykchdy") in line <f68r1.S.28> are labels.

# V. Grid

For the grid the transcription by Takeshi Takahashi was used [see Takahashi].

## daiin- or n-series

**[daiin]**

| | | | | | |
|---|---|---|---|---|---|
| daiin (863) | aiin (469) | dain (211) | ain ( 89) | daiiin ( 17) | aiiin ( 41) |
| daiir ( 23) | aiir ( 23) | dair (106) | air ( 74) | | |
| daiim ( 5) | aiim ( 3) | daim ( 11) | aim ( 7) | | |
| daiis ( 5) | aiis ( 3) | dais ( 4) | ais ( 1) | | |
| daiil ( 1) | aiil ( 1) | dail ( 2) | ail ( 5) | | |
| doiin ( 19) | oiin ( 33) | doin ( 4) | oin ( 4) | doiiin ( 3) | oiiin ( 10) |
| doiir ( 8) | oiir ( 3) | doir (---)[84] | oir ( 2) | | |
| daiiral ( 2) | aiiral ( 1) | dairal ( 5) | airal ( 2) | | |
| daiiral (---) | aiiral (---) | dairar ( 2) | airar ( 1) | | |
| daiirol ( 1) | aiirol ( 1) | dairol ( 2) | airol ( 1) | | |
| odaiin ( 60) | oaiin ( 26) | odain ( 18) | oain ( 11) | odaiiin ( 4) | |
| odaiir ( 2) | oaiir ( 4) | odair ( 5) | oair ( 3) | | |
| odaiil ( 1) | oaiil ( 1) | | | | |
| odaiim ( 1) | oaiim (---) | | | | |
| dodaiin ( 1) | doaiin ( 3) | | | | |
| ydaiin ( 21) | | ydain ( 5) | | | |
| ydaiir (---) | | ydair ( 2) | | | |
| ydaiil ( 1) | | ydail (---) | | | |
| saiin (144) | | sain ( 68) | | saiiin ( 1) | |
| soiin ( 21) | | soin ( 4) | | soiiin ( 6) | |
| saiir ( 6) | | sair ( 28) | | | |
| soiir ( 1) | | soir ( 1) | | | |
| saiim ( 1) | | saim ( 2) | | | |
| oraiin ( 38) | | orain ( 27) | | oraiiin ( 4) | |
| osaiin ( 8) | | osain ( 3) | | osaiiin ( 1) | |
| oraiir (---) | | orair ( 5) | | | |
| ysaiin ( 4) | | ysain (---) | | | |
| yraiin ( 1) | | yrain (---) | | | |
| raiin ( 75) | | rain ( 26) | | raiiin ( 1) | |
| raiir (---) | | rair ( 6) | | | |
| raraiin (---) | | rarain ( 6) | | raraiiin ( 1) | |
| roraiin (---) | | rorain ( 1) | | | |
| soraiin ( 7) | | sorain ( 2) | | | |
| saraiin ( 4) | | sarain ( 3) | | | |
| soaiin ( 4) | | soain ( 1) | | | |
| daiidy ( 7) | aiidy ( 2) | daidy (---)[85] | aidy (---) | daiiidy ( 1) | |
| daiiny ( 1) | aiiny ( 4) | dainy ( 1) | ainy ( 2) | daiiiny ( 1) | |
| daiiry ( 1) | aiiry ( 1) | dairy ( 4) | airy ( 1) | | |
| daiindy ( 3) | | daindy (---) | | | |
| dan ( 20) | da ( 9) | an ( 7) | | | |
| don (---) | do ( 16) | on (---) | | | |

[84] "---" stands for a glyph group which does not exist within the VMS.

[85] There is one instance of daidy ("daidy") in line <f27v.P.2>. Takahashi transcribes this glyph group as da dy ("da dy").

**[ch + daiin]**

| | | | | |
|---|---|---|---|---|
| chaiin ( 45) | chain ( 18) | chaiiin ( 1) | choiin ( 13) | |
| chodaiin ( 44) | chodain ( 9) | | pchodaiin ( 5) | pchodain ( 3) |
| chokaiin ( 17) | chokain ( 11) | | | |
| chotaiin ( 9) | chotain ( 4) | | | |
| choraiin ( 5) | chorain ( 2) | | | |
| chosaiin ( 4) | chosain (---) | | | |
| choaiin ( 7) | choain (---) | | | |
| cheaiin ( 3) | cheain ( 2) | | | |
| chedaiin ( 32) | chedain ( 19) | chedaiiin ( 2) | | qochedaiin (1) |
| chekaiin ( 8) | chekain ( 7) | chekaiiin ( 1) | | |
| chetaiin ( 3) | chetain ( 5) | | | |
| chepaiin ( 1) | chepain (---) | | | |
| pchedaiin ( 4) | pchedain ( 2) | | opchedaiin ( 4) | qopchedaiin (1) |
| kchedaiin ( 1) | kchedain (---) | | okchedaiin ( 1) | |
| tchedaiin ( 1) | tchedain (---) | | otchedain ( 1) | |
| pchdaiin ( 3) | kchdain (---) | | opchdain ( 1) | |
| kchdaiin ( 1) | | | | qokchdaiin (1) |
| tchdaiin ( 2) | | | otchdaiin ( 1) | qotchdain (1) |
| cheedaiin ( 4) | cheedain ( 2) | | | |
| cheodaiin ( 11) | cheodain ( 8) | | | |
| cheekaiin ( 2) | cheekain ( 3) | | | |
| cheokaiin ( 2) | cheokain ( 4) | | | |
| cheotaiin ( 1) | cheotain ( 1) | | | |
| cheepaiin ( 1) | cheepain (---) | | | |
| cheopaiin ( 1) | cheopain (---) | | | |
| cheoaiin ( 2) | cheoain ( 1) | | | |
| chdaiin ( 16) | chdain ( 9) | | | |
| chkaiin ( 18) | chkain ( 12) | | | |
| chtaiin ( 4) | chtain ( 3) | | | |
| chpaiin ( 1) | chpain (---) | | | |
| chfaiin ( 1) | chfain (---) | | | |
| ckhaiin ( 3) | ckhain ( 1) | ckhoiin ( 1) | chckhaiin ( 3) | |
| cthaiin ( 13) | cthain ( 4) | cthoiin ( 2) | chcthaiin ( 1) | |
| cphaiin ( 7) | cphain ( 1) | | | |
| dchaiin ( 5) | dchain ( 1) | | | |
| kchaiin ( 3) | kchain ( 1) | | | |
| tchaiin ( 3) | tchain ( 1) | | | |
| pchaiin ( 1) | pchain (---) | | | |
| shaiin ( 20) | shain ( 8) | shaiiin ( 1) | shoiin ( 6) | |
| sheaiin ( 9) | sheain ( 2) | | | |
| shoaiin ( 4) | shoain (---) | | | |
| shedaiin ( 15) | shedain ( 11) | shedaiiin ( 1) | | |
| shekaiin ( 2) | shekain ( 5) | shekaiiin ( 1) | | |
| shodaiin ( 23) | shodain ( 5) | | | |
| shokaiin ( 3) | shokain ( 1) | | | |
| shotaiin ( 2) | shotain (---) | | | |
| shosaiin ( 5) | shosain (---) | | | |
| shkaiin ( 4) | shkain ( 3) | | | |
| shdaiin ( 3) | shdain (---) | | | |
| shtaiin ( 1) | shtain (---) | | | |
| kshaiin ( 1) | kshain (---) | | | |
| tshaiin ( 1) | tshain (---) | | | |
| sheodaiin ( 5) | sheodain (---) | | | |

**[qo + daiin]**

```
qokain (279) | qokaiin (262) | gokaiiin (  2) | qolkain ( 5) | qolkaiin ( 2)
qotain ( 64) | qotaiin ( 79) | qotaiiin (  2)
qodain ( 11) | qodaiin ( 42) | qodaiiin (  2)
qorain (  4) | qoraiin (  1) | qoraiiin (---)
qokair ( 17) | qokaiir (  3)
qotair (  6) | qotaiir (---)
qodair (  3) | qodaiir (---)
 qoain (  7) |  qoaiin ( 23)
 qoair (  4) |  qoaiir (  1)
  qoin (  1) |   qoiin (  3) |   qoiiin (  4)
 okain (144) |  okaiin (212) |  okaiiin (  4)
 otain ( 96) |  otaiin (154) |  otaiiin (  1)
 opain (  2) |  opaiin ( 13) |                   qopaiin (  6)
                ofaiin (  5) |                   qofaiin (  1)
                okoiin (  9) |                   qokoiin (  6)
 otoin (  1) |  otoiin (  3) |                   qotoiin (  3)
                opoiin (  1) |                   qopoiin (  1)
                ofoiin (---) |                   qofoiin (  1)
                ykoiin (  4)
                ytoiin (  4) |   ytoiiin (  1)
 okair ( 22) |  okaiir (  6) |     okais (  1) |  okaiis (  1) | okail (  1)
 otair ( 21) |  otaiir (  4) |     otais (  4) |                 otail (  1)
 opair (  4) |  opaiir (  1) |                    opaiis (  1) | opail (  1)
 ofair (  1) |  ofaiir (---) |     ofais (  1) |  ofaiis (  1)
 ykain ( 10) |  ykaiin ( 45) |   ykaiiin (  1) |    yaiin (  6)
 ytain ( 13) |  ytaiin ( 43)
 ykair (  8) |  ykaiir (  2) |                      yaiir (  2) |  yair (  2)
 ytair (  3) |  ytaiir (  1)
 ypair (  3) |  ypaiir (---)
 yfair (  1) |  yfaiir (---)
  kain ( 48) |   kaiin ( 65) |    kaiiin (  3) |    koiin (  5)
  tain ( 16) |   taiin ( 42) |    taiiin (  1) |    toiin (  2)
  pain (---) |   paiin (  8)
  kair ( 14) |   kaiir (---)
  tair ( 13) |   taiir (---)
  pair (  2) |   paiir (  2)
 skain (  1) |  skaiin (  3)
 spain (---) |  spaiin (  1)
kodain (  1) | kodaiin (  3) |  okodaiin (  2)
todain (  2) | todaiin (  9) |  otodaiin (  5) |   teodaiin (  2)
podain (---) | podaiin (  5) |  opodaiin (  2)
tedain (---) | tedaiin (  3) |  otedaiin (  3) |  oteodaiin (  8)
             qokedain (  4) | qokedaiin (  3)
                              qotedaiin (  3) | qoteodaiin (  1)
                              qokodaiin (  1)
                              qotodaiin (  1)
podair (  2) | podaiir (  2) |                                 pchdair ( 4)
korain (  2) | koraiin (  2) |  okoraiin (  2)
torain (  1) | toraiin (---) |  otoraiin (  1)
porain (  2) | poraiin (  3) |  oporaiin (  2)
 akain (  1) |  akaiin (  1)
 atain (---) |  ataiin (  1)
 arain (  1) |  araiin ( 10)
```

**[l + daiin]**

```
  olaiin ( 52) |   olain ( 13) |   olaiiin ( 1) | cholaiin ( 4) | polaiin ( 8)
  oloiin (  5) |   oloin (  1)
  lkaiin ( 49) |   lkain ( 35) |   lkaiiin ( 1)
  ldaiin (  3) |   ldain (  2) |   ldaiiin ( 1)
  lraiin (  1) |   lrain (  1)
  ltaiin (  1) |   ltain (  2)
  lkaiir (  2) |   lkair (  4)
   laiin ( 13) |    lain (  5) |    laiiin ( 2)
   loiin (  4) |    loin (---) |    loiiin ( 1)
   roiin (  4) |    roin (---)
 olkaiin ( 31) |  olkain ( 33) | cholkaiin ( 4) | cholkain ( 2)
 oltaiin (  2) |  oltain (  2) | choltaiin ( 1) |               cheoltain ( 1)
 oldaiin (  9) |  oldain (  2) | choldaiin ( 1) |               cheoldain ( 1)
 olkaiir (  1) |  olkair (  4)
  alaiin (  4) |   alain (  4) | alaiiin ( 2)
 alkaiin (  5) |  alkain (  7) |                  chalkain ( 1)
 aldaiin (  3) |  aldain (---)
```

## ol- or l-series

```
[ol]
   ol (537) |    al (260) | dol (117) |   dal (253) | dl ( 20) | odl (  4)
   or (363) |    ar (350) | dor ( 73) |   dar (318) | dr (  1) | odr (  1)
   om ( 22) |    am ( 88) | dom (  7) |   dam ( 98) | dm (  2)
   os ( 29) |    as (  5) | dos (  1) |   das (  4) | ds (  2) | ods (  1)
   od (  5) |               dod (  1) |               dd (  1)
  sol ( 75) |   sal ( 55) | osol (---) |  osal (  3) | sl (  1)
  sor ( 57) |   sar ( 84) | osor (---) |  osar (  2)
  sos (  8) |   sas (  2) | osos (  1) |  osas (---)
  som (  1) |   sam ( 10) | osom (---) |  osam (---)

[ch + ol]
  chol (396) |    cheol (172) |  cheeol (  9) |  cholo (  3)
  chor (219) |    cheor (100) |  cheeor ( 14)
  chos ( 38) |    cheos ( 33) |  cheeos (  7)
  chom ( 15) |    cheom ( 10) |  cheeom (---)
 dchor ( 24) |   dcheor (  4) |   dchol ( 26) | dcheol (  8) | dcho ( 5) | dcheo ( 7)
 kchor ( 20) |   kcheor (  4) |   kchol ( 21) | kcheol (  5) | kcho ( 5) | kcheo ( 4)
 tchor ( 19) |   tcheor (  3) |   tchol ( 13) | tcheol (  6) | tcho ( 7) | tcheo ( 7)
 pchor ( 12) |   pcheor (  5) |   pchol (  8) | pcheol ( 11) | pcho ( 2) | pcheo ( 3)
 fchor (  3) |   fcheor (---) |   fchol (  3) | fcheol (  1) | fcho ( 1) | fcheo (--)
 ochor (  6) |   ocheor (  1) |   ochol (  5) | ocheol (  1)
 ychor ( 16) |   ycheor (  9) |   ychol ( 12) | ycheol ( 14)
 kchos (  3) |   kcheos (---) |  okchos (  1)
 tchos (  4) |   tcheos (  2) |  otchos (  4)
                 ocheos (  2) |   ochos (  1)
 chdar ( 20) |   chedar ( 30) |   chdal ( 18) | chedal ( 24) | chda ( 1) | cheda ( 1)
 chkar ( 12) |   chekar (  8) |   chkal ( 13) | chekal ( 12) | chka ( 1) | cheka (--)
 chtar (  3) |   chetar (  6) |   chtal (  6) | chetal (  1)
 chpar (  1) |   chepar (  4) |   chpal (  2) | chepal (---)
                cheedar (  5)
 chkam (  3) |   chekam (  1) |  chokam (  4)
 chtam (  1) |   chetam (  1) |  chotam (  5)
tchdar (---) |  tchedar (  2) |  tchdal (---) | tchedal (  2)
kchdar (  1) |  kchedar (  1)
kchdal (  2) |  kchedal (  2)
pchdar (  5) |  pchedar ( 11) |  pchdal (  1) | pchedal (  5) |  pchdam ( 2)
                ychedar (  4) |                 ychedal (  2)
                ochedar (  1) |                 ochedal (  1)
 choar (  3) |   cheoar (  5) |   choal (  1) |  cheoal (  2)
 choor (  2) |   cheoor (  1) |   chool (  2) |  cheool (---)
                 chodar ( 14) |  chodor (---) |  chodal (  7) |  chodol ( 2)
                 chokar (  7) |  chokor (  6) |  chokal (  9) |  chokol ( 4)
                 chotar ( 11) |  chotor (  4) |  chotal (  9) |  chotol ( 7)
                               chokeor (  1) |                 chokeol ( 5)
                               choteor (---) |                 choteol ( 3)
                               choteos (  1) |                 chokeos ( 2)
                tchodar (  2) |                 tchodal (---) | tchodol ( 1)
                pchodar (  2) |                 pchodal (  1) | pchodol ( 2)
                cheodar (  4) |                 cheodal (  7)
                cheokar (  1) |                 cheokal (  2)
                cheotar (  1) |                 cheotal ( 11)
               tcheodar (  1) |                tcheodal (  1)
               kcheodar (  1) |                kcheodal (---)
               pcheodar (  3) |                pcheodal (  1)
 chdam ( 10) |   chodam (  1)
chedam (  6) |  cheodam (  2)
 chdor (  8) |   chedor (  2) |   chdol (  2) | chedol (  6) | chdo ( 1) | chedo ( 2)
 chkor (  1) |   chekor (  1) |   chkol (  3) | chekol (  3) | chko ( 1) | cheko (--)
 chtor (  2) |   chetor (---) |   chtol (  5) |  chetol (---)
 chpor (  2) |   chepor (---) |   chpol (---) |  chepol (  1)
otchor ( 17) |  otcheor (  4) |  otchol ( 28) | otcheol (  1)
okchor ( 20) |  okcheor (  2) |  okchol ( 14) | okcheol (  1)
opchor (  6) |  opcheor (  2) |  opchol (  6) | opcheol (  7)
otchar (  6) |                   otchal (  4) |                otcham (  6)
okchar (  4) |                   okchal (  3)
opchar (  2) |                   opchal (  3)
```

```
ofchar (  2) |                   ofchal (---)
ytchor ( 13) |                   ytchol (  6)
ykchor (  5) |  ykcheor (  3) |  ykchol (  6)
ypchor (  2) |                   ypchol (  3)
yfchor (  1) |                   yfchol (---)
qotchor (14) | qotcheor (---) | qotchol ( 13)
qokchor (11) | qokcheor (  2) | qokchol ( 17) | chokchor (  1) | chokchol (  4)
qopchor ( 3) | qopcheor (---) | qopchol (  6)
qofchor ( 1) | qofcheor (---) | qofchol (  2)
qotcho ( 11) |  qotcheo (  4)
qokcho ( 10) |  qokcheo (  3)
   cho ( 68) |     cheo ( 65) |   cheeo ( 16)
   chl ( 26) |     chel (---) |   cheel (  1)
   chr (  9) |     cher (  5) |   cheer (  2)
   chs ( 18) |     ches ( 36) |   chees ( 33) | cheees (  1)
   chm (---) |     chem (  1) |   cheem (  1)
    ch (  4) |      che (  2) |    chee (  1)
  ycho (  5) |    ycheo ( 15) |  ycheeo (  8)
  ocho (  2) |    ocheo (---) |  ocheeo (---)
 schol (  5) |   scheol (  2) | scheeol (  1)
 schor (  3) |   scheor (  2)
 schos (  1) |   scheos (  1)
 schom (---) |   scheom (  1)
```

**[sh + ol]**

```
  shol (186) |    sheol (114) |  sheeol ( 14) |  sholo (  1)
  shor ( 97) |    sheor ( 51) |  sheeor (  9)
  shos ( 10) |    sheos ( 17) |  sheeos (  1)
  shom (  4) |    sheom (  4)
 dshor ( 14) |   dsheor (  3) |   dshol (  5) | dsheol (  9) | dsho ( 9) | dsheo ( 2)
 kshor (  2) |   ksheor (---) |   kshol (  3) | ksheol (  2) | ksho ( 8) | ksheo ( 4)
 tshor (  4) |   tsheor (---) |   tshol (  6) | tsheol (---) | tsho ( 5) | tsheo (--)
                                  pshol (  5)
 oshor (  2) |   osheor (---) |   oshol (  1) | osheol (  1)
 yshor (  2) |   ysheor (  4) |   yshol (  2) | ysheol (  1)
 shdar (  9) |   shedar (  7) |   shdal (  4) | shedal ( 11) | shda ( 1)
 shkar (  2) |   shekar (  2) |   shkal (  1) | shekal (  4) | shka ( 1)
                 shokar (  3) |                 shokal (  4)
 shtar (  1) |   shetar (  1) |   shtal (  3) | shetal (---)
 shkam (---) |   shekam (  2)
 shtam (---) |   shetam (  1)
tshdar (  1) |  tshedar (  2) |  tshdal (---) | tshedal (  1)
pshdar (  2) |  pshedar (  1) |  pshdal (  1)
 shoar (  3) |  shodar (---) |   shoal (---) |  shodal (  3)
sheoar (---) |  sheodar (  1) |  sheoal (  1) | sheodal (  4)
               tsheodar (  1) |               tsheodal (  1)
 shdam (  1) |   shodam (  1)
shedam (  2) |  sheodam (  1)
 shdor (  1) |   shedor (  1) |   shdol (  1) | shedol (  1) | shdo ( 1)
 shkor (---) |   shekor (  1) |   shkol (  2) | shekol (  1) | shko ( 1)
 shtor (  1) |   shetor (---) |   shtol (  1) | shetol (---)
   sho (130) |     sheo ( 47) |   sheeo (  8)
   shs (  4) |     shes ( 13) |   shees (  9) | sheees (  2)
   shl (  3) |     shel (---)
   shr (  2) |     sher (  3)
    sh ( 15) |      she ( 25) |    shee ( 13) |     cs (  4)[86]
  ysho (  3) |    ysheo (  2) |  ysheeo (  2)
 sshol (  1) |   ssheol (  1)
 sshor (  1) |   ssheor (---)
```

---

[86] Takahashi transcribes ("sh") as "cs" if the additional -stroke is at the end of the -glyph.

**[ch + al]**

```
  chal ( 48) |  cheal ( 30) |  cheeal (  2)
  char ( 72) |  chear ( 51) |  cheear (  1)
  cham ( 20) |  cheam (  5) |                 charam (  4)
  chan ( 11) |  chean (  2)
  chas (  1) |  cheas (  1)
   cha (  2) |   chea (---) |   cheea (  1)
 dchar (  4) |  dchal (  2)
 kchar (  2) |  kchal (  1)
 tchar (  4) |  tchal (  2)
 pchar (  3) |  pchal (  3)
  shar ( 34) |  shear ( 21) |  sheear (  2)
  shal ( 15) |  sheal ( 19) |  sheeal (  1)
  sham (  7) |  sheam (  2)
  shan (  5) |  shean (---)
  shas (---) |  sheas (  1)
   sha (  1) |   shea (  1)
 kshar (  4)
 tshar (  1)
 pshar (  1)
```

**[cth + ol]**

```
  cthol (60) |   cthor ( 45) |   cthom (  9) |   cthos (  1) | ctho ( 15) | cth (  5)
  ckhol (22) |   ckhor (  9) |   ckhom (---) |   ckhos (  3) | ckho (  4)
  cphol (15) |   cphor (  6) |   cphom (  1) |   cphos (---) | cpho (  2)
  cfhol ( 6) |   cfhor (---)
  cthod ( 4)
  ckhod ( 1)
 ctheol (10) |  ctheor (  6)
 ckheol ( 7) |  ckheor (  1)
 cpheol ( 3) |  cpheor (  4)
 cfheol ( 2) |  cfheor (---)
 ctholy ( 4) |  cthory (  2)
 ckholy ( 1) |  ckhory (  1)
chcthol (--) | chcthor (  1)
chckhol ( 2) | chckhor (  2) | chckhom (  1)
chcphol ( 1) | chcphor (---)
chcfhol ( 1) | chcfhor (  1)
```

**[cth + al]**

```
  cthal (  7) |   cthar ( 20) |   ctham (  1)
  ckhal (  4) |   ckhar (  3) |   ckham (  3)
  cphal (  2) |   cphar (  4)
chcthal (  2) | chcthar (  1) | chctham (  2)
chckhal (  5) | chckhar (---) | chckham (  1)
                 cthear (  1)
                 ckhear (  1)
```

**[qo + os]**

```
qokos (  1) |  qokeos (  5) | qokeeos (  4) | qokees (  8) | qokeees (  3)
qotos (  1) |  qoteos (  1) | qoteeos (  3) | qotees (  4) | qoteees (  2)
 okos (  8) |   okeos ( 14) |  okeeos (  6) |  okees ( 16) |    okes (  3)
 otos (  4) |   oteos ( 29) |  oteeos ( 10) |  otees ( 10) |    otes (  5)
                 oeos (  1) |   oeeos (  2) |   oees (  9) |   oeees (  9)
                                                  ees (  6) |    eees (  9)
 ykos (---) |   ykeos (  3) |  ykeeos (  3) |  ykees (  4)
 ytos (  1) |   yteos (  3) |  yteeos (  1) |  ytees (  1)
  kos (  3) |    keos (  1) |   keeos (  4)
  tos (  4) |    teos (  3) |   teeos (  3)
  kas (  4)
  tas (  1)
```

**[qo + ol]**

```
    qol (151) |      qo ( 29) |     qoly (  7)
    qor ( 22) |     qod (  8) |      qos (  4)
    qok (  9) |     qot (  7) |      qop (  4) |   qof (  1)
  qokol (104) |  qokeol ( 52) |  qokeeol ( 11) | qokeo (  7) | qokeeo (23) | qoko (9)
  qotol ( 47) |  qoteol ( 12) |  qoteeol (  5) | qoteo (  5) | qoteeo ( 3) | qoto (3)
  qokor ( 36) |  qokeor ( 21) |  qokeeor ( 10)
  qotor ( 29) |  qoteor (  5) |  qoteeor (---)
  qopol (  6)
  qopor (  4)
  qofol (  2)
                qockhol (  7) | qockheol (  3) | ockhol (  1)
                qockhor (  1) | qockheor (  1) | ockhor (  1)
                qocthol (  2) | qoctheol (  1) | octhol (  1)
                qocthor (  1) | qoctheor (  1) | octhor (  1)
                qockhos (  1) | qockheos (  1) |                 octhos ( 1)
   okol ( 82) |   okeol ( 66) |   okeeol ( 18) |     ok (  4)
   okor ( 34) |   okeor ( 22) |   okeeor ( 14)
   otol ( 86) |   oteol ( 42) |   oteeol (  9) |     ot (  9)
   otor ( 46) |   oteor ( 12) |   oteeor (  4)
   opol (  4)
   opor (  8)
   ofor (  2)
   ofol (  1)
   ykor ( 10) |   ykeor (  8) |   ykeeor (  4)
   ykol ( 14) |   ykeol ( 14) |   ykeeol ( 13)
   ytor ( 14) |   yteor (  3) |   yteeor (  1) |    yto (  5)
   ytol ( 12) |   yteol (  6) |   yteeol (  1)
   yfor (  1)
    kol ( 37) |    keol ( 20) |    keeol ( 13)
    kor ( 26) |    keor ( 10) |    keeor (  8)
    tol ( 48) |    teol ( 15) |    teeol (  5)
    tor ( 23) |    teor (  4) |    teeor (  1)
    pol ( 24)
    por (  8)
    fol (  3)
   odol (  2) |   odeol (---) |                  qodol (  1)
   odor (  8) |   odeor (  1) |                  qodor (  2)
   okom (  7) |   okeom (  6) |                  qokom (  2) |  qokeom ( 2)
   otom (  1) |   oteom (  2) |                  qotom (  1)
qotedor (  2) |  otedor (  1)
qotedol (  1) |  otedol (  3)
qokedor (---) |  okedor (  3) | okeodor (  1) |  ykedor (  1)
qokedol (  1) |  okedol (---)
qololal (---) |                 qokoral (  1) | qokorar (  2)
qokolol (---) |                                 qokoror (  1)
qotolol (  1)
qofolol (---) |                                                          qoforom (1)
 okolal (---) |  okolar (  2) |  okoral (  1)
 otolal (---) |                                 otorar (  1) | otolam ( 2) |otoram (1)
 opolal (---) |                                 oporar (  1)
 ofolal (---) |                                                           oforam (1)
 otolol (  1) |  otolor (  2) |                              otolom ( 1)
 ykolol (---) |  ykolor (  1)
 ypolol (  1)
```

**[qo + al]**

```
  qokal (191) |  qokeal (  4) |  qokeeal (  1) |   qokl (  9)
  qokar (152) |  qokear (  6) |  qokeear (  2)
  qotal ( 59) |  qoteal (  2) |  qoteeal (  1) |   qotl (  2)
  qotar ( 63) |  qotear (  2) |  qoteear (---)
  qopal (  2)
  qopar (  5)
   okal (138) |   okeal ( 12) |   okeeal (  2)
   okar (129) |   okear (  7) |   okeear (  4) |   okas (  4)
   otal (143) |   oteal (  6) |   oteeal (  2)
   otar (141) |   otear (  4) |   oteear (---) |   otas (  4)
   opal (  9)
   opar ( 10)
   ofal (  4)
   ofar (  4) |                                   ofas (  1)
   opam (  4)
   ykal ( 16) |   ykeal (  1) |   ykeeal (---)
   ykar ( 36) |   ykear (---) |   ykeear (  2)
   ytal ( 19) |   yteal (  1) |   yteeal (---)
   ytar ( 26) |   ytear (  1) |   yteear (---)
   ypal (  2) |    ypar (  4) |     yfal (  1)
    kal ( 23) |    keal (---) |    keeal (---)
    kar ( 52) |    kear (---) |    keear (  3)
    tal ( 20) |    teal (---) |    teeal (  1)
    tar ( 43) |    tear (  2) |    teear (  1)
    pal (  2) |     par (  5) |      far (  3)
  qokam ( 25) |    okam ( 26) |     ykam (  5)
  qotam ( 12) |    otam ( 47) |     ytam ( 13)
  qokan (  8) |    okan (  5) |     ykan (  1)
  qotan (  2) |    otan (  5) |     ytan (---)
  qodal (  7) |    odal ( 13) |     ydal (  3) |   adal (  1) |  sodal ( 5)
  qodar ( 11) |    odar ( 24) |     ydar (  2) |   adar (  1) |  sodar ( 6)
  qodam (  3) |    odam (  6) |     ydam (  1) |   adam (  2)
    kam (  9) |   olkam (  9) |     lkam (  7) |  alkam (  6)
    tam (  5) |   oltam (  1) |     ltam (  1)
    kan (  3) |   olkan (---) |     lkan (  1)
    tan (  1) |   oltan (---) |     ltan (---)
   qoar ( 12) |   qoear (  3) |   qoeear (  2) |  eear (  1) |  qeear ( 1) | qear (1)
   qoor (  8) |   qoeor (  2) |   qoeeor (  3) |  eeor (  4) |  qeeor ( 1)
   qoal (  4) |   qoeal (---) |   qoeeal (---) |                qeeal ( 1)
   qool (  4) |   qoeol (  5) |   qoeeol (  2) |  eeol (  2) |               qeol (1)
   qoos (  3) |   qoeos (---) |   qoeeos (---) |  eeos (  1)
   okod (  3) |   okeod (  6) |   okeeod (---)
    oar ( 16) |    oear (---) |    oeear (  3)
    oal (  3) |    oeal (---) |    oeeal (  2)
    oor (  3) |    oeor (  2) |    oeeor (  2)
qotedar (  3) |  otedar ( 11) |  oteodar (  7) | ytedar (  3)
qotedal (  3) |  otedal (  4) |  oteodal (  6)
qokedar (  8) |  okedar (  6) |  okeodar (  3) | ykedar (  1) | ykeodar ( 2)
qokedal (  3) |  okedal (  7) |  okeodal (  3)
qoteedar ( 3) | oteedar (  3) | oteeodar (  1)
qoteedal (--) | oteedal (  2) | oteeodal (---)
qokeedar ( 6) | okeedar (  2) | okeeodar (  1)
qokeedal ( 3) | okeedal (  3) | okeeodal (  1)
                  tedar (  1) |   teodar (  3) |   teoar (  1)
                  tedal (  1) |   teodal (  3)
                  kedar (  3) |   keodar (  2) |   keoar (  1)
                  kedal (  1) |   keodal (  1)
qokalal (---) | qokalar (  1) |                            qokalam (1) |qokaram (2)
qotalal (  1)
 okalal (  6) |  okalar (  6) |  okaral (  5) | okarar ( 1) | okalam (2) | okaram (3)
 otalal (  3) |  otalar (  3) |  otaral (  3) | otarar ( 5) | otalam (3) | otaram (4)
 opalal (  2) |  opalar (  1) |  oparal (  1) | oparar (--) | opalam (1) | oparam (1)
 ofalal (---) |  ofalar (  1) |  ofaral (  1) | ofarar (--) | ofalam (-) | ofaram (1)
 okalol (  3) |  okalor (  3) |  okarol (  2)
 otalol (---) |  otalor (  4) |                              otalom (1)
 opalol (---) |  opalor (  2)
 ytalal (---) |  ytalar (  1) |                 ytarar (  1)
```

**[l + ol]**

```
   lol ( 44) |   olol ( 18) |   alol (  9) |   dalol (  7) | dolol (  1)
   lor ( 43) |   olor ( 31) |   alor (  7) |   dalor (  8) | dolor (  2)
   rol ( 20) |   orol ( 15) |   arol ( 12) |   darol (  3) | dorol (---)
   ror ( 17) |   oror (  5) |   aror (  6) |   daror (  3) | doror (  1)
   los (  5) |   olos (  2) |   alos (  2)
   ros (---) |   oros (---) |   aros (  1)
   rom (  4) |   orom (  5) |   arom (  1) |   darom (  3)
   lom (  5) |   olom (  3) |   alom (  6) |   dalom (  2)
   oly ( 57) |    aly ( 29) |   doly (  3) |    daly ( 30) |    ly ( 14)
   ory ( 17) |    ary ( 26) |   dory (  4) |    dary ( 24) |    ry ( 13)
   olo (  5) |    alo (  3) |   dolo (  1) |    dalo (  3) |    lo ( 15)
   oro (  5) |    aro (  1) |   doro (---) |    daro (  3) |    ro ( 10)
  loly (  7) |  ololy (  1) |  aloly (  3)
  lory (  2) |  olory (---) |  alory (  1)
  roly (  3) |  oroly (  1) |                   raly (  2)
  rory (  2) |  orory (  1) |                   rary (  4)
  rosy (  1) |  orosy (---)
  lkol (  5) |  olkol (  5) |  alkol (  1)
  lkor (  4) |  olkor (  4)
 lkeol (  5) | olkeol (  7) | lkeeol (  4) | olkeeol (  1)
 lkeor (  1) | olkeor (  1) | lkeeor (  2) | olkeeor (  2)
                              lkeeos (  1) | olkeeos (  1)
  lkeo (  3) |  olkeo (---) |  lkeeo (  4) |  olkeeo (  4)
  soly (  3) |   saly (  5)
  sory (  5) |   sary (  8) |                 osary (  1) | osaro (  1)
  somy (  1) |   samy (---)
    lr ( 12) |    olr (  6) |   alr (---) |    dolr (  1) |  dalr (  2) |  dlr (  1)
    ls ( 10) |    ols ( 18) |   als (  4) |    dols (---) |  dals (  6) |  dls (  1)
    rl (  2) |    orl (---) |   arl (  4) |    dorl (  1) |  darl (---)
 lchor (  6) | lcheor (  2) | lshor (  1)
 lchol (  4) | lcheol (  8) | lshol (---) |  lsheol  (  2)
  lcho (  3) |  lcheo (  4) |  lsho (  2) |   lsheo  (  1)
olchor (  3) |olcheor (  3) |olshor (---) |  olsheor (  2)
olchol (---) |olcheol (  7) |olshol (---) |  olsheol (  4)
 olcho (  1) | olcheo (  2) | olsho (  3) |   olsheo (---)
```

**[l + al]**

```
   rar ( 21) |   orar (  7) |  arar (  7) |  darar (  3) | dorar (  2)
   ral ( 17) |   oral ( 10) |  aral ( 16) |  daral (  2) | doral (  1)
   lar (  6) |   olar (---) |  alar (  6) |  dalar (  5) | dolar (  2)
   lal (  7) |   olal (  7) |  alal (  5) |  dalal (  5)
   ram ( 14) |   oram ( 10) |  aram ( 12) |  daram (  7)
   lam (  6) |   olam (  6) |  alam (  8) |  dalam (  7)
 sorol (  1) |  soral (  3) | sarol (  4) |  saral (  6)
 solor (  1) |  solar (---) | salor (---) |  salar (  2)
 solol (  1) |  solal (  1) | salol (  2) |  salal (  2)
 soror (---) |  sorar (---) | saror (---) |  sarar (  4)
 rorol (  3) |  roral (  1) | rarol (---) |  raral (  2)
  lkar ( 30) |  olkar ( 19) | alkar (  4) | dalkar (  3)
  lkal (  5) |  olkal ( 11) | alkal ( 11) | dalkal (  1) |   lkl (  9)
  ltar (  1) |  oltar (  3)
  ltal (---) |  oltal (  2)
  ldar (  5) |  oldar (  6)
  ldal (---) |  oldal (  3)
  ldam (---) |  oldam (  6) |               doldam (  1)
 lchar (  3) | lchear (  1) | lshar (  1)
 lchal (  6) | lcheal (  2)
olchar (  2) |olchear (  2)
olchal (  1) |olcheal (---)
 olcha (  1) | olchea (---)
 chlal (  1) | cholal (  4) | chalal (  1)
 chlar (  2) | cholar (  2)
 chrar (---) | chorar (  1)
```

## chedy- or y-series

**[chedy]**

| | | | | | |
|---|---|---|---|---|---|
| chedy (501) | cheedy (59) | shedy (426) | sheedy (84) | csedy (6)[87] | |
| chey (344) | cheey (174) | shey (283) | sheey (144) | cheeey (9) | sheeey (6) |
| chy (155) | chdy (150) | shy (104) | shdy (46) | chd (7) | shd (7) |
| chchy (2) | shchy (5) | chshy (2) | shshy (1) | | |
| chody (94) | choy (13) | shody (55) | shoy (2) | cheoy (4) | sheoy (4) |
| | coy (1) | sody (5) | soy (4) | | |
| choly (15) | chaly (5) | sholy (4) | shaly (1) | cholor (5) | cholody (5) |
| chory (12) | chary (6) | shory (6) | shary (---) | choror (3) | chorody (1) |
| cheody (89) | cheeody (12) | sheody (50) | sheeody (3) | | |
| cheoly (5) | cheeoly (---) | sheoly (1) | sheeoly (1) | | |
| cheky (65) | cheeky (24) | sheky (36) | sheeky (14) | eky (1) | |
| chety (25) | cheety (25) | shety (9) | sheety (8) | ety (7) | |
| chepy (7) | cheepy (---) | shepy (1) | sheepy (---) | | |
| dchy (30) | dchdy (8) | dshy (8) | dshdy (2) | | |
| ched (23) | cheed (2) | shed (18) | sheed (6) | | |
| chod (9) | cheod (5) | shod (11) | sheod (11) | | |
| chok (5) | cheok (2) | shok (5) | sheok (---) | | |
| chot (4) | cheot (1) | shot (3) | sheot (1) | | |
| chop (2) | cheop (1) | shop (---) | sheop (---) | | |
| chof (1) | cheof (---) | shof (---) | sheof (---) | | |
| chek (3) | cheek (---) | shek (11) | sheek (3) | | |
| chet (1) | cheet (1) | shet (1) | sheet (3) | | |
| chep (4) | cheep (---) | shep (1) | sheep (1) | | |
| chef (1) | cheef (2) | shef (---) | sheef (---) | | |
| choldy (10) | cheoldy (5) | sholdy (8) | sheoldy (2) | | |
| ychey (17) | ycheey (24) | yshey (12) | ysheey (10) | yrchey (1) | |
| ochey (8) | ocheey (3) | oshey (7) | osheey (2) | | |
| ychedy (13) | ycheedy (7) | yshedy (10) | ysheedy (6) | | |
| ochedy (8) | ocheedy (1) | oshedy (3) | osheedy (1) | | |
| ychy (4) | ychdy (2) | yshy (1) | yshdy (---) | | |
| ochy (5) | ochdy (1) | oshy (---) | oshdy (1) | | |
| dchey (18) | dcheey (13) | dshey (14) | dsheey (8) | | |
| dchedy (26) | dcheedy (4) | dshedy (36) | dsheedy (4) | | |
| dchody (2) | dcheody (2) | dshody (2) | dsheody (2) | | |
| chekchy (5) | chokchy (16) | shekchy (5) | shokchy (9) | | |
| chetchy (4) | chotchy (12) | shetchy (2) | shotchy (3) | | |
| chedchy (1) | chodchy (4) | shedchy (1) | | | |
| chepchy (4) | chopchy (5) | shepchy (2) | shopchy (2) | | |
| chefchy (3) | chofchy (2) | | | | |
| chechy (1) | chochy (4) | shechy (2) | shochy (1) | | |

[87] Takahashi transcribes shedy ("shedy") as "csedy" if the additional '-stroke is at the end of the sh-glyph.

**[ch + chedy]**

```
 chckhy (140) | chckhdy ( 13) | chckhedy ( 11) |    ckhedy (  4)
 chcthy ( 79) | chcthdy (  7) | chcthedy (  7) |    cthedy ( 10)
 chcphy ( 11) | chcphdy (  2) | chcphedy (  3) |    cphedy (  8)
 chcfhy (  2) | chcfhdy (  1) | chcfhedy (---) |    cfhedy (---)
                chckhhy (  9) |                      ckhhy (  5)
                chcthhy (  7) |                      cthhy (  3)
                chcphhy (  3) |                      cphhy (  1)
                chcfhhy (  3) |                      cfhhy (---)
 shckhy ( 60) | shckhdy (  2) | shckhedy (  6) | sheckhedy (  4) | checkhedy (  1)
 shcthy ( 31) | shcthdy (  1) | shcthedy (  1) | shecthedy (  1) | checthedy (  2)
 shcphy (  2) | shcphdy (---) | shcphedy (---) | shecphedy (---) | checphedy (  1)
  ckhey (131) |   cthey ( 50) |    cphey (  6) |     cfhey (  2)
 ckheey ( 11) |  ctheey ( 13) |   cpheey (  2) |    cfheey (  1)
qockhey ( 18) | qocthey (  5) |  qocphey (  1) |   qocfhey (  1) |  qockheey (  4)
 qockhy ( 19) |  qocthy (  7) |   qocphy (  2)
 ockhey (  7) |  octhey (  6) |   ocphey (  1) |    ocfhey (  1)
chockhey(  5) |chocthey (  6)
shockhey(  3) |shocthey (  1) |                    shocfhey (  1)
  ckhoy (  1) |   cthoy (---) |    cphoy (  2) |     cfhoy (  1)
  ockhy ( 13) |   octhy ( 10) |    ocphy (  3) |     ocfhy (---)
  aikhy (  2) |   aithy (  5) |    aiphy (  1) |     aifhy (  1)[88]
   cthy (111) |   cthdy (  8) |   cthody ( 18) |   ctheody (  6) |  cty (5) | cto (4)
   ckhy ( 39) |   ckhdy (  4) |   ckhody (  4) |   ckheody (  5) |  cky (2) | cko (-)
   cphy ( 16) |   cphdy (---) |   cphody (  2) |   cpheody (  3) |  cpy (1) | cpo (-)
   cfhy (  6) |   cfhdy (  1) |   cfhody (---) |   cfheody (  3) |  cfy (-) | cfo (-)
  chckh (  3) |  chckhd (  4) |  chckhod (  1)
  chcth (  3) |  chcthd (  1) |  chcthod (  1)
checkhy ( 47) | chckhey ( 30) |  chockhy ( 21) |   checkhdy (  2) | checkhey ( 10)
checthy ( 28) | chcthey (  7) |  chocthy ( 18) |   checthdy (  1) | checthey (  4)
                chcphey (  3) |  chocphy (  3) |   checphdy (  2) | checphey (  4)
                chcfhey (  1) |  chocfhy (  1) |   checfhdy (---)
sheckhy ( 35) | shckhey ( 12) |  shockhy (  5) |   sheckhdy (  1) | sheckhey (  4)
shecthy ( 20) | shcthey (  7) |  shocthy ( 12) |   shecthdy (---) | shecthey (  1)
shecphy (  3) | shcphey (---) |  shocphy (---) |  shecphhdy (  1)
shecfhy (---) | shcfhey (---) |  shocfhy (  2) |   shecfhdy (---)
                                 choikhy (  2)[87]
                                 choiphy (  1)
                                 shoikhy (  4)
                                 shoifhy (  1)
```

[88] Takahashi transcribes 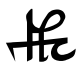 ("ckh") as "ikh", 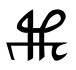 ("cth") as "ith", "cph" as "iph" and "cfh" as "ifh" if the  ("c") is drawn as a plain quill stroke .

**[qo + chedy]**

| | | | | |
|---|---|---|---|---|
| qokeedy (305) | qokedy (272) | qokeeedy ( 5) | | |
| qoteedy ( 74) | qotedy ( 91) | qoteeedy ( 3) | | |
| qodeedy ( 3) | qodedy (---) | | | |
| qokeed ( 15) | qoked ( 7) | | | |
| qoteed ( 3) | qoted ( 4) | | | |
| okeed ( 3) | oked ( 2) | | | |
| oteed ( 8) | oted ( 1) | | | |
| okeedy (105) | okedy (118) | okeeedy ( 9) | chokeedy ( 3) | chokedy ( 3) |
| oteedy (100) | otedy (155) | oteeedy ( 3) | choteedy ( 1) | chotedy ( 2) |
| keedy ( 53) | kedy ( 44) | keeedy ( 3) | chkeedy ( 2) | chkedy ( 5) |
| teedy ( 13) | tedy ( 42) | teeedy ( 4) | chteedy (---) | chtedy ( 2) |
| ekeedy (---) | ekedy ( 1) | | chekeedy ( 4) | chekedy ( 4) |
| eteedy ( 1) | etedy ( 1) | | cheteedy ( 1) | chetedy ( 3) |
| eedy ( 6) | edy (---) | eeedy ( 8) | | |
| oeedy ( 7) | oedy ( 2) | oeeedy ( 1) | | |
| deey ( 7) | dey ( 1) | deeey ( 1) | | |
| oeey ( 6) | oey (---) | oeeey ( 3) | | |
| oeeo ( 3) | oeo ( 1) | oeeeo ( 1) | | |
| ykeedy ( 30) | ykedy ( 23) | ykeeedy ( 4) | | |
| yteedy ( 28) | ytedy ( 24) | yteeedy ( 1) | | |
| qokeey (308) | qokey (107) | qokeeey ( 26) | | |
| qoteey ( 42) | qotey ( 24) | qoteeey ( 4) | | |
| qokeody ( 32) | qokody ( 9) | qokeeody ( 13) | qokoy ( 2) | |
| qoteody ( 12) | qotody ( 11) | qoteeody ( 6) | qotoy ( 2) | |
| qokeod ( 8) | qokod ( 7) | qokeeod ( 2) | | |
| qoteod (---) | qotod ( 2) | qoteeod ( 1) | | |
| qoeey ( 15) | qoey (---) | qoeeey ( 7) | qochey ( 6) | |
| qoeedy ( 20) | qoedy ( 4) | qoeeedy ( 3) | | |
| qoeody (---) | qoody ( 1) | qoeeody ( 3) | | |
| okeody ( 37) | okody ( 16) | okeeody ( 16) | chokeody ( 5) | shokeody (---) |
| oteody ( 39) | otody ( 14) | oteeody ( 11) | choteody ( 1) | shoteody (---) |
| ykeody ( 16) | ykody ( 2) | ykeeody ( 12) | chykeody (---) | shykeody ( 1) |
| yteody ( 14) | ytody ( 9) | yteeody ( 9) | chyteody ( 1) | shyteody (---) |
| oteotey ( 4) | oteoteotsho (1) | | | |
| okeoky ( 1)[89] | okeokeokeody (1) | | | |
| okeoy ( 2) | okoy ( 4) | koy (---) | ykoy (---) | |
| oteoy ( 1) | otoy ( 3) | toy ( 5) | ytoy ( 2) | |
| keody ( 21) | kody ( 7) | keeody ( 8) | chkeody ( 1) | shkeody ( 1) |
| teody ( 8) | tody ( 8) | teeody ( 4) | chteody ( 1) | shteody ( 1) |
| okeey (177) | okey ( 63) | okeeey ( 27) | chokeey ( 11) | chokey ( 7) |
| oteey (140) | otey ( 57) | oteeey ( 8) | choteey ( 7) | chotey ( 9) |
| ekeey ( 1) | ekey ( 2) | ekeeey (---) | chekeey ( 6) | chekey ( 7) |
| eteey (---) | etey (---) | eteeey (---) | cheteey ( 3) | chetey ( 5) |
| | | | shokeey ( 3) | shokey ( 5) |
| | | | | shotey ( 1) |
| | | | shekeey ( 6) | shekey ( 5) |
| | | | | shetey ( 1) |
| | | | cheekey ( 6) | sheekey ( 1) |
| | | | cheetey ( 3) | sheetey ( 2) |
| | | | cheekeey ( 1) | sheekeey ( 2) |
| | | | cheeteey ( 4) | sheeteey ( 1) |
| | | | cheokey ( 3) | sheokey ( 4) |
| | | | cheokeey ( 5) | sheokeey ( 2) |
| keey ( 44) | key ( 14) | keeey ( 11) | chkeey ( 13) | chkey ( 8) |
| teey ( 20) | tey ( 11) | teeey ( 1) | chteey ( 3) | chtey ( 1) |
| ykeey ( 58) | ykey ( 8) | ykeeey ( 6) | chykey ( 2) | |
| yteey ( 28) | ytey ( 13) | yteeey ( 3) | | |
| ykeeo ( 7) | ykeo ( 3) | | | |
| yteeo ( 3) | yteo ( 1) | | | |
| okeeo ( 15) | okeo ( 14) | okeeeo ( 2) | chokeeo ( 1) | chokeo ( 2) |
| oteeo ( 12) | oteo ( 13) | oteeeo ( 1) | | |
| keeo ( 9) | keo ( 4) | keeeo ( 1) | | |
| teeo ( 1) | teo ( 2) | teeeo (---) | | |

[89] ("oteody") occurs on page <f71r> in line <f71r.R1.1> side by side with ("okeoky") and above of ("oteotey") in line <f71r.S1.9>.

**[qo + k + dy]**

| | | | | |
|---|---|---|---|---|
| dy (270) | sy ( 35) | so ( 5) | sa ( 4) | |
| ody ( 46) | odchy ( 1) | oy ( 6) | | |
| qody ( 17) | qodchy ( 2) | qoy ( 9) | | |
| qoky (147) | qoty ( 87) | qopy ( 6) | | |
| qokdy ( 4) | qotdy (---) | qopdy ( 1) | | |
| loky (---) | loty ( 4) | lopy (---) | | |
| oky (102) | okchy ( 39) | choky ( 39) | cheoky ( 10) | cheockhy ( 10) |
| oty (115) | otchy ( 48) | choty ( 37) | cheoty ( 5) | cheocthy ( 5) |
| opy ( 7) | opchy ( 15) | chopy ( 3) | cheopy ( 1) | cheocphy ( 1) |
| ofy ( 2) | ofchy ( 3) | chofy (---) | cheofy (---) | cheocfhy (---) |
| | okshy ( 10) | shoky ( 8) | sheoky ( 6) | sheockhy ( 2) |
| | otshy ( 4) | shoty ( 8) | sheoty ( 1) | sheocthy ( 2) |
| | opshy (---) | shopy (---) | sheopy (---) | sheocphy ( 2) |
| yty ( 24) | ytchy ( 29) | chyty ( 4) | chytchy ( 1) | ytcho ( 2) |
| yky ( 18) | ykchy ( 19) | chyky ( 6) | chykchy ( 3) | ykcho ( 6) |
| | ypchy ( 4) | chypy (---) | chypchy ( 1) | ypcho (---) |
| | ytshy ( 3) | shyty (---) | shytchy ( 2) | yksho ( 1) |
| | ykshy ( 2) | shyky (---) | shykchy ( 2) | ytsho ( 2) |
| dyky ( 5) | dyty ( 2) | | | |
| ydy ( 8) | | chydy ( 2) | | |
| oko ( 8) | okcho ( 9) | choko ( 2) | chokcho ( 2) | |
| oto ( 9) | otcho ( 11) | choto ( 3) | chotcho ( 1) | |
| | oksho ( 4) | shoko ( 1) | shokcho (---) | |
| | otsho ( 2) | shoto ( 1) | shotcho ( 1) | |
| ky ( 25) | kchy ( 29) | chky ( 18) | chkchy ( 6) | yk ( 5) |
| ty ( 16) | tchy ( 24) | chty ( 13) | chtchy ( 2) | |
| py ( 2) | pchy ( 2) | chpy ( 1) | chpchy ( 4) | |
| | kshy ( 5) | shky ( 13) | chkshy ( 1) | |
| | tshy ( 5) | shty ( 4) | chtshy ( 1) | |
| | pshy ( 1) | | | |
| qokaly ( 18) | okaly ( 24) | okealy ( 1) | kaly ( 2) | chkaly ( 2) |
| qotaly ( 5) | otaly ( 19) | otealy (---) | taly ( 1) | chtaly ( 1) |
| qokary ( 1) | okary ( 11) | okeary (---) | kary ( 5) | |
| qotary ( 1) | otary ( 5) | oteary (---) | tary (---) | |
| qokoly ( 1) | okoly ( 5) | okeoly ( 6) | okeeoly ( 2) | |
| qotoly ( 1) | otoly ( 11) | oteoly ( 2) | oteeoly ( 1) | |
| qokory ( 1) | okory ( 3) | okeory ( 1) | | |
| qotory ( 1) | otory ( 4) | oteory ( 1) | | |
| | ykaly ( 6) | | | |
| | ytaly ( 5) | | | |

**[qo + k + chedy]**

| | | | | |
|---|---|---|---|---|
| qokchedy (39) | qokchdy (56) | qokshedy (11) | qokshdy ( 4) | qokcheedy ( 2) |
| qotchedy (24) | qotchdy (23) | qotshedy ( 3) | qotshdy ( 3) | qotcheedy ( 4) |
| qopchedy (32) | qopchdy (15) | qopshedy ( 4) | qopshdy (--) | qopcheedy ( 2) |
| qofchedy ( 8) | qofchdy ( 3) | | | |
| qotchody ( 3) | qokchody ( 2) | qopchody ( 1) | | |
| qotchod ( 2) | qokchod ( 4) | | | |
| qokechedy ( 4) | qokechdy ( 3) | qokeshedy ( 1) | qokeshdy ( 1) | |
| okechedy ( 4) | okechdy ( 4) | okeshedy ( 1) | | |
| otechedy ( 1) | otechdy ( 5) | | | |
| qockhedy ( 4) | qockhdy ( 1) | qockhody ( 1) | | |
| qocthedy ( 3) | qocthdy ( 1) | qocthody ( 1) | | |
| qocphedy (--) | qocphdy ( 1) | qocphody ( 1) | | |
| okchedy (25) | okchdy (21) | okshedy ( 3) | okshdy ( 1) | |
| otchedy (34) | otchdy (30) | otshedy (13) | otshdy ( 3) | otcheedy ( 1) |
| opchedy (50) | opchdy (19) | opshedy ( 2) | opshdy ( 1) | opcheedy ( 3) |
| ofchedy ( 7) | ofchdy ( 5) | ofshedy (--) | ofshdy ( 1) | ofcheedy (--) |
| otchody ( 6) | | otshody (--) | | |
| okchody ( 3) | | okshody ( 1) | | |
| opchody ( 2) | | opshody ( 1) | | |
| ockhody ( 6) | | ockhody ( 2) | | |
| octhedy ( 2) | octhdy ( 2) | octhody ( 4) | | |
| ocphedy ( 1) | | | | |
| ytchedy (10) | ytchdy (10) | ytshedy ( 2) | | |
| ykchedy ( 7) | ykchdy ( 8) | ykshedy ( 1) | | |
| ypchedy (12) | ypchdy ( 6) | ypshedy ( 1) | | ypcheedy ( 1) |
| yfchedy ( 1) | yfchdy ( 1) | yfshedy (--) | | |
| ytchody ( 5) | | | | |
| ykchody ( 4) | | | | |
| kchedy (22) | kchdy (20) | kshedy ( 6) | kshdy ( 5) | kcheedy ( 1) |
| tchedy (33) | tchdy (15) | tshedy ( 8) | tshdy ( 4) | tcheedy (--) |
| pchedy (34) | pchdy (11) | pshedy ( 3) | pshdy ( 1) | pcheedy ( 2) |
| fchedy (11) | fchdy ( 4) | fshedy ( 2) | | |
| kechdy ( 4) | | keshdy ( 1) | | |
| kchody ( 6) | kcheody ( 5) | kshody ( 4) | ksheody ( 5) | |
| tchody ( 8) | tcheody ( 6) | tshody (--) | tsheody ( 2) | |
| pchody ( 5) | pcheody ( 7) | pshody ( 1) | psheody ( 5) | |
| fchody (--) | fcheody ( 2) | fshody ( 1) | fsheody (--) | |
| okched (--) | okchd ( 6) | okshed ( 1) | okshd ( 2) | |
| qokchey (30) | qokchy (69) | qokshey ( 8) | qokshy (10) | |
| qotchey (19) | qotchy (63) | qotshey ( 3) | qotshy ( 5) | |
| qopchey (10) | qopchy (11) | | | qopcheey ( 4) |
| qofchey ( 1) | qofchy ( 2) | qofshey ( 1) | qofshy ( 1) | |
| qokechy (13) | | qokeshy ( 1) | | |
| qokeechy ( 6) | | qokeeshy ( 1) | | |
| qotechy ( 2) | | qoteshy ( 1) | | |
| okchey (32) | | okshey ( 9) | | okcheey ( 7) |
| otchey (31) | | otshey ( 7) | | otcheey ( 3) |
| opchey (29) | | opshey ( 1) | | opcheey ( 5) |
| ofchey ( 5) | | ofshey (--) | | ofcheey (--) |
| okechy ( 6) | | okeshy ( 3) | | |
| okeechy ( 1) | | okeeshy ( 2) | | |
| otechy ( 4) | | oteshy ( 1) | | |
| ykchey ( 6) | | | | |
| ytchey (12) | | | | |
| ypchey ( 5) | | | | |
| yfchey ( 3) | | | | |
| ykechy (--) | | ykeshy ( 1) | | |
| ytechy ( 1) | | | | |
| ykeechy ( 3) | | ykeeshy ( 2) | | |
| kchey (21) | | kshey ( 6) | | kcheey ( 5) |
| tchey (22) | | tshey ( 9) | | tcheey ( 6) |
| pchey (12) | | | | pcheey ( 4) |
| fchey ( 2) | | | | fcheey ( 4) |
| kechy ( 4) | | | | keechy ( 3) |
| techy ( 1) | | | | teechy (--) |
| qokched ( 1) | qokchd ( 6) | qokshed (--) | qokshd ( 3) | |
| qotched ( 5) | qotchd ( 4) | qotshed (--) | qotshd (--) | |

**[l + chedy]**

```
  lchedy (119) |  olchedy ( 38) |  qolchedy ( 10) |  alchedy (  4)
  rchedy ( 11) |  orchedy (  2)
  schedy (  7)
 lcheedy (  9) | olcheedy (  3) | qolcheedy (  2)
                 solchedy (  8) |                   salchedy (  2)
                 dolchedy (  3) |                   dalchedy (  4)
                 polchedy (  6) |                   palchedy (---)
                 tolchedy (  1) |                   talchedy (---)
                 kolchedy (  1) |                   kalchedy (  1)
   lchdy ( 18) |   olchdy (  8) |   qolchdy (  1) |   alchdy (  2) |  dalchdy (  5)
   lchey ( 45) |   olchey ( 29) |   qolchey ( 11) |   alchey (  4) |  solchey (  4)
   rchey (  9) |   orchey (  6)
   schey (  5)
  lcheey ( 13) |  olcheey ( 13) |  qolcheey (  1) |  alcheey (  1)
  rcheey (  4) |  orcheey (  2)
  scheey (  2)
  lchody (  3) |  olchody (---)
 lcheody (  3) | olcheody (  2)
    lchy (  8) |    olchy (  8) |    qolchy (  2) |    alchy (  1)
    rchy (  4)
     ldy ( 25) |     oldy ( 27) |                        aldy ( 14)
                    doldy (  2) |                       daldy ( 17)
                     dody (  7) |                        dady (  5)
      ld (  4) |      old (  3) |                         ald (  3)
   lched (  6) |   olched (  3) |                      alched (  1)
    lody (  3) |    olody (  1) |    qolody (  1) |    alody (  6)
    rody (  2) |    orody (---) |    qorody (---) |    arody ( 13)
 qokoldy (  3) |   okoldy (  8) |   qokaldy (  9) |   okaldy (  9) |  okalody (  3)
 qotoldy (  1) |   otoldy (  1) |   qotaldy (  1) |   otaldy (  7) |  otalody (  1)
                   opaldy (  1)
                   ykoldy (  1) |                      ykaldy (  1)
                   ytoldy (  5)
                    koldy (  7) |                       kaldy (  4)
                    toldy (  2)
                    poldy (  2)
 lkchedy ( 16) | olkchedy (  6) |   lkshedy (  4) | olkshedy (  1) |   olkshdy (  1)
 ltchedy (---) | oltchedy (  1) |                   oltshedy (  1)
 lpchedy (  6) | olpchedy (  4) |                   olpshedy (  1)
 lfchedy (  2) | olfchedy (  4)
  lkchdy (  5) |  olkchdy (  4)
  ltchdy (  1) |  oltchdy (  1)
  lpchdy (  2) |  olpchdy (  1)
```

**[l + shedy]**

```
  lshedy ( 42) |  olshedy (23) |  qolshedy (  2) |  alshedy (  1)
  rshedy (  7) |  orshedy (--)
  sshedy (  5)
 lsheedy (  6) | olsheedy ( 4) | qolsheedy (  2)
 rsheedy (  3) | orsheedy ( 1)
 lsheody (  3) | olsheody ( 1)
  lshody (  1)
   lshdy (  2) |   olshdy ( 4) |   qolshdy (  1) |   alshdy (  1)
  lsheey (  8) |  olsheey ( 7) |  qolsheey (  1)
   lshey ( 18) |   olshey (11) |   qolshey (  1) |   alshey (  4)
   rshey (  2)
   sshey (  6)
    lshy (  3) |    olshy ( 3) |    qolshy (  2) |    alshy (  2)
  lkeedy ( 41) |  olkeedy (42) |  qolkeedy (  7) |                solkeedy (  5)
  lteedy (  5) |  olteedy ( 3) |  qolteedy (  1) |                solteedy (  1)
  lkeody (  4) |  olkeody ( 2)
  lteody (  1) |  olteody ( 2)
                  alkeedy ( 4)
 lkeeody (  6) | olkeeody ( 7)
   lkeey ( 41) |   olkeey (40) |   qolkeey (  6) |                 solkeey (  4)
   lteey (  2) |   olteey ( 2) |   qolteey (---)
   lkedy ( 29) |   olkedy (27) |   qolkedy (  1)
   ltedy (  7) |   oltedy ( 5) |   qoltedy (  1)
    lkey (  7) |    olkey (12) |    qolkey (  1)
    ltey (---) |    oltey ( 1) |    qoltey (---)
 lkeeedy (  4) | olkeeedy ( 3)
 lteeedy (---) | olteeedy ( 1)
  lkeeey (  8) |  olkeeey ( 9)
  lkchey (  8) |  olkchey ( 4) |      lkchy (  1) |   olkchy (  4)
  ltchey (  1) |  oltchey ( 2) |      ltchy (---) |   oltchy (  2)
  lpchey (  1) |  olpchey ( 2)
     lky ( 17) |     olky (22) |      qolky (  4) |   cholky (  4)
     lty (  2) |     olty ( 1) |                     cholty (  1)
```

## VI. Statistics for graph 2

| | repeated words | compared words | same words in % |
|---|---|---|---|
| Line 0 | 2523 | 35821 | 6.8% |
| Line -1 | 2250 | 35812 | 6.3% |
| Line -2 | 2090 | 35803 | 5.8% |
| Line -3 | 1823 | 35794 | 5.1% |
| Line -4 | 1746 | 35785 | 4.9% |
| Line -5 | 1710 | 35780 | 4.8% |
| Line -6 | 1712 | 35779 | 4.8% |
| Line -7 | 1634 | 35772 | 4.6% |
| Line -8 | 1646 | 35763 | 4.6% |
| Line -9 | 1562 | 35759 | 4.4% |
| Line -10 | 1568 | 35756 | 4.4% |

**Table 1: repeated words within the previous 20 lines for each line of the VMS**

## VII. Statistics for graph 3

| | similar words | compared words | similar words in % |
|---|---|---|---|
| Line 0 | 9105 | 35821 | 25.4% |
| Line -1 | 8771 | 35812 | 24.5% |
| Line -2 | 7818 | 35803 | 21.8% |
| Line -3 | 7360 | 35794 | 20.6% |
| Line -4 | 6823 | 35785 | 19.1% |
| Line -5 | 6728 | 35780 | 18.8% |
| Line -6 | 6649 | 35779 | 18.6% |
| Line -7 | 6646 | 35772 | 18.6% |
| Line -8 | 6242 | 35763 | 17.5% |
| Line -9 | 6321 | 35759 | 17.7% |
| Line -10 | 6233 | 35756 | 17.4% |

**Table 2: similar words within the previous 20 lines for each line of the VMS**

## VIII. Statistics for graph 4

| | similar words | compared words | similar words in % |
|---|---|---|---|
| Line 0 | 921 | 63036 | 1.5% |
| Line -1 | 824 | 63028 | 1.3% |
| Line -2 | 791 | 63023 | 1.3% |
| Line -3 | 844 | 63016 | 1.3% |
| Line -4 | 851 | 63009 | 1.4% |
| Line -5 | 839 | 63001 | 1.3% |
| Line -6 | 792 | 62995 | 1.3% |
| Line -7 | 859 | 62989 | 1.4% |
| Line -8 | 811 | 62982 | 1.4% |
| Line -9 | 797 | 62975 | 1.4% |
| Line -10 | 824 | 62969 | 1.4% |

Table 3: similar words within the previous 20 lines for each line of "The Aeneid" by Virgil 80-19 BC[90]

| | similar words | compared words | similar words in % |
|---|---|---|---|
| Line 0 | 3948 | 102357 | 3.9% |
| Line -1 | 6470 | 102348 | 6.6% |
| Line -2 | 6783 | 102343 | 6.6% |
| Line -3 | 6135 | 102336 | 6.0% |
| Line -4 | 6370 | 102327 | 6.2% |
| Line -5 | 5995 | 102319 | 5.9% |
| Line -6 | 6224 | 102311 | 6.1% |
| Line -7 | 5937 | 102304 | 5.8% |
| Line -8 | 6043 | 102397 | 5.9% |
| Line -9 | 5808 | 102289 | 5.7% |
| Line -10 | 5965 | 102282 | 5.8% |

Table 4: similar words within the previous 20 lines for each line of "The Aeneid"[91]

90 The Aeneid by Virgil 80-19 BC [Virgil]

91 The Aeneid by Virgil translated by Dryden in 1697 [Dryden]

# IX. Glyph group statistics for all pages

| | Paragraph initial | Position in line: Beginning | Middle | End | Currier: A | B | ? |
|---|---|---|---|---|---|---|---|
| p/f/k/t+ | **617** | 284 | 1711 | 153 | 916 | 1617 | 232 |
| y/o/d/s+ | 55 | **2239** | 14099 | 1740 | 5778 | 9437 | 2916 |
| +m | (7)[92] | (10)[93] | 386 | **658** | 370 | 511 | 180 |
| other | 44 | 746 | 13778 | 1430 | 4374 | 10029 | 134 |

Table 5: glyph group statistics for all pages

# X. Glyph group statistics for pages in Currier A

| | Paragraph initial | Position in line: Beginning | Middle | End | Σ |
|---|---|---|---|---|---|
| p/f/k/t+ | **199** | 98 | 545 | 74 | 916 |
| y/o/d/s+ | 18 | **900** | 4048 | 812 | 5778 |
| +m | (4) | (9) | 169 | **188** | 370 |
| other | 22 | 289 | 3595 | 459 | 4374 |

Table 6: glyph group statistics for pages in Currier A

# XI. Glyph group statistics for pages in Currier B

| | Paragraph initial | Position in line: Beginning | Middle | End | Σ |
|---|---|---|---|---|---|
| p/f/k/t+ | **389** | 163 | 992 | 73 | 1617 |
| y/o/d/s+ | 32 | **1212** | 7349 | 846 | 9439 |
| +m | (3) | (1) | 106 | **401** | 511 |
| other | 20 | 414 | 8686 | 909 | 10029 |

Table 7: glyph group statistics for pages in Currier B

---

92 This groups are <f3r.P.15> tsheoarom, <f3r.P.18> pcheoldom, <f23r.P.1> pydchdom, <f53r.P.1> kodam, <f103r.P.21> pcheam, <f112r.P.37> pom and <f113v.P.1> polorarom. Since they start with p, f, k or t they are also counted for p/f/k/t+.

93 This groups are <f3v.P.3> okom, <f17v.P.3> oldaim, <f24r.P.16> sham, <f31v.P.7> oltcheedam, <f58r.P.14> sharam, <f89r1.P2.7> om, <f89r1.P2.11> daim, <f93r.P.2> ycham and <f93r.P.9> shodam. Since they all start with y, o, s or d they are also counted for y/o/s/d+.

## XII. Statistics for chody ("chody") and chedy ("chedy")

| | Paragraph initial | Position in line | | | Currier | | |
|---|---|---|---|---|---|---|---|
| | | Beginning | Middle | End | A | B | ? |
| chody | - | 1 | **84** | 9 | **44** | **42** | 8 |
| k/t/ckh/cth+ | **5** | 2 | 15 | 3 | 14 | 8 | 3 |
| y/o/d/r+ | - | **8** | 24 | 2 | 20 | 6 | 8 |
| ?+chody+? | - | - | 11 | 6 | 10 | 5 | 2 |
| chedy | - | 6 | **459** | 36 | 2 | **470** | 29 |
| k/t/ckh/cth+ | **19** | 25 | 71 | 3 | - | 109 | 9 |
| y/o/d/r+ | 2 | **55** | 232 | 22 | 3 | 267 | 41 |
| ?+chedy+? | - | 16 | 265 | 34 | 2 | 300 | 13 |

Table 8: statistics for chody ("chody") and chedy ("chedy")

## XIII. Statistics for chodar ("chodar") and chedar ("chedar")

| | Paragraph initial | Position in line | | | Currier | | |
|---|---|---|---|---|---|---|---|
| | | Beginning | Middle | End | A | B | ? |
| chodar | - | 1 | **10** | 3 | **7** | **6** | 1 |
| k/t/ckh/cth+ | **2** | - | 1 | 1 | 3 | 1 | - |
| y/o/d/r+ | - | - | 5 | - | 2 | 1 | 2 |
| +m | - | - | - | - | - | - | - |
| ?+chodar+? | - | - | - | - | - | - | - |
| chedar | - | - | **30** | - | - | **27** | 3 |
| k/t/ckh/cth+ | **7** | 3 | 4 | - | - | 11 | 3 |
| y/o/d/r+ | - | **4** | 3 | 1 | - | 8 | - |
| +m | - | - | - | **1** | - | 1 | - |
| ?+chedar+? | - | - | 6 | - | - | 5 | 1 |

Table 9: statistics for chodar ("chodar") and chedar ("chedar")

## XIV. Statistics for chodal ("chodal") and chedal ("chedal")

| | Paragraph initial | Position in line | | | Currier | | |
|---|---|---|---|---|---|---|---|
| | | Beginning | Middle | End | A | B | ? |
| chodal | – | – | 6 | 1 | 1 | **4** | 2 |
| f/p/k/t+ | – | – | – | 1 | 1 | – | – |
| y/o/d/s+ | – | – | 2 | – | – | – | 2 |
| ?+chodal+? | – | – | 8 | 1 | **6** | 1 | 2 |
| chedal | – | 1 | **22** | 1 | 1 | **21** | 2 |
| f/p/k/t+ | **4** | 2 | 1 | – | – | 7 | – |
| y/o/d/s+ | – | **2** | 6 | 1 | – | 7 | 2 |
| ?+chedal+? | – | – | 6 | 1 | – | 4 | 3 |

Table 10: statistics for chodal ("chodal") and chedal ("chedal")

## XV. Statistics for qokeody ("qokeody") and qokeedy ("qokeedy")

| | Paragraph initial | Position in line | | | Currier | | |
|---|---|---|---|---|---|---|---|
| | | Beginning | Middle | End | A | B | ? |
| qokeody | – | 2 | 30 | – | **15** | **13** | 4 |
| qokeedy | 3 | 29 | 268 | 5 | – | **301** | 4 |

Table 11: statistics for qokeody ("qokeody") and qokeedy ("qokeedy")

## XVI. Statistics for od ("od") and ed ("ed")

| | Currier | | |
|---|---|---|---|
| | A | B | ? |
| ?+od+? | **940** | **906** | 449 |
| ?+ed+? | 36 | **4603** | 362 |

Table 12: statistics for od ("od") and ed ("ed")

## XVII. Statistics for dam ("dam")

| | Paragraph initial | Position in line: Beginning | Middle | End | Currier: A | B | ? |
|---|---|---|---|---|---|---|---|
| dam | - | - | 40 | **58** | **47** | **35** | 16 |
| k/t/ckh/cth+ | **1** | - | 1 | 10 | 2 | 9 | 1 |
| y/o/d/s+ | - | **2** | 19 | 31 | 17 | 26 | 9 |
| ?+dam+? | - | - | 17 | 24 | 10 | 29 | 2 |

Table 13: statistics for dam ("dam")

## XVIII. Statistics for daiin ("daiin")

| | Paragraph initial | Position in line: Beginning | Middle | End | Currier: A | B | ? |
|---|---|---|---|---|---|---|---|
| daiin | 2 | 146 | **587** | 128 | **511** | **292** | 60 |
| k/t/ckh/cth+ | **34** | 12 | 24 | 2 | 25 | 35 | 12 |
| y/o/d/s+ | - | **43** | 170 | 32 | 104 | 94 | 47 |
| ?+daiin+? | - | 13 | 184 | 16 | 63 | 128 | 22 |

Table 14: statistics for daiin ("daiin")

## XIX. Statistics for ol ("ol")

| | Paragraph initial | Position in line: Beginning | Middle | End | Currier: A | B | ? |
|---|---|---|---|---|---|---|---|
| ol | 1 | 29 | 462 | 45 | 101 | **386** | 50 |
| k/t/ckh/cth+ | **162** | 51 | 245 | 23 | 208 | 246 | 27 |
| y/o/d/s+ | 10 | **319** | 2024 | 295 | 938 | 1355 | 355 |
| +m | 2 | 2 | 21 | **38** | 26 | 24 | 13 |
| ?+ol+? | 7 | 106 | 1590 | 154 | 871 | 852 | 134 |

Table 15: statistics for ol ("ol")

## XX. Statistics for dol ("dol"), dal ("dal"), dor ("dor") and dar ("dar")

| | Paragraph initial | Position in line | | | Currier | | |
|---|---|---|---|---|---|---|---|
| | | Beginning | Middle | End | A | B | ? |
| dol | - | 16 | **97** | 5 | **64** | **49** | 5 |
| f/p/k/t+ | **7** | - | 6 | 1 | 5 | 9 | - |
| y/o/d/s+ | - | **4** | 47 | 9 | 28 | 24 | 8 |
| +m | - | - | 2 | 1 | 2 | - | 1 |
| ?+dol+? | - | - | 23 | 6 | 12 | 16 | 1 |
| dal | - | 4 | **201** | 48 | 85 | **123** | 45 |
| f/p/k/t+ | **24** | 9 | 20 | 3 | 14 | 38 | 4 |
| y/o/d/s+ | 1 | **21** | 233 | 51 | 84 | 158 | 64 |
| +m | - | - | 3 | **10** | 2 | 6 | 5 |
| ?+dal+? | - | 2 | 112 | 14 | 29 | 76 | 23 |
| dor | - | **22** | **48** | 4 | **51** | **22** | 1 |
| f/p/k/t+ | **7** | - | 2 | - | 1 | 8 | - |
| y/o/d/s+ | 3 | **5** | 41 | 6 | 31 | 15 | 9 |
| ?+dor+? | 1 | 1 | 20 | 2 | 5 | 18 | 1 |
| dar | 1 | **32** | **244** | 41 | 94 | **161** | 62 |
| f/p/k/t+ | **35** | 5 | 29 | 4 | 9 | 53 | 11 |
| y/o/d/s+ | 1 | **28** | 187 | 34 | 61 | 128 | 61 |
| +m | - | - | 4 | **13** | 5 | 10 | 2 |
| ?+dar+? | - | 3 | 139 | 14 | 26 | 111 | 19 |

Table 16: statistics for dol ("dol"), dal ("dal"), dor ("dor") and dar ("dar")

## XXI. Similar consecutive glyph groups

| position in line | $G_n = G_{n+1}$ | $G_{n+1}$ part of $G_n$ | $G_n$ part of $G_{n+1}$ | compared groups |
|---|---|---|---|---|
| n=1 | 15 (0.4%) | 100 (2.6%) | 30 (0.8%) | 3881 |
| n>1 | 275 (1.1%) | 333 (1.3%) | 407 (1.6%) | 25410 |

Table 17: similar consecutive glyph groups grouped by line position and by the type of similarity

| position in line | $L_n < L_{n+1}$ | $L_n = L_{n+1}$ | $L_n > L_{n+1}$ | compared groups |
|---|---|---|---|---|
| n=1 | 1240 | 775 | 1866 | 3881 |
| n>1 | 9907 | 5004 | 10229 | 25410 |

Table 18: similar consecutive glyph groups grouped by line position and by their length difference

| length($G_n$) | $G_n$ part of $G_{n+1}$ | $G_n = G_{n+1}$ | $G_{n+1}$ part of $G_n$ |
|---|---|---|---|
| 2 | 162 | 30 | - |
| 3 | 61 | 24 | 41 |
| 4 | 102 | 62 | 75 |
| 5 | 88 | 80 | 106 |
| 6 | 20 | 61 | 104 |
| 7 | 2 | 27 | 68 |
| 8 | 2 | 6 | 23 |
| 9 | 0 | 0 | 10 |
| 10 | 0 | 0 | 2 |
| 11 | 0 | 0 | 1 |
| 12 | 0 | 0 | 2 |
| 13 | 0 | 0 | 0 |
| 14 | 0 | 0 | 1 |

**Table 19: sequence count grouped by the length of the first glyph group and by the type of similarity**

| length($G_{n+1}$) | $G_{n+1}$ part of $G_n$ | $G_n = G_{n+1}$ | $G_n$ part of $G_{n+1}$ |
|---|---|---|---|
| 2 | 215 | 30 | - |
| 3 | 47 | 24 | 18 |
| 4 | 88 | 62 | 64 |
| 5 | 70 | 80 | 88 |
| 6 | 11 | 61 | 130 |
| 7 | 1 | 27 | 80 |
| 8 | 1 | 6 | 36 |
| 9 | 0 | 0 | 15 |
| 10 | 0 | 0 | 4 |
| 11 | 0 | 0 | 1 |
| 12 | 0 | 0 | 1 |

**Table 20: sequence count grouped by the length of the second glyph group and by the type of similarity**

| | $G_n$ part of $G_{n+1}$ | $G_n = G_{n+1}$ | $G_{n+1}$ part of $G_n$ |
|---|---|---|---|
| avg($l_n$) | 3.4 | 4.8 | 5.5 |
| avg($l_{n+1}$) | 5.9 | 4.8 | 2.9 |

**Table 21: average glyph group length for consecutive groups which are similar to each other**

| first word (n=1) prefix | count | in line (n>1) prefix | count |
|---|---|---|---|
| y | 8 | q | 42 |
| d | 6 | ch | 20 |
| s | 6 | o | 18 |
| t | 6 | ot | 14 |
| o | 4 | ok | 14 |
| q | 4 | d | 13 |
| qot | 3 | che | 10 |
| yt | 3 | t | 8 |
| qok | 3 | k | 7 |
| yk | 2 | l | 7 |

Table 22: most common prefixes for $G_n$ which are missing for $G_{n+1}$

| first word (n=1) prefix | count | in line (n>1) prefix | count |
|---|---|---|---|
| ch | 5 | q | 42 |
| che | 3 | ot | 17 |
| she | 3 | l | 15 |
| y | 2 | ch | 12 |
| | | ok | 12 |
| | | d | 11 |
| | | o | 11 |
| | | k | 10 |
| | | ol | 9 |
| | | che | 7 |

Table 23: most common prefixes added in $G_{n+1}$

## XXII. Distribution of glyphs before and after f, p, t and k

| before | f | p | t | k |
|---|---|---|---|---|
| space | 124 | 545 | 995 | 1158 |
| o | 158 | 579 | 3884 | 6124 |
| y | 28 | 93 | 607 | 747 |
| l | 39 | 40 | 107 | 1080 |
| e | 48 | 81 | 214 | 491 |
| ch | 80 | 223 | 988 | 935 |
| r | 7 | 32 | 104 | 215 |
| other | 21 | 37 | 64 | 191 |
| all | 505 | 1630 | 6964 | 10941 |

Table 24: distribution of glyph in front of f, p, t and k

| after | f | p | t | k |
|---|---|---|---|---|
| a | 79 | 192 | 1572 | 3018 |
| o | 64 | 245 | 793 | 815 |
| y | 32 | 71 | 494 | 763 |
| e | 4 | 5 | 1823 | 3917 |
| ch | 81 | 227 | 977 | 954 |
| ee | 193 | 755 | 999 | 1094 |
| eee | 20 | 72 | 180 | 216 |
| other | 32 | 63 | 126 | 164 |
| all | 505 | 1630 | 6964 | 10941 |

Table 25: distribution of glyphs following to f, p, t and k

# XXIII. Glyph group length distribuition

| length($G_n$) | VMS line 1 | VMS line 2 | VMS all lines |
|---|---|---|---|
| 1 | 69 | 126 | 561 |
| 2 | 307 | 432 | 2009 |
| 3 | 508 | 642 | 3178 |
| 4 | 1069 | 1363 | 6168 |
| 5 | 1537 | 1802 | 8759 |
| 6 | 1452 | 1437 | 6897 |
| 7 | 1037 | 822 | 4044 |
| 8 | 477 | 262 | 1546 |
| 9 | 219 | 88 | 562 |
| 10 | 68 | 19 | 165 |
| 11 | 13 | 9 | 40 |
| 12 | 7 | 4 | 18 |
| 13 | 3 | 2 | 6 |
| average | 5.45 | 5.00 | 5.07 |

Table 26: glyph group length distribution for all paragraphs

| length($G_n$) | Currier A line 1 | Currier A line 2 | Currier A all lines |
|---|---|---|---|
| 1 | 52 | 71 | 341 |
| 2 | 85 | 109 | 560 |
| 3 | 172 | 251 | 1261 |
| 4 | 368 | 471 | 2372 |
| 5 | 461 | 544 | 2871 |
| 6 | 354 | 349 | 1871 |
| 7 | 254 | 205 | 1093 |
| 8 | 128 | 70 | 475 |
| 9 | 59 | 16 | 177 |
| 10 | 19 | 7 | 53 |
| 11 | 7 | 1 | 11 |
| 12 | 1 | 1 | 3 |
| 13 | 1 | 1 | 2 |
| average | 5.24 | 4.76 | 4.88 |

Table 27: glyph group length distribution for paragraphs in Currier A

| length($G_n$) | Currier B line 1 | Currier B line 2 | Currier B all |
|---|---|---|---|
| 1 | 16 | 54 | 205 |
| 2 | 202 | 297 | 1300 |
| 3 | 315 | 360 | 1758 |
| 4 | 638 | 821 | 3417 |
| 5 | 1015 | 1174 | 5418 |
| 6 | 1024 | 1020 | 4640 |
| 7 | 733 | 573 | 2705 |
| 8 | 321 | 171 | 976 |
| 9 | 149 | 65 | 350 |
| 10 | 143 | 12 | 100 |
| 11 | 6 | 6 | 25 |
| 12 | 6 | 3 | 15 |
| 13 | 2 | 0 | 3 |
| average | 5.54 | 5.09 | 5.17 |

Table 28: glyph group length distribution for paragraphs in Currier B

| line | Currier A | | Currier B | | VMS | |
|---|---|---|---|---|---|---|
| | all | gallow | all | gallow | all | gallow |
| 1 | 1962 | 1088 | 4471 | 2997 | 6699 | 4290 |
| 2 | 2096 | 952 | 4556 | 2304 | 6882 | 3448 |
| 3 | 1890 | 872 | 3459 | 1728 | 5562 | 2759 |
| 4 | 1507 | 668 | 2259 | 1138 | 3934 | 1932 |
| 5 | 1039 | 456 | 1521 | 771 | 2691 | 1322 |
| 6 | 734 | 330 | 1033 | 514 | 1859 | 912 |
| 7 | 478 | 210 | 676 | 352 | 1197 | 598 |
| 8 | 333 | 151 | 509 | 244 | 888 | 428 |
| 9 | 223 | 103 | 383 | 185 | 648 | 316 |
| 10 | 169 | 83 | 285 | 138 | 500 | 242 |

Table 29: usage of gallow glyphs for paragraphs in Currier A, Currier B and the VMS

# XXIV. Similarities for paragraph initial glyph groups

| line | possible source group | possible copy | line |
|---|---|---|---|
| <f3r.P.1> | tsheos | pcheol | <f3r.P.11> |
| <f3r.P.15> | tsheoarom | pcheoldom | <f3r.P.18> |
| <f11r.P.1> | tshol | tchol | <f11r.P.5> |
| <f15v.P.1> | poror | toror | <***f16r***.P2.10> |
| <f22r.P.7> | kchol shol | kchol chor | <f22r.P.10> |
| <f22r.P.10> | kchol chor | tshol shor | <***f23v***.P.6> |
| <f26v.P.1> | pchedar | pchdar | <f26v.P.3> |
| <f28v.P1.1> | kshol | pchol | <f28v.P2.7> |
| <f36v.P.1> | pcharasy | tchar | <f36v.P.5> |
| <f49v.P.1> | kshor shol | pchorchor | <f49v.P.14> |
| <f95r1.P.6> | tchdor | kshdor | <f95r1.P.1> |
| <f104v.P.1> | pchdar | kchdal | <f104v.P.5> |
| <f104v.P.5> | kchdal | tchdol | <f104v.P.12> |
| <f104v.P.1> | pchdar | tshedar | <f104v.P.22> |
| <f104v.P.22> | tshedar | pchedar | <f104v.P.27> |
| <f111v.P.26> | polanar | polar | <f111v.P.34> |
| <f111v.P.31> | poechey | polchey | <f111v.P.37> |
| <f111v.P.37> | polchey | polchey | <***f112r***.P.1> |
| <f112v.P.11> | tchor aiin | pchodaiin | <f112v.P.15> |
| <f112v.P.11> | tchor aiin | pchoraiin | <f112v.P.20> |
| <f112v.P.15> | pchodaiin | pchdaiin | <f112v.P.25> |
| <f112v.P.25> | pchdaiin | pchodaiin | <f112v.P.27> |
| <f114r.P1.22> | tchedair | tchedaiin | <f114r.P1.24> |
| <f114r.P1.22> | tchedair | pchedairr | <f114r.P1.28> |
| <f114r.P1.24> | tchedaiin | tcheodaiin | <f114r.P1.32> |

Table 30: similarities for paragraph initial glyph groups

# XXV. Code table as described by Trithemius

| | Polygraphiae III-5 | Polygraphiae III-15 |
|---|---|---|
| a | paſa | maſtra |
| b | paſe | maſtre |
| c | paſi | maſtri |
| d | paſo | maſtro |
| e | paſu | maſtru |
| f | paſan | maſtran |
| g | paſen | maſtren |
| h | paſin | maſtrin |
| i | paſon | maſtron |
| k | paſun | maſtrun |
| l | paſas | maſtral |
| m | paſes | maſtrel |
| n | paſis | maſtril |
| o | paſos | maſtrol |
| p | paſus | maſtrul |
| q | paſal | maſtras |
| r | paſel | maſtres |
| s | paſil | maſtris |
| t | paſol | maſtros |
| u | paſul | maſtrus |
| x | paſar | maſtraſſ |
| y | paſer | maſtreſſ |
| z | paſir | maſtriſſ |
| w | paſor | maſtroſſ |

Table 31: code table as described by Trithemius [Hermes: p. 143]